\documentclass[10pt]{article}

\usepackage{color}
\definecolor{rd}{rgb}{.33,0,0}
\definecolor{or}{rgb}{.66,0,0}
\definecolor{pi}{rgb}{.66,.33,.33}
\definecolor{gn}{rgb}{0,.33,0}
\definecolor{be}{rgb}{0,0,.66}
\definecolor{ma}{rgb}{.33,0,.33}
\definecolor{vi}{rgb}{.33,0,.66}
\definecolor{gy}{rgb}{0,.33,.66}
\definecolor{ye}{rgb}{.66,.33,0}
\definecolor{bk}{rgb}{0,0,0}

\def\citeA{\cite}
\newcommand{\institute}[1]{\date{\normalsize #1}}

\newcommand{\Title}[1]{\title{\bf #1}}

\begin{document}

\newcommand{\be}{\begin{equation}} 
\newcommand{\fe}{\end{equation}}
\newcommand{\eqn}{\label}
\newcommand{\bel}{\begin{equation}\label}

\def\thf{\baselineskip=\normalbaselineskip\multiply\baselineskip
by 7\divide\baselineskip by 6}

%\thf  

\def\fff{\baselineskip=\normalbaselineskip}

%\lta and \gta produce > and < signs with twiddle underneath

\def\spose#1{\hbox to 0pt{#1\hss}}\def\lta{\mathrel{\spose{\lower 3pt\hbox
{$\mathchar"218$}}\raise 2.0pt\hbox{$\mathchar"13C$}}}  \def\gta{\mathrel
{\spose{\lower 3pt\hbox{$\mathchar"218$}}\raise 2.0pt\hbox{$\mathchar"13E$}}} 

\def\Libra{\spose {--} {\cal L}}
\def\Diam{\spose {\raise 0.3pt\hbox{+}} {\diamondsuit}  }

\font\fiverm=cmr5
\def\d{\delta}
\def\dL{\spose {\lower 5.0pt\hbox{\fiverm L} } {\delta}}
\def\dE{\spose {\lower 5.0pt\hbox{\fiverm E} } {\delta}}
\def\DL{\spose {\lower 5.0pt\hbox{\, \fiverm L} } {\Delta}}
\def\DE{\spose {\lower 5.0pt\hbox{\, \fiverm E} } {\Delta}}
\def\xiI{\spose {\raise 3.0pt\hbox{$\ \acute{\ }$}} {\xi}}
\def\xiII{\spose {\raise 3.0pt\hbox{$\, \acute{\ }\!\acute{\ }$}} {\xi}}
\def\xiJ{\spose {\raise 3.0pt\hbox{$\ \grave{\ }$}} {\xi}}
\def\xiJI{\spose {\raise 3.0pt\hbox{$\,\grave{\ }\!\acute{\ }$}} {\xi}}
\def\xiJJ{\spose {\raise 3.0pt\hbox{$\, \grave{\ }\!\grave{\ }$}} {\xi}}
\def\xiIJ{\spose {\raise 3.0pt\hbox{$\acute{\ }\grave{\ }$} } {\xi}}

\def\eqdef{\fff\ \vbox{\hbox{$_{_{\rm def}}$} \hbox{$=$} }\ \thf }

\def\ov{\overline}  

\def\uth{{^{\,_{(3)}}}\!}  \def\utw{{^{\,_{(2)}}}\!}
\def\ud{{^{\,_{(\rm d)}}}\!}  \def\udi{{^{\,_{(\rm d-1)}}}\!}
\def\up{{^{\,_{(\rm p)}}}\!}  \def\udp{{^{\,_{(\rm d)}}}\!}

\def\Df{{\cal D}}    \def\Ff{{\cal F}}
\def\Kf{{\cal H}}    \def\Xf{{\cal X}}
\def\Af{{\cal A}}    \def\af{{\alpha}} 
\def\Qf{{\cal Q}}    \def\Pf{{\cal P}} 
\def\kil{k}          \def\el{\ell}     \def\rad{r}

%BLUE symbols
\def\bg{{\color{be}g}} \def\nabl{{\color{be}\nabla\!}}
\def\bepsilon{{\color{be}\varepsilon}}
\def\calR{{\color{be}{\cal R}}} \def\calW{{\color{be}{\cal W}}}
\def\calS{{\color{be}{\cal S}}}

%MAGENTA symbols External Physical fields
\def\bPhi{{\color{ma}\mit\Phi}} \def\bF{{\color{ma} F}} \def\bN{{\color{ma} N}}
\def\bD{{\color{ma} D}}

%PINK symbols for external gauge fields
\def\bA{{\color{pi}A}} \def\bB{{\color{pi}B}}

%VIOLET symbols for mixed internal-external physical fields
\def\oD{{\color{vi}\ov D}}
\def\hlag{{\color{vi}\hat{\cal L}}} \def\hT{{\color{vi}\hat T}}
\def\hj{{\color{vi}\hat j}}  \def\hW{{\color{vi}\hat w}} \def\hf{{\color{vi}\hat f}}
\def\vv{{\color{vi}v}}     \def\ie{{\color{vi}e}}
\def\oP{{\color{vi}\ov P}} \def\calP{{\color{vi}\cal P}}
\def\caloP{{\color{vi}\ov{\cal P}}}
\def\Xx{{\color{vi}X}}   \def\Upsilo{{\color{vi}\mit\Upsilon}}
\def\Ha{{\color{vi}H}} \def\mPi{{\color{vi}\mit\Pi}}  \def\tu{{\color{vi}\tau}}

\def\Beta{{\color{vi}\cal B}} \def\bbeta{{\color{vi}\beta}}
\def\sima{{\color{vi}\sigma}}
\def\Ja{{\color{vi}J}} \def\Ma{{\color{vi}M}} \def\Sa{{\color{vi}S}}
\def\calJ{{\color{vi}{\cal J}}} \def\calM{{\color{vi}{\cal M}}}
\def\psA{{\color{vi}{\cal A}}} \def\psF{{\color{vi}{\cal F}}}

%GREEN symbols for purely geometric fields on worldsheet
\def\ag{{\color{gn}\perp\!}}    \def\og{{\color{gn}\eta}}
\def\onab{{\color{gn}\ov\nabla\!}}
\def\Ke{{\color{gn}K}} \def\Re{{\color{gn}R}}
\def\Fe{{\color{gn}F}}  \def\Ce{{\color{gn}C}}
\def\Ome{{\color{gn}\mit\Omega}} \def\Xe{{\color{gn}\mit\Xi}} 
\def\ae{{\color{gn}a}} 
\def\oDf{{\color{gn}\ov{\cal D}}}  
\def\stah{{\color{gn}\star}} 
\def\calE{{\color{gn}\cal E}} \def\oS{\ov{\color{gn}\cal S}}
\def\hg{{\color{gn}\gamma}}

%SKYBLUE symbols for reference system on worldsheet)
\def\ii{{\color{gy}i}} \def\ji{{\color{gy}j}} 
\def\nauti{{\color{gy}_0}} \def\wuni{{\color{gy}_1}} 
\def\prim{{\color{gy}\prime}} 
\def\dit{{\color{gy}\dot{\,}\!\!}} \def\ddit{{\color{gy}\ddot{\,}\!\!}}
\def\sigme{{\color{gy}\sigma}} \def\Kie{{\color{gy}K}}
\def\rhoe{{\color{gy}\rho}} \def\pome{{\color{gy}\varpi}}
\def\lambde{{\color{gy}\lambda}} \def\iote{{\color{gy}\iota}} 
\def\A{{\color{gy}_A}} \def\B{{\color{gy}_B}}
\def\X{{\color{gy}_X}} \def\Y{{\color{gy}_Y}}

%BEIGE symbols for gauge dependent fields on worldsheet
\def\phie{{\color{ye}\varphi}} \def\psie{{\color{ye}\psi}}

%BROWN symbols for physical fields on worldsheet
\def\olag{{\color{rd}\ov{\cal L}}} 
\def\Lambde{{\color{rd}\Lambda}} \def\calK{{\color{rd}\cal K}}
\def\of{{\color{rd}\ov f}} 
 \def\tf{{\color{rd}f}} \def\cf{{\color{rd}\check f}}
\def\oj{{\color{rd}\ov j}} 
 \def\oW{{\color{rd}\ov w}}
 \def\bet{{\color{rd}\beta}}
  \def\elle{{\color{rd}\ell}} \def\chie{{\color{rd}\chi}}
\def\ch{{\color{rd}\chi}} \def\Te{{\color{rd}T}} \def\Ue{{\color{rd}U}} 
\def\Pe{{\color{rd}P}} 
\def\oT{{\color{rd}\ov T}} 
\def\cE{{\color{rd}c_{_{\rm E}}}} \def\cL{{\color{rd}c_{_{\rm L}}}}
\def\lamb{{\color{rd}\lambda}} \def\varthe{{\color{rd}\vartheta}}
\def\mue{{\color{rd}\mu}} \def\nue{{\color{rd}\nu}}
\def\ue{{\color{rd}u}} \def\ve{{\color{rd}v}}
\def\calT{{\color{rd}\cal T}} \def \otT{{\color{rd}\widetilde T}} 
\def\caltT{{\color{rd}\widetilde{\cal T}}} \def\tT{{\color{rd}\ov t}}
 \def\caloT{{\color{rd}\ov{\cal T}}} 
\def\pe{{\color{rd}p}}  \def\ce{{\color{rd}c}}  \def\ze{{\color{rd}z}}
\def\kag{{\color{rd}\mit\Sigma}} \def\crec{{\color{rd}\epsilon}}

%RED symbols for global quantities and constants:
\def\hbah{{\color{or}\hbar}}  
\def\okappa{{\color{or}\ov  h}}  \def\NG{{\color{or}G}} 
\def\Uzero{{\color{or} U_{_0}}}  \def\Ye{{\color{or}Y}}
\def\Qe{{\color{or}Q}} \def\kappazero{{\color{or}\kappa_{_0}}} 
\def\ee{{\color{or}e}}  \def\qe{{\color{or}q}}  \def\she{{\color{or}\sharp}}
\def\mast{{\color{or}m_\ast}} \def\rast{{\color{or}r_\ast}}
\def\mag{{\color{or}m}}  \def\bag{{\color{or}{\rm b}}}   \def\cag{{\color{or}{\rm c}}} 

\def\Phie{{\color{or}\Phi}}    \def\Ac{{\color{or}\cal I}}
\def\AcI{\spose {\raise 3.0pt\hbox{$\ \acute{\ }$} } {\Ac}}
 \def\Ze{{\color{or}Z}} \def\Ne{{\color{or}N}}

\Title{{\color{rd}Brane Dynamics for Treatment of 
Cosmic Strings and Vortons}}

\author{{\color{gn}Brandon Carter}}

\institute{{\color{be} D.A.R.C., (UPR 176, CNRS),
\\ Observatoire de Paris, 92 Meudon, France }\\
email: carter@groseille.obspm.fr}

\maketitle

{\bf Abstract: } This course provides a self contained introduction to
the general theory of relativistic brane models, of the category that
includes point particle, string, and membrane representations for
phenomena that can be considered as being confined to a worldsheet of
the corresponding dimension (respectively one, two, and three) in a
thin limit approximation. The first part of the course is concerned
with purely kinematic aspects: it is shown how, to second differential
order, the geometry (and in particular the inner and outer curvature)
of a brane worldsheet of arbitrary dimension is describable in terms of
the first, second, and third fundamental tensor; the extension to a
foliation by a congruence of such worldsheets is also briefly
discussed. In the next part, it is shown how  -- to lowest order in the
thin limit -- the evolution of such a brane worldsheet will always be
governed by a simple tensorial equation of motion whose left hand side
is the contraction of the relevant surface stress tensor
$\oT{^{\mu\nu}}$ with the (geometrically defined) second fundamental
tensor $\Ke_{\mu\nu}{^\rho}$, while the right hand side will simply
vanish in the case of free motion and will otherwise be just the
orthogonal projection of any external force density that may happen to
act on the brane.  (Allowance for first order deviations from such a
thin limit treatment would require evolution equations of a more
complicated kind of which a prototype example is presented.) The last
part of the course concentrates on the case of a string: in a four
dimensional spacetime background, this is only case that is non trivial
in the strong sense of having a worldsheet with both dimension  and
codimension greater than one. This case is of particular importance for
cosmology because it applies to the cosmic strings that arise as vortex
defects of the vacuum in many kinds of field theory, and that may
conceivably play an important role in the evolution of the universe. It
is shown how to set up a Lagrangian providing the complete system of
(both internal and external) string evolution equations governing the
large scale motion of vacuum vortices of both the simple
Nielsen-Olesen-Kibble type and of Witten's superconducting type,
including allowance for conservative external forces of both
electromagnetic and Kalb-Ramond axion type, but neglecting the
dissipative effects that are likely to be important soon after string
formation but not later on. The last part of the course deals with
closed loops under typical circumstances in which not only dissipative
but also conservative (electromagnetic or axion type) external forces
can be neglected in a lowest order approximation, so that the evolution
equations will be homogeneous, and thus easily soluble both for dynamic
circular configurations and also for the stationary (centrifugally
supported) configurations known as vortons, which, if they are
sufficiently stable, may be of considerable cosmological significance.

\bigskip\bigskip

As an aide memoire for readers equiped for viewing colour, warmer hues
will be used for physical quantities of a material nature, as
contrasted with essentially geometric quantities for which cooler hues
will be reserved. Among the latter scalar and tensorial fields
determined just by the background spacetime geometry itself will be
indivated by {\color{be} \it blue symbols} while background coordinate
or frame dependent quantities will be indicated by {\bf \it black
symbols}.  Similarly scalar and tensorial fields determined on the
brane worldsheet just by the embedding geometry will be indicated by
{\color{gn}\it green symbols} quantities that are fully determined as
scalars or tensors just by the embedding geometry, while internal
coordinate or frame dependent quantities on the brane worldsheet will
be indicated by {\color{gy}\it skyblue symbols}. As far as more
essentially physical quantities are concerned, {\color{pi} \it pink
symbols} will be used for background gauge fields while {\color{ma} \it
magenta symbols} will be used for physical background fields. On the
other hand {\color{ye} \it beige symbols} will be used for worldsheet
supported gauge fields while {\color{rd} \it brown symbols} will be
used for  physical quantities confined to and intrinsic with respect to
the worldsheet, and {\color{vi}\it violet symbols} will be used for
quantities of mixed type, including physical fields confined to the
worldsheet but defined with respect to an external reference system, as
well as external fields that may be defined with respect to fields on
the world sheet. Finally {\color{or} \it red symbols} will be used for
global integrals and coupling constants.

\bigskip\bigskip\bigskip

\centerline {CONTENTS}
\medskip

{\it Optional (non-essential) branch subsections marked *.}

\bigskip

\ref{0-1}  Introduction.

\medskip\noindent
{\bf \ref{Section1}  Worldsheet curvature analysis.}

\ref{1-1} The first fundamental tensor.

\ref{1-2} The inner and outer curvature tensors.

\ref{1-3} The second fundamental tensor.

\ref{1-4} The extrinsic curvature vector and the 
conformation tensor.

\ref{1-5} *The Codazzi, Gauss, and Schouten identities.

\ref{1-6} *Extension to a foliation.

\ref{1-7} *The adapted connection and curvature of a foliation. 

\ref{1-8} *The special case of a string worldsheet in 4-dimensions.

 \medskip\noindent
{\bf \ref{Section2} Laws of motion for a regular pure brane complex.}

\ref{2-1} Regular and distributional formulations of a brane action.

\ref{2-2} Current, vorticity, and stress-energy tensor.

\ref{2-3} Conservation of current and vorticity.

\ref{2-4} Force and stress balance equation.

\ref{2-5} Equations of extrinsic motion.

\medskip\noindent
{\bf \ref{Section3} Perturbations and curvature effects beyond
the pure brane limit.}

\ref{3-1} First order perturbations and the extrinsic characteristic equation.

\ref{3-2} *Higher order displacements by ``straight'' transportation.

\ref{3-3} *``Straight'' differentiation of covariant fields.

\ref{3-4} *Perturbed Goto Nambu Dirac action.

\ref{3-5} *Higher order geodynamic models.

\medskip\noindent
{\bf \ref{Section4} Strings and other spacially isotropic brane models.}

\ref{4-1} The general category of ``perfect'' brane models.

\ref{4-2} The special case of ``barotropic'' brane models.

\medskip\noindent
{\bf \ref{Section5} Essentials of elastic string dynamics.}

\ref{5-1} *Bicharacteristic formulation of extrinsic equations.

\ref{5-2} Preferred orthonormal diad for nondegenerate state.

\ref{5-3} Duality in elastic string models.

\ref{5-4} *The Hookean prototype example of an elastic string model.

\ref{5-5} General elastic string  models.

\ref{5-6} Standard normalisation of current and charge.

\ref{5-7} Analytic string model for Witten vortex.

\ref{5-8} *The self-dual transonic string model.

\ref{5-9} *Integrability and application of the transonic 
string model.

\medskip\noindent
{\bf \ref{Section6} Symmetric configurations including rings and their vorton 
equilibrium states.}

\ref{6-1} Energy-momentum flux conservation laws.

\ref{6-2} Bernoulli constants for a symmetric string configuration.

\ref{6-3} Generating tangent vector field for symmetric solutions.

\ref{6-4} Hamiltonian for world sheet generator.

\ref{6-5} Free evolution of circular loops.

\ref{6-6} Vorton equilibrium states.

\ref{6-7} *Allowance for electromagnetic ``spring'' effect.

\ref{6-8} *Vortons in cosmology.

\vfill\eject

\subsection{Introduction}
\label{0-1}

The present course is an updated and extended version of lectures
written up a couple of years ago for a summer school on ``Formation and
Interactions of Topological Defects'' \citeA{Carter95} at the Newton Institute
in Cambridge. In preparation for the more specific study of strings in
the later sections, the first part of this course is intended as an
introduction to the systematic study, in a classical relativistic
framework, of ``branes", meaning physical models in which the relevant
fields are confined to supporting worldsheets of lower dimension than
the background spacetime. While not entirely new \citeA{Dirac62}, \citeA{HoweTucker77},
this subject is still at a rather early stage of development (compared
with the corresponding quantum theory \citeA{Achucarroetal87} which has been
stimulated by the  rise of ``superstring theory"), the main motivation
for recent work \citeA{Carter90} on classical relativistic brane theory being
its application to vacuum defects produced by the Kibble
mechanism \citeA{Kibble76}, particularly when of composite type as in the case
of cosmic strings attached to external domain walls \citeA{VilenkinEverett82} and of
cosmic strings carrying internal currents of the kind whose likely
existence was first proposed by Witten \citeA{Witten85}.  A propos of the
latter, a noteworthy example of the progress that has been achieved
since the previous version \citeA{Dirac62} of these notes was written is the
construction (see Subsection \ref{5-7}) for the first time of a class
of reasonably simple {\it analytically explicit} string
models \citeA{CarterPeter95} that are capable of describing the macroscopic
behaviour of such Witten vortices in a manner that is qualitatively
satisfactory and quantitatively adequate not just when the current is
very small but even for the largest physically admissable values. This
makes it possible to provide an analytic description (see Subsection
\ref{6-5}) of the dynamics of the circular ring configurations whose
equilibrium states (see Subsection \ref{6-6}) are the simplest examples
of what are known as vortons.

Before the presentation in Section \ref{Section2} of the dynamic laws
governing the evolution of a brane worldsheet,  Section \ref{Section1}
provides  a recapitulation of the essential differential geometric
machinery \citeA{Carter92a}, \citeA{Carter92b} needed for the analysis of a timelike
worldsheet of dimension d say in a background space time manifold of
dimension n. At this stage no restriction will be imposed on the
curvature of the metric -- which will as usual be represented with
respect to local background coordinates $x^\mu$ ($\mu$= 0, ..., n--1)
by its components $\bg_{\mu\nu}$ -- though it will be postulated to be
flat, or at least stationary or conformally flat, in many of the
applications to be discussed later.

\section{ Worldsheet curvature analysis}
\label{Section1}

\subsection{The first fundamental tensor}
\label{1-1}

The development of geometrical intuition and of computationally efficient
methods for use in string and membrane theory has been hampered by a tradition
of publishing results in untidy, highly gauge dependent, notation (one of the
causes being the undue influence still exercised by Eisenhart's obsolete
treatise ``Riemannian Geometry"  \citeA{Eisenhart26}). For the intermediate steps in
particular calculations it is of course frequently useful and often
indispensible to introduce specifically adapted auxiliary structures, such as
curvilinear worldsheet coordinates $\sigme^\ii$ ($\ii$= 0, ..., d--1) and the
associated bitensorial derivatives
\be x^\mu_{\ ,\ii}={\partial x^\mu\over\partial\sigme^\ii}\, ,\eqn{1.1}\fe
or specially adapted orthonormal frame vectors, consisting of an internal 
subset of vectors $\iote_{\!\A}{^\mu}$ ({ $\A$}= 0,  ... , d--1) 
tangential to the worldsheet and an external subset of vectors 
$\lambde_{\X}{^\mu}$ ({ $\X$} = 1, ... , n--d) orthogonal to the 
worldsheet, as characterised by
\be
\iote_{\!\A}{^\mu}\iote_{\!\B\mu}=\eta_{\A\B} \, ,\hskip 1 cm 
\iota_{\!\A}{^\mu}\lambde_{\X\mu}=0\ ,\hskip 1 cm 
\lambde_{\X}{^\mu}\lambde_{\Y\mu}=\delta_{\X\Y}\, ,\eqn{1.2}\fe
where $\eta_{\A\B}$ is a fixed d-dimensional Minkowski metric and the
Kronecker matrix $\delta_{\X\Y}$ is a fixed (n--d)-dimensional Cartesion
metric. Even in the most recent literature there are still (under
Eisenhart's uninspiring influence) many examples of insufficient effort to 
sort out the messy clutter of indices of different kinds (Greek or Latin,
early or late, small or capital) that arise in this way by grouping the
various contributions into simple tensorially covariant combinations. Another
inconvenient feature of many publications is that results have been left 
in a form that depends on some particular gauge choice (such as the 
conformal gauge for internal string coordinates) which obscures the 
relationship with other results concerning the same system but in a 
different gauge.

The strategy adopted here \citeA{Stachel80} aims at minimising such problems
(they can never be entirely eliminated) by working as far as possible with a
single kind of tensor index, which must of course be the one that is most
fundamental, namely that of the background coordinates, $x^\mu$. Thus, to
avoid dependence on the internal frame index {$\A$} (which is
lowered and raised by contraction with the fixed d-dimensional Minkowski
metric $\eta_{\A\B}$ and its inverse $\eta^{\A\B}$) and on the external
frame index {$\X$} (which is lowered and raised by contraction with
the fixed (n-d)-dimensional Cartesian metric $\delta_{\X\Y}$ and its inverse
$\delta^{\X\Y}$), the separate internal frame vectors $\iote_{\!\A}{^\mu}$
and external frame vectors $\lambde_{\X}{^\mu}$ will as far as possible be
eliminated in favour of the frame gauge independent combinations
\be
\og^\mu_{\ \nu}=\iote_{\!\A}{^\mu}\iote^{\A}{_\nu} \, ,\hskip 1 cm
\ag^{\!\mu}_{\,\nu}=\lambde_{\X}{^\mu}\lambde^{\!\X}{_\nu}\, ,\eqn{1.3}\fe
of which the former, $\og^\mu_{\ \nu}$, is what will be referred to as the
(first) {\it fundamental tensor} of the metric, which acts as the (rank d)
operator of tangential projection onto the world sheet, while the latter,
$\ag^{\!\mu}_{\,\nu}$, is the complementary  (rank n--d) operator of 
projection orthogonal to the world sheet.

The same principle applies to the avoidance of unnecessary involvement of
the internal coordinate indices which are lowered and raised by contraction
with the induced metric on the worldsheet as given by
\be \hg_{\ii\ji}=\bg_{\mu\nu} x^\mu_{\ ,\ii} x^\nu_{\ ,\ji}\, , \eqn{1.4}\fe
and with its contravariant inverse $\hg^{\ii\ji}$.
After being cast (by index raising if necessary) into its contravariant form,
any internal coordinate tensor can be directly projected onto a corresponding
background tensor in the manner exemplified by the intrinsic metric
itself, which gives 
\be \og^{\mu\nu}= \hg^{\ii\ji} x^\mu_{\ ,\ii} x^\nu_{\ ,\ji} \, ,\eqn{1.5}\fe
thus providing an alternative (more direct) prescription for the fundamental
tensor that was previously introduced via the use of the internal frame in 
(\ref{1.3}). This approach also provides a direct prescription for the orthogonal
projector that was introduced via the use of an external frame in
(\ref{1.3}) but that is also obtainable immediately from (\ref{1.5}) as
\be \ag^{\!\mu}_{\,\nu}=\bg^\mu_{\ \nu}-\og^\mu_{\ \nu}\, .\eqn{1.6}\fe

As well as having the separate operator properties 
\be \og^\mu_{\ \rho}\,\og^\rho_{\ \nu}=\og^\mu_{\ \nu} \ , \hskip 1.2 cm
\ag^{\!\mu}_{\ \rho}\ag^{\!\rho}_{\ \nu}=\ag^{\!\mu}_{\ \nu}
\label{1.6a}\fe 
the tensors defined by (\ref{1.5}) and (\ref{1.6}) will evidently be related
by the conditions
\be \og^\mu_{\ \rho}\ag^{\!\rho}_{\ \nu}\,=\,0\,=\ag^{\!\mu}_{\ \rho}
\og^\rho_{\ \nu} \, . \label{1.6b}\fe

\subsection{The inner and outer curvature tensors}
\label{1-2}

In so far as we are concerned with tensor fields such as the frame vectors
whose support is confined to the d-dimensional world sheet, the effect of
Riemannian covariant differentation $\nabl_\mu$ along an arbitrary
directions on the background spacetime  will not be well defined, only the
corresponding tangentially projected differentiation operation
\be \onab_\mu\eqdef\og\,^\nu_{\ \mu}\nabl_\nu\, , \eqn{1.7}\fe
being meaningful for them, as for instance in the case of a scalar field
$\phie$ for which the tangentially projected gradient is given in
terms of internal coordinate differentiation simply by
$\onab{^\mu}\phie=\hg^{\ii\ji} x^\mu{_{\!,\ii}}\,\phie_{,ji}\,$.

An irreducible basis for the various possible covariant derivatives of 
the frame vectors consists of the {\it internal rotation} pseudo-tensor
$\rhoe_{\mu\ \rho}^{\,\ \nu}$ and the {\it external rotation} (or ``twist")
pseudo-tensor $\pome_{\mu\ \rho}^{\,\ \nu}$ as given by
\be \rhoe_{\mu\ \rho}^{\,\ \nu}=\og^\nu\!{_\sigma}\,\iote^{\A}{_\rho}
\onab_\mu\, \iote_{\!\A}{^\sigma}=-\rhoe_{\mu\rho}{^\nu}\, ,\hskip 1 cm
\pome_{\mu\ \rho}^{\,\ \nu}=\ag^{\!\nu}_{\,\sigma}\,\lambde^{\!\X}{_\rho}
\onab_\mu \lambde_{\X}{^\sigma}=-\pome_{\mu\rho}{^\nu}\, , \eqn{1.8}\fe
together with their {\it mixed} analogue $\Ke_{\mu\nu}{^\rho}$ which is 
obtainable in a pair of equivalent alternative forms given by
\be \Ke_{\mu\nu}{^\rho}=\ag^{\!\rho}_{\,\sigma}\,\iote^{\A}{_\nu}
\onab_\mu\, \iote_{\!\A}{^\sigma}=-\og^\sigma\!{_\nu}\,\lambde_{\X}{^\rho}
\onab_\mu \lambde^{\!\X}{_\sigma}\, .\eqn{1.9}\fe

The reason for qualifying the fields (\ref{1.8}) as ``pseudo" tensors is that
although they are tensorial in the ordinary sense with respect to changes
of the background coordinates $x^\mu$ they are not geometrically well 
defined just by the geometry of the world sheet but are gauge dependent
in the sense of being functions of the choice of the internal and external
frames $\iote_{\!\A}{^\mu}$ and $\lambde_{\X}{^\mu}$. 
The gauge dependence of $\rhoe_{\mu\ \rho}^{\,\ \nu}$ and $\pome_{\mu\
\rho}^{\,\ \nu}$ means that both of them can  be set to zero at any chosen
point on the worldsheet by choice of the relevant frames in its vicinity.
However the condition for it to be possible to set these pseudo-tensors to
zero throughout an open neigbourhood is the vanishing of the curvatures of the
corresponding frame bundles as characterised with respect to the respective
invariance subgroups SO(1,d--1) and SO(n--d) into which the full Lorentz
invariance group SO(1,n--1) is broken by the specification of the
d-dimensional world sheet orientation. The {\it inner curvature} that needs to
vanish for it to be possible for $\rhoe_{\mu\ \rho}^{\,\ \nu}$ to be set to
zero in an open neighbourhood is of Riemannian type, is obtainable (by a
calculation of the type originally developed by Cartan that was made familiar
to physicists by Yang Mills theory) as \citeA{Carter92a}
\be \Re{_{\kappa\lambda}}{^\mu}{_\nu}=2\og^\mu{_{\!\sigma}}\og^\tau{_{\!\mu}}
\og^\pi{_{[\lambda}}\onab_{\kappa]}\rhoe_{\pi\ \tau}^{\,\ \sigma}+2
\rhoe_{[\kappa}{^{\mu\pi}}\rhoe_{\lambda]\pi\nu}\, , \eqn{1.11}\fe
while the {\it outer curvature} that needs to vanish for it to be possible
for the ``twist" tensor $\pome_{\mu\ \rho}^{\,\ \nu}$ to be set to zero in 
an open neighbourhood is of a less familiar type that is given \citeA{Carter92a} by
\be \Ome{_{\kappa\lambda}}{^\mu}{_\nu}=2\ag^{\!\mu}_{\,\sigma}\ag^{\!\tau}
_{\,\mu}\,\og^\pi{_{[\lambda}}\onab_{\kappa]}\pome_{\pi\ \tau}^{\,\ \sigma}
+2\pome_{[\kappa}{^{\mu\pi}}\pome_{\lambda]\pi\nu}\, .\eqn{1.12}\fe
The frame gauge invariance of the expressions (\ref{1.11}) and (\ref{1.12}),
-- which means that  $\Re{_{\kappa\lambda}}{^\mu}{_\nu}$ and 
$\Ome{_{\kappa\lambda}}{^\mu}{_\nu}$ are unambiguously well defined as
tensors in the strictest sense of the word -- is not immediately obvious from
the foregoing formulae, but it is made manifest in the the alternative
expressions given in Subsection \ref{1-5}.

\subsection{The second fundamental tensor}
\label{1-3}

Another, even more fundamentally important, gauge invariance property that is
not immediately obvious from the traditional approach -- as recapitulated in
the preceeding subsection is -- that of the entity  $ \Ke_{\mu\nu}{^\rho}$
defined by the mixed analogue (\ref{1.9}) of (\ref{1.8}), which  (unlike
$\rhoe_{\mu\ \rho}^{\,\ \nu}$ and $\pome_{\mu\ \rho}^{\,\ \nu}$, but like
$\Re{_{\kappa\lambda}}{^\mu}{_\nu}$ and $\Ome{_{\kappa\lambda}}{^\mu}{_\nu}$) is
in fact a geometrically well defined tensor in the strict sense. To see that
the formula (\ref{1.9}) does indeed give a result that is frame gauge
independent, it suffices to verify that it agrees with the alternative --
manifestly gauge independent definition  \citeA{Carter90}
\be \Ke_{\mu\nu}{^\rho} \eqdef \og\,^\sigma_{\ \nu}\onab_\mu
\og\,^\rho_{\ \sigma} \, .\eqn{1.10}\fe
whereby the entity that we refer to as the {\it  
second fundamental tensor} is constructed directly from the
the first fundamental tensor $\og^{\mu\nu}$ as given by (\ref{1.5}).

Since this second fundamental tensor, $\Ke_{\mu\nu}{^\rho}$ will play a very
important role throughout the work that follows, it is worthwhile to linger
over its essential properties. To start with it is to be noticed that a
formula of the form (\ref{1.10}) could of course be meaningfully meaningful
applied not only to the fundamental projection tensor of a d-surface, but also
to any (smooth) field of rank-d projection operators $\og\,^\mu_{\ \nu}$ as
specified by a field of arbitrarily orientated d-surface elements. What
distinguishes the integrable case, i.e. that in which the elements mesh
together to form a well defined d-surface through the point under
consideration, is the condition that the tensor defined by (\ref{1.10}) should
also satisfy the {\it Weingarten identity} 
\be \Ke_{[\mu\nu]}{^\rho} =0 \eqn{1.13}\fe 
(where the square brackets denote antisymmetrisation), this
symmetry property of the second fundamental tensor being
derivable \citeA{Carter90}, \citeA{Carter92a} as a version of the well known Frobenius
theorem. In addition to this non-trivial symmetry property, the second
fundamental tensor is also obviously tangential on the first two indices and
almost as obviously orthogonal on the last, i.e. 
\be\ag^{\!\sigma}_{\,\mu}\Ke_{\sigma\nu}{^\rho}=\Ke_{\mu\nu}{^\sigma}
\og_\sigma{^\rho}=0   \, . \eqn{1.14}\fe
The second fundamental tensor $\Ke_{\mu\nu}{^\rho}$ has the property of fully
determining the tangential derivatives of the first fundamental tensor
$\og\,^\mu_{\ \nu}$ by the formula
\be \onab_\mu\og{_{\nu\rho}}=2\Ke_{\mu(\nu\rho)} \eqn{1.15}\fe
(using round brackets to denote symmetrisation) and it can be seen to be
characterisable by the condition that the orthogonal projection of the
acceleration of any tangential unit vector field $\ue^\mu$ will be given by
\be \ue^\mu \ue^\nu \Ke_{\mu\nu}{^\rho}=\ag^{\!\rho}_{\,\mu}\dot \ue^\mu\ ,
\hskip 1 cm \dot \ue^\mu={\ue^\nu\nabl{_\nu} \ue^\mu} \, . \eqn{1.16}\fe

\subsection{The extrinsic curvature vector and the 
conformation tensor} \label{1-4}

It is very practical for a great many purposes to introduce the {\it
extrinsic curvature vector} $K^\mu$, defined as the trace of the second
fundamental tensor, which is automatically orthogonal to the worldsheet, 
\be \Ke^\mu\eqdef \Ke^\nu_{\ \nu}{^\mu}
\ , \hskip 1 cm  \og^\mu_{\ \nu}\Ke {^\nu}=0 \, .\eqn{1.17}\fe
It is useful for many specific purposes to work this out in terms of the
intrinsic metric $\hg_{\ii\ji}$ and its determinant $\vert\hg\vert $. 
It suffices to use the simple expression 
$\onab^{\,\mu}\phie=\hg^{\ii\ji} x^\mu{_{,\ii}}\phie_{,\ji}$ 
for the tangentially projected gradient of a scalar field $\phie$ on the
worldsheet, but for a tensorial field (unless one is using Minkowski
coordinates in a flat spacetime) there will also be contributions involving
the background Riemann Christoffel connection 
\be\Gamma_{\mu\ \rho}^{\,\ \nu}=\bg^{\nu\sigma}\big(\bg_{\sigma(\mu,\rho)}
-{_1\over^2}\bg_{\mu\rho,\sigma}\big)\, .\eqn{1.18}\fe
The curvature vector is thus obtained in explicit detail as
\be \Ke^\nu=\onab_\mu\og^{\mu\nu}={1\over\sqrt{\Vert \hg\Vert}}\Big(
\sqrt{\Vert\hg\Vert}\hg^{\ii\ji}x^\nu_{\, ,\ii}\Big){_{,\ji}}+
\hg^{\ii\ji}x^\mu_{\, ,\ii}x^\rho_{\, ,\ji}\Gamma_{\mu\ \rho}^{\,\ \nu}\, .
\eqn{1.19}\fe
This last expression is  technically useful for certain specific
computational purposes, but it must be remarked that much of the 
literature on cosmic string dynamics has been made unnecessarily heavy 
to read by a tradition of working all the time with long strings of non
tensorial terms such as those on the right of (\ref{1.19}) rather than taking
advantage of such more succinct tensorial expressions as the preceeding 
formula $\onab_\mu\og^{\mu\nu}$. As an alternative to the universally 
applicable tensorial approach advocated here, there is of course another more 
commonly used method of achieving succinctness in particular circumstances,
which is to sacrifice gauge covariance by using specialised kinds of 
coordinate system. In particular for the case of a string, i.e. for a 
2-dimensional worldsheet, it is standard practise to use conformal 
coordinates $\sigme^{\nauti}$ and $\sigme^{\wuni}$ so that the corresponding 
tangent vectors $\dit x^\mu=x^\mu_{\, ,\nauti}$ and $x^{\prim\mu}= 
x^\mu_{\, ,\wuni}$ satisfy the restrictions $\dit x^\mu x^\prim_{\, \mu}=0$, 
$\dit x^\mu\dit x_\mu+x^{\prim\mu}x^\prim_{\,\mu}=0$, which implies 
$\sqrt{\Vert \hg\Vert}=x^{\prim\mu}x^\prim_{\,\mu}=-\dit x^\mu\dit x_\mu$ 
so that (\ref{1.19}) simply gives $\sqrt{\Vert\hg \Vert}\,\Ke^\nu=$ 
$x^{\prim\prim\nu}-\ddit x^\nu + (x^{\prim\mu}x^{\prim\rho}
-\dit x^\mu\dit x^\rho)\Gamma_{\mu\ \rho}^{\,\ \nu}$.

The physical specification of the extrinsic curvature vector (\ref{1.17}) for
a timelike d-surface in a dynamic theory provides what can be taken as the
equations of extrinsic motion of the d-surface \citeA{Carter90}, \citeA{Carter92b}, the
simplest possibility being the ``harmonic" condition $\Ke^\mu=0$ that is
obtained (as will be shown in the following sections) from a surface measure
variational principle such as that of the Dirac membrane model \citeA{Dirac62}, or
of the Goto-Nambu string model \citeA{Kibble76} whose dynamic equations in a flat
background are therefore expressible with respect to a standard conformal
gauge in the familiar form $x^{\prim\prim\mu}-\ddit x^\mu=0$. 

There is a certain analogy between the Einstein vacuum equations, which
impose the vanishing of the trace $\calR_{\mu\nu}$ of the background
spacetime curvature $\calR_{\lambda\mu}{^\rho}{_\nu}$, and the
Dirac-Gotu-Nambu equations, which impose the vanishing of the trace
$\Ke^\nu$ of the second fundamental tensor $\Ke_{\lambda\mu}{^\nu}$.
Just as it is useful to separate out the Weyl tensor
\citeA{Schouten54}, i.e. the trace free part of the Ricci background
curvature which is the only part that remains when the Einstein vacuum
equations are satisfied, so also analogously, it is useful to separate
out the the trace free part of the second fundamental tensor, namely
the extrinsic conformation tensor \citeA{Carter92a}, which is the only
part that remains when equations of motion of the Dirac - Goto - Nambu
type are satisfied. Explicitly, the trace free {\it extrinsic
conformation} tensor $\Ce_{\mu\nu}{^\rho}$ of a d-dimensional imbedding
is defined \citeA{Carter92a} in terms of the corresponding first and
second fundamental tensors $\eta_{\mu\nu}$ and $\Ke_{\mu\nu}{^\rho}$ as
\be \Ce_{\mu\nu}{^\rho}\eqdef \Ke_{\mu\nu}{^\rho}-{1\over{\rm d}}\og{_{\mu\nu}}
\Ke^\rho \ , \hskip 1 cm     \Ce^\nu_{\ \nu}{^\mu}=0 \ .\eqn{1.20}\fe
Like the Weyl tensor $\calW_{\lambda\mu}{^\rho}{_\nu}$ of the background metric
(whose  definition is given implicitly by (\ref{1.25}) below) this
conformation tensor has the noteworthy property of being invariant with
respect to conformal modifications of the background metric:
\be \bg_{\mu\nu}\mapsto {\rm e}^{2\alpha}\bg_{\mu\nu}\, ,\hskip 0.6 cm
\Rightarrow\hskip 0.6 cm \Ke_{\mu\nu}{^\rho}\mapsto \Ke_{\mu\nu}{^\rho}
+\eta_{\mu\nu}\ag^{\!\rho\sigma}\nabl_\sigma\alpha\, ,\hskip 0.6 cm 
 \Ce_{\mu\nu}{^\rho}\mapsto \Ce_{\mu\nu}{^\rho}\, .\eqn{1.21}\fe
This formula is useful \citeA{Carteretal94} for calculations of the kind undertaken by
Vilenkin \citeA{Vilenkin91} in a standard Robertson-Walker type cosmological
background, which can be obtained from a flat auxiliary spacetime metric by a
conformal transformation for which ${\rm e}^\alpha$ is a time dependent Hubble
expansion factor. 

\subsection{*The Codazzi, Gauss, and Schouten identities}
\label{1-5}

As the higher order analogue of (\ref{1.10}) we can go on to introduce the 
{\it third} fundamental tensor$ \citeA{Carter90}$ as
\be \Xe_{\lambda\mu\nu}{^\rho} \eqdef\og\,^\sigma_{\ \mu}\og\,^\tau_{\ \nu}
\ag^{\!\rho}_{\,\alpha}\onab_\lambda \Ke_{\sigma\tau}{^\alpha} \, , 
\eqn{1.22}\fe
which  by construction is obviously symmetric between the second and third
indices and tangential on all  the first three indices.  In a spacetime 
background that is flat (or of constant curvature as is the case for the 
DeSitter universe model) this third fundamental tensor is fully symmetric 
over all the first three indices by what is interpretable as the {\it 
generalised Codazzi identity} which is expressible \citeA{Carter92a} in a 
background with arbitrary Riemann curvature $\calR_{\lambda\mu}{^\rho}{_\sigma}$ 
as 
\be \Xe_{\lambda\mu\nu}{^\rho}= \Xe_{(\lambda\mu\nu)}{^\rho} +{_2\over^3}
\og\,^\sigma_{\ \lambda}\og\,^\tau_{\ {(\mu}}  \og\,^\alpha_{\ {\nu)}}
\calR_{\sigma\tau}{^\beta}{_\alpha}\ag^{\!\rho}_{\,\beta}
\  \eqn{1.23}\fe
It is to be noted that a script symbol $\calR$ is used here in order to
distinguish the (n - dimensional) background Riemann curvature tensor from the
intrinsic curvature tensor (\ref{1.11}) of the (d - dimensional) worldship to
which the ordinary symbol $\Re$ has already allocated.

For many of the applications that will follow it will be sufficient just to
treat the background spacetime as flat, i.e. to take
$\calR_{\sigma\tau}{^\beta}{_\alpha}=0$. At this stage however, we shall allow
for an unrestricted background curvature. For n$>2$ this will be decomposible
in terms of its trace free  Weyl part $\calW_{\mu\nu}{^\rho}{_\sigma}$ (which as
remarked above is conformally invariant) and the corresponding background
Ricci tensor and its scalar trace,
\be \calR_{\mu\nu}= \calR_{\rho\mu}{^\rho}{_\nu}  \, , \hskip 1 cm
 \calR=\calR^\nu_{\ \nu}\, , \eqn{1.24}\fe
in the form \citeA{Schouten54}
\be\calR_{\mu\nu}{^{\rho\sigma}}=\calW_{\mu\nu}{^{\rho\sigma}} +{_4\over^{ n-2} }
\bg^{[\rho}_{\ [\mu}\calR^{\sigma]}_{\ \nu]}-{_2\over ^{(n-1)(n-2)} } \calR
\bg^{[\rho}_{\ [\mu}\bg^{\sigma]}_{\ \nu]} \, ,\eqn{1.25}\fe
(in which the Weyl contribution can be non zero only for n$\geq$ 4). In terms
of the tangential projection of this background curvature, one can evaluate
the corresponding {\it internal} curvature tensor (\ref{1.11}) in the form
\be  \Re{_{\mu\nu}}{^\rho}{_\sigma}= 2\Ke^\rho{_{[\mu}}{^\tau}
\Ke_{\nu]\sigma\tau}+ \og\,^\kappa_{\ \mu} \og\,^\lambda_{\ \nu}
\calR_{\kappa\lambda}{^\alpha}{_\tau} \og\,^\rho_{\ \alpha}\og\,^\tau_{\ \sigma}
  \, , \eqn{1.26}\fe
which is the translation into the present scheme of what is well known in
other schemes as the {\it generalised Gauss identity}. The much less well
known analogue for the (identically trace free and conformally invariant) {\it
outer} curvature (\ref{1.12}) (for which the most historically appropriate
name might be argued to be that of Schouten \citeA{Schouten54}) is given \citeA{Carter92a}
in terms of the corresponding projection of the background Weyl tensor by the
expression 
\be \Ome{_{\mu\nu}}{^\rho}{_\sigma}= 2\Ce_{[\mu}{^{\tau\rho}}\Ce_{\nu]\tau\sigma}
+ \og\,^\kappa_{\ \mu} \og\,^\lambda_{\ \nu}\calW_{\kappa\lambda}{^\alpha}{_\tau}
\ag^{\!\rho}_{\,\alpha}\ag^{\!\tau}_{\,\sigma}
 \, . \eqn{1.27}\fe
It follows from this last identity  that in a background that is flat or 
conformally flat (for which it is necessary, and for n$\geq 4$ sufficient, that
the Weyl tensor should vanish) the vanishing of the extrinsic conformation
tensor $\Ce_{\mu\nu}{^\rho}$ will be sufficient (independently of the
behaviour of the extrinsic curvature vector $\Ke^\mu$) for vanishing of the
outer curvature tensor $\Ome{_{\mu\nu}}{^\rho}{_\sigma}$, which is the
condition for it to be possible to construct fields of vectors $\lambde^\mu$
orthogonal to the surface and such as to satisfy the generalised
Fermi-Walker propagation  condition to the effect that $\ag^{\!\rho}_{\,\mu}
\onab_\nu\lambde_\rho$ should vanish. It can also be shown \citeA{Carter92a}
(taking special trouble for the case d=3 )  that in a conformally flat
background (of arbitrary dimension n) the vanishing of the conformation
tensor $\Ce_{\mu\nu}{^\rho}$ is always sufficient (though by no means
necessary) for conformal flatness of the induced geometry in the imbedding.

\subsection{*Extension to a foliation} 
\label{1-6}

It is useful for many purposes to extend the analysis of a single embedded
d-surface (as described in the preceeding subsections) to the more general
context of a smooth foliation by a congruence of such surfaces. Even when the
object of physical interest is just a single embedded surface it is
geometrically instructive to consider its relation to neighbouring surfaces of
a purely mathematical nature, and of course there are also many applications
involving modes in which a foliation by congruence of surface is a physically
essential constituent structure (for example the congruence of flow
trajectories in an ordinary fluid model \citeA{HawkingEllis73}, or the congruence of
string-like vortex lines in the kind of model \citeA{CarterLanglois95} appropriate for the
macroscopic description of a rotating superfluid.)

In the case of such a foliation, the first fundamental tensor $\og^{\mu\nu}$
as specified by (\ref{1.5}) and of the corresponding orthogonal projector
$\ag^{\!\mu}_{\,\nu}$ as specified by (\ref{1.6}) will no longer have 
support confined to a single d-surface, but will be well defined
as ordinary tensor fields over at least an open neighbourhood of the
background spacetime. This means that  they will have well defined (Riemannian or
pseudo-Riemannian) covariant derivatives of the ordinary unrestricted kind
(not just tangential covariant derivatives of the kind described in the
preceeding sections). These covariant derivatives given  will be
fully determined by the specification of a certain (first) {\it deformation
tensor}, $\Kf_{\mu\ \rho}^{\ \nu}$ say, via an expression of the form 
\be \nabl_{\!\mu}\,\og^\nu_{\ \rho}= -\nabl_{\!\mu}\ag^{\!\nu}_{\,\rho}
=\Kf_{\mu\ \rho}^{\ \nu}+\Kf_{\mu\nu}^{\ \ \rho} \, . \eqn{F.1}\fe

It can easily be seen from (\ref{1.6a}) that the required deformation tensor
will be given simply by 
\be \Kf_{\mu\ \nu}^{\ \rho}=\og^\rho_{\ \sigma}\,\nabl_{\!\mu}\,
\og^\sigma_{\ \nu}=-\ag^{\!\sigma}_{\,\nu}\nabl_{\!\mu}
\!\ag^{\!\rho}_{\,\sigma} \, .\eqn{F.2}\fe 
The middle and last indices of this tensor will evidently have
the respective properties of tangentiality and orthogonality that are
expressible as 
\be \ag^{\!\sigma}_{\,\nu} \Kf_{\mu\ \sigma}^{\ \rho}= 0 \, ,\hskip 1.2 cm 
\Kf_{\mu\ \sigma}^{\ \rho}\,\og^\sigma_{\ \nu}=0 \ .\eqn{F.3}\fe 
There is no automatic tangentiality or orthogonality property
for the first index of the deformation tensor (\ref{F.2}), which is thus
reducible with respect to the tangential and orthogonally lateral projections
(\ref{1.5})(\ref{1.6}) to  a sum 
\be \Kf_{\mu\ \nu}^{\ \rho}=\Ke_{\mu\ \nu}^{\ \rho}
-L_{\mu\nu}^{\,\ \ \rho}\ \eqn{F.4}\fe 
in which such a property is obtained for each of the parts 
\be \Ke_{\mu\ \nu}^{\ \rho}=\og^\sigma_{\ \mu}
\Kf_{\sigma\ \nu}^{\ \rho}\ , \hskip 1 cm L_{\mu\nu}^{\,\ \
\rho}=-\ag^{\!\sigma}_{\,\mu} \Kf_{\sigma\ \nu}^{\ \rho}\ , \eqn{F.5}\fe
which satisfy the conditions 
\be \ag^{\!\sigma}_{\,\mu}\Ke_{\sigma\ \nu}^{\
\rho}= \ag^{\!\rho}_{\,\sigma} \Ke_{\mu\ \nu}^{\ \sigma}
=0=\Ke_{\mu\ \sigma}^{\ \rho}\,\og^\sigma_{\ \nu}\ . \eqn{F.6}\fe 
and 
\be \og^{\sigma}_{\ \mu}L_{\sigma\ \nu}^{\,\ \rho}= 
\og^{\rho}_{\ \sigma} L_{\mu\ \nu}^{\,\ \sigma}=0=
L_{\mu\ \sigma}^{\,\ \rho}\!\ag^{\!\sigma} _{\,\nu}\ .
\eqn{F.7}\fe

The  first of these decomposed parts, which might be described as the {\it
tangential turning} tensor, is expressible explicitly as
\be  \Ke_{\mu\ \nu}^{\ \rho}=\og^\rho_{\ \sigma}\og^\tau_{\ \mu}
\nabl_{\!\tau}\og^\sigma_{\ \nu} \, , 
\eqn{F.8}\fe
and is thus evidently identifiable with the {\it second fundamental tensor} as
defined by (\ref{1.10}). The corresponding explicit formula for the second
part of the decomposition (\ref{F.4}) is
\be L_{\mu\ \nu}^{\ \rho}=\ag^{\!\rho}_{\,\sigma}
\ag^{\!\tau}_{\, \mu} \nabl_{\!\tau}\ag^{\!\sigma}_{\, \nu} \, . 
\eqn{F.9}\fe

Unlike  $\Ke_{\mu\rho}^{\ \ \nu}$, which by the Weingarten identity (\ref{1.13})
is automatically symmetric on its first two (tangential) indices, the {\it
lateral turning tensor}, $L_{\mu\nu}^{\,\ \ \rho}$ will in general have an
antisymmetric part,  $\omega_{\mu\nu}^{\,\ \ \rho}$ say, as well as a
symmetric part,   $\theta_{\mu\rho}^{\,\ \ \nu}$ say, with respect to its
first two (orthogonal) indices. It will be decomposible in terms of these
parts in the form
\be L_{\mu\nu}^{\,\ \ \rho} = \omega_{\mu\nu}^{\,\ \ \rho}
+ \theta_{\mu\nu}^{\,\ \ \rho} \, ,\eqn{F.10}\fe
with
\be \omega_{(\mu\nu)}^{\ \ \ \rho} =0\, ,\hskip 1 cm
\theta_{[\mu\nu]}^{\ \ \ \rho} =0\, .\eqn{F.11}\fe

Just as the Wiengarten symmetry property (\ref{1.13}) of $\Ke_{\mu\nu}^{\ \
\rho}$ is interpretable as the integrability condition for the d-surfaces
under consideration, so by analogy, it is evident that the special condition
that is necessary and sufficient for the existence of  a complementary
orthogonal foliation by (n-d)-surfaces is that $L_{\mu\nu}^{\,\ \ \rho}$
should also be symmetric, i.e. that the rotation tensor $\omega_{\mu\nu}^{\,\
\ \rho}$ should vanish. 

The  part of  $L_{\mu\nu}^{\, \ \rho}$ that remains even if the foliation is
(n-d) surface orthogonal is the symmetric part, $\theta_{\mu\nu}^{\, \ \
\rho}$, which is the natural generalisation of the usual two index divergence
tensor $\theta_{\mu\nu}$ of an ordinary fluid flow. For a 1-dimensional
timelike foliation, which will have a unique future directed unit tangent
vector $u^\mu$, the contraction $\theta_{\mu\nu}= \theta_{\mu\nu}^{\, \ \
\rho} u_\rho$ gives the usual divergence tensor, whose trace $\theta_\nu^{\
\nu}=\nabl_\mu u^\mu$ is the ordinary scalar divergence of the flow. The
evolution of this divergence scalar is governed by the Raychaudhuri identity
whose generalisation to an evolution equation for $\theta_{\mu\nu}$ was given
by Hawking and Ellis \citeA{HawkingEllis73}. In the next subsection it will be shown how
(following Capovilla and Guven \citeA{CapovillaGuven95}) one can obtain a further
extension (\ref{F.56}) of the Raychaudhuri identity that provides an evolution
equation for the three index generalised divergence tensor
$\theta_{\mu\nu}^{\, \ \ \rho}$ which (unlike the ordinary divergence tensor
$\theta_{\mu\nu}$) is always well defined whatever the dimension of the
foliation.

Just as the first order derivatives of the fundamental tensor $\og^\mu_{\
\nu}$ of a foliation are specifiable by the first deformation tensor
$\Kf_{\mu\ \rho}^{\ \nu}$ defined in the previous section, so analogously the
derivatives of the next order will be fully determined by the further
specification of a corresponding {\it second deformation tensor},
$\Xf_{\lambda\mu\ \rho}^{\ \ \nu}$  say, via an expression of the
slightly less simple form
\be \nabl_{\!\lambda}\,\Kf_{\mu\ \rho}^{\ \nu}=\Xf_{\lambda\mu\ \rho}^{\ \ \nu}
+\Kf_{\lambda\sigma}^{\ \ \nu}\Kf_{\mu\ \rho}^{\ \sigma}
-\Kf_{\lambda\rho}^{\ \ \sigma}\Kf_{\mu\ \sigma}^{\ \nu}\, ,\eqn{F.12}\fe
that is easily derivable from the definition
\be \Xf_{\lambda\mu\ \rho}^{\ \ \nu}=
\og^\nu_{\ \sigma}\ag^{\!\tau}_{\,\rho}\nabl_{\!\lambda}\,\Kf_{\mu\ \tau}
^{\ \sigma} \, ,\eqn{F.13}\fe
using the projection properties (\ref{F.3}).

Just as the second fundamental tensor of an individual surface was obtained
from the first deformation tensor of the foliation by tangential projection
according to (\ref{F.4}), so analogously the corresponding {\it third fundamental
tensor}, $\Xe_{\lambda\mu\ \rho}^{\ \ \ \nu}$, as defined by (\ref{1.22}),
will be obtainable from the second deformation tensor by the slightly less
simple tangential projection operation
\be \Xe_{\lambda\mu\ \rho}^{\ \ \ \nu}=\og^\alpha_{\ \lambda}\,\og^\beta_{\ \mu}
\big(\Xf_{\alpha\beta\ \rho}^{\,\ \ \nu}+\Kf_{\alpha\beta}^{\,\ \ \sigma}
\Kf_{\sigma\ \rho}^{\ \nu}\big) \, .\eqn{F.14}\fe
It can be seen to follow from the Weingarten integrability property
(\ref{1.13}) that the antisymmetric part of this third fundamental tensor will
be given directly by the simple projection operation
\be \Xe_{[\lambda\mu]\ \rho}^{\ \ \ \ \nu}=\og^\alpha_{\ \lambda}\,
\og^\beta_{\ \mu}\Xf_{[\alpha\beta]\ \rho}^{\,\ \ \ \nu}\, .\eqn{F.15}\fe
It is easy to use the projection properties (\ref{F.3}) to verify that the
tensor obtained by this procedure does indeed satisfy the original definition
(\ref{1.22}) of the third fundamental tensor, which can thereby be seen to be
unambigously determined, like the second fundamental tensor, just by the
specification of an individual imbedded surface, independently of any
extension to a space covering foliation. 

The higher order differential information contained in the second deformation
tensor and the third fundamental tensor is not needed for the evaluation of
the curvature tensors discussed in the following sections, even though they
are dependent on the second derivatives of the background metric itself.
However a useful biproduct of the curvature analysis that follows is the
derivation of of a separation identity (interpretable as a generalisation of
the historic Codazzi identity) that expresses the antisymmetric part of the
second deformation tensor -- and hence by (\ref{F.15}) that of the third
fundamental tensor -- directly as the corresponding projection of the
background spacetime curvature $\calR_{\lambda\mu\ \rho}^{\,\ \ \nu}$.

\subsection{*The adapted connection and curvature of a foliation. }
\label{1-7}

Due to existence of the decomposition whereby a background spacetime
vector, with components $\xi^\mu$ say, is split up 
as the sum of its surface tangential part $\og^\mu_{\ \nu}\,
\xi^\nu$ and its surface orthogonal part $\ag^{\!\mu}_{\,\nu}\!\xi^\nu$,
there will be a  corresponding adaptation of the ordinary concept of parallel
propagation with respect to the background connection $\Gamma_{\mu\ \rho}
^{\,\ \nu}$. The principle of the adapted propagation concept is to
follow up an ordinary operation of infinitesimal parallel propagation
by the projection adjustment that is needed to ensure that purely tangential
vectors propagate onto purely tangential vectors while purely orthogonal
vectors propagate onto purely orthogonal vectors. Thus for purely tangential
vectors, the effect of the adapted propagation is equivalent to that of
ordinary internal parallel propagation with respect to the induced metric
in the imbedded surface, while for purely orthogonal vectors it is
interpretable as the natural generalisation of the standard concept of
Fermi-Walker propagation. For an infinitesimal displacement $dx^\mu$
the deviation between the actual component variation $(dx^\nu)\partial_\nu
\xi^\mu$ and the variation that would be obtained by the corresponding
adapted propagation law will be expressible in the form $(dx^\nu)\Df_\nu
\xi^\mu$ where $\Df$ denotes the corresponding {\it adapted differentiation}
operator, whose effect will evidently be given by
\be \Df_\nu\,\xi^\mu=\og^\mu_{\ \rho}\nabl_{\!\mu}\big(\og^\rho_{\ \sigma}
\xi^\sigma\big)+\ag^{\!\mu}_{\,\rho}\!\nabl_{\!\mu}\big(\!\ag^{\!\rho}
_{\,\sigma}\!\xi^\sigma\big) \, .\eqn{F.16}\fe
It can thus be seen that this operation will be expressible in the
form
\be \Df_\nu\,\xi^\mu=\nabl_{\!\mu}\xi^\nu+\af_{\nu\ \sigma}^{\,\ \mu}\xi^\sigma
=\partial_\nu\,\xi^\nu+\Af_{\nu\ \sigma}^{\,\ \mu}\xi^\sigma\, ,\eqn{F.17}\fe
where the adapted {\it foliation connection}
components $\Af_{\mu\ \rho}^{\,\ \nu}$ are given by the formula
\be \Af_{\mu\ \rho}^{\,\ \nu}=\Gamma_{\mu\ \rho}^{\,\ \nu}
+\af_{\mu\ \rho}^{\,\ \nu}\, ,\eqn{F.18}\fe
in which the $\af_{\mu\ \rho}^{\,\ \nu}$ are the components of the relevant
{\it adaptation tensor}, whose components can be seen from (\ref{F.16})
to be given by
\be \af_{\mu\ \rho}^{\,\ \nu}=\big(\og^\nu_{\ \sigma}-\!\ag^{\!\nu}_{\,\sigma}
\big)\nabl_{\!\mu}\,\og^\sigma_{\ \rho}\, .\eqn{F.19}\fe
The entirely covariant index lowered version of this adaptation tensor
can thus be seen to be expressible in terms of the deformation tensor
defined by (\ref{F.2}) in the form
\be \af_{\mu\nu\rho}=2\Kf_{\mu[\nu\rho]} \, .\eqn{F.20}\fe

The fact that this last expression is manifestly antisymmetric with
respect to the last two indices of the adaptation tensor makes it
evident that, like the usual Riemannian differentiation operator $\nabl$,
the adapted differentiation operator $\Df$ will commute with index
raising or lowering, since the metric remains invariant under adapted
propagation:
\be \Df_\mu\, \bg_{\nu\rho}=0\, .\eqn{F.21}\fe
However, unlike $\nabl$, the adapted differentiation operator has the
convenient property of also commuting with tangential and orthogonal
projection, since it can be seen to follow from (\ref{F.1}) and (\ref{F.3})
that the corresponding operators also remain invariant under adapted
propagation:
\be \Df_\mu\, \og^\nu_{\ \rho}=0\ ,\hskip 1 cm \Df_\mu\!\ag^{\!\nu}_{\,\rho}=0
\, .\eqn{F.22}\fe

There is of course a price to be paid in order to obtain this considerable
advantage of $\Df$ over $\nabl$, but it is not exhorbitant: all that has to 
be sacrificed is the analogue of the symmetry property 
\be \Gamma_{[\mu\ \rho]}^{\ \ \nu}=0 \, .\eqn{F.22a}\fe 
expressing the absence of torsion in the Riemannian case. For the adapted
connection $\Af_{\mu\ \rho}^{\,\ \nu}$, the torsion tensor defined by
\be \Theta_{\mu\ \rho}^{\,\ \nu}=2\Af_{[\mu\ \rho]}^{\ \ \nu}=
2\af_{[\mu\ \rho]}^{\ \ \nu}\, ,\eqn{F.23}\fe
will not in general be zero. 

When one is dealing not with an extended foliation but just with an
individual imbedded surface, the general adapted differentiation operator
$\Df$ will not be well defined, but adapted differentiation in tangential 
directions will still be specified by the corresponding unambiguously 
defined restricted differentiation operator $\oDf$ whose tensorial 
component representation is given by the adapted analogue of (\ref{1.7}), 
namely
\be \oDf_\mu  \eqdef \og^\nu_{\ \mu} \Df_\nu\, . \eqn{F.24}\fe
Restriction to a single imbedded surface will not invalidate the properties of
commuting with index raising and tangential or orthogonal projection as
expressed in the general case by (\ref{F.21}) and (\ref{F.22}), so in abstract
symbolical notation we shall have not only $\big[\,\oDf, \bg\,\big]=0$, but
also $\big[\,\oDf, \og\,\big] =0$, and $\big[\,\oDf,\!\ag \big] =0$. It is
evident from (\ref{F.17}) and(\ref{F.23}) that the effect of this tangentially
restricted adapted differentiation operator on any surface supported (but not
necessarily surface tangential) vector field with components $\xi^\mu$ will be
given by
\be \oDf_\mu\,\xi^\nu=\onab_{\!\mu}\xi^\mu
+\ae_{\mu\ \rho}^{\,\  \nu}\xi^\rho\ ,\hskip 1 cm
\ae_{\mu\ \rho}^{\,\  \nu}=\og^\sigma_{\ \mu}\af_{\sigma\ \rho}^{\,\ \nu}
\, ,\eqn{F.25}\fe
where the relevant tangentially restricted part of the adaptation tensor,
which  may naturally described as the {\it acceleration tensor},
is given explicitly as function only of the ordinary (restricted)
second fundamental tensor of the imbedded surface by the simple formula
\be \ae_{\mu\nu\rho}=2\Ke_{\mu[\nu\rho]} \, .\eqn{F.26}\fe
As a simple illustration, it is to be remarked that in the particular
case for which the imbedded surface is a particle world-line (with d=1)
having a timelike tangent vector $ue^\mu$ satisfying the unit normalisation 
condition $\ue^\nu \ue_\nu=-1$, for which the first and second fundamental tensors
will be given respectively by  $\og^{\mu\nu}= -\ue^\mu \ue^\nu$
and $\Ke_{\mu\ \rho}^{\ \nu}= \ue_\mu \ue^\nu\dot \ue_\rho$ where $\dot\ue^\rho
=\ue^\sigma\nabl_{\!\sigma}\ue^\rho$ are the components of the ordinary
acceleration vector, it follows that the corresponding acceleration tensor 
will be given by $\ae_{\mu\nu\rho}= 2\ue_\mu \ue_{[\nu}\dot \ue_{\rho]}$.

The curvature associated with the adapted connection (\ref{F.17}) can be read
out from the ensuing commutator formula, which, for an abitrary vector field
with components $\xi^\mu$, will take the standard form
\be \Df_{[\mu}\Df_{\nu]}\xi^\rho=\Ff_{\mu\nu\ \sigma}^{\,\ \ \rho}\xi^\sigma
-\Theta_{\mu\ \nu}^{\,\ \sigma}\Df_\sigma\xi^\rho\, ,\eqn{F.27}\fe
in which the torsion tensor components $\Theta_{\mu\ \nu}^{\,\ \sigma}$ are as
defined by (\ref{F.23}) while the components $\Ff_{\mu\nu\ \sigma}^{\,\ \
\rho}$ are defined by a Yang-Mills type curvature formula of the form
\be\Ff_{\mu\nu\ \sigma}^{\,\ \ \rho} =2\partial_{[\mu}\Af_{\nu]\ \sigma}^{\ \ 
\rho}+2\Af_{[\mu}^{\,\ \rho\tau}\Af_{\nu]\tau\sigma} \, .\eqn{F.28}\fe
Although the connection components $\Af_{\mu\ \rho}^{\, \nu}$ from which it
is constructed are not of tensorial type, the resulting curvature components
are of course strictly tensorial. This is made evident by evaluating
the components (\ref{F.28}) of this {\it amalgamated foliation curvature} 
in terms of the background curvature tensor 
\be\calR_{\mu\nu\ \sigma}^{\,\ \ \rho}=2\partial_{[\mu}\Gamma_{\nu]\ \sigma}^{\ \ 
\rho}+\Gamma_{\mu\ \tau}^{\ \rho}\Gamma_{\nu\ \sigma}^{\ \tau}- 
\Gamma_{\nu\ \tau}^{\ \rho}\Gamma_{\mu\ \sigma}^{\ \tau}\, ,\eqn{F.29}\fe
and the adaptation tensor $\af_{\mu\ \rho}^{\,\ \nu}$ given by (\ref{F.20}), 
which gives the manifestly tensorial expression
\be \Ff_{\mu\nu\ \sigma}^{\,\ \ \rho} = \calR_{\mu\nu\ \sigma}^{\,\ \ \rho}+
2\nabl_{\![\mu}\af_{\nu]\ \sigma}^{\ \ 
\rho}+2\af_{[\mu}^{\,\ \rho\tau}\af_{\nu]\tau\sigma} \, .\eqn{F.30}\fe

Although it does not share the full set of symmetries of the Riemann tensor,
the foliation curvature obtained in this way will evidently be antisymmetric
in both its first and last pairs of indices: \be
\Ff_{\mu\nu\rho\sigma}=\Ff_{[\mu\nu][\rho\sigma]} \, .\eqn{F.31}\fe Using the
formula (\ref{F.20}), it can be see from (\ref{F.3}) and (\ref{F.12}), that
the difference between this adapted curvature and the ordinary background
Riemann curvature will be given by
\be \Ff_{\mu\nu}^{\,\ \ \rho\sigma} - \calR_{\mu\nu}^{\,\ \ \rho\sigma}
=4\Xf_{[\mu\nu]}{^{[\rho\sigma]}} +2\Kf_{[\mu}^{\,\ \sigma\tau}
\Kf_{\nu]\ \tau}^{\,\ \rho} +2\Kf_{[\mu}^{\,\ \tau\rho}
\Kf_{\nu]\tau}^{\,\ \ \sigma}  \, ,\eqn{F.32}\fe
which superficially appears to depend on the higher order derivatives
involved in the second deformation tensor $\Xf_{\mu\nu}^{\ \ \rho\sigma}$.

This appearance is however deceptive. The reason for qualifying the
foliation curvature tensor (\ref{F.28}) as ``amalgamated" is that because the
adapted derivation operator has been constructed in such a way as to map
purely tangential vector fields onto purely tangential vector fields, and
purely orthogonal vector fields onto purely orthogonal vector fields, it
follows that the same applies to the corresponding curvature, which will
therefore consist of an additive amalgamation of two separate parts having
the form
\be \Ff_{\mu\nu\ \sigma}^{\,\ \ \rho}=\Pf_{\mu\nu\  \sigma}^{\,\ \ \rho}
+\Qf_{\mu\nu\ \sigma}^{\,\ \ \rho}\, ,\eqn{F.33}\fe
in which the ``inner" curvature acting on purely tangential vectors 
is given by a doubly tangential projection as
\be \Pf_{\mu\nu\ \sigma}^{\,\ \ \rho}=\Ff_{\mu\nu\ \lambda}^{\,\ \ \kappa}\,
\og^\rho_{\ \kappa}\,\og^\lambda_{\ \sigma} \, , \eqn{F.34}\fe
while  the ``outer" curvature acting on purely orthogonal vectors is given 
by a doubly othogonal projection as
\be \Qf_{\mu\nu\ \sigma}^{\,\ \ \rho}=\Ff_{\mu\nu\ \lambda}^{\,\ \ \kappa}
\!\ag^{\!\rho}_{\,\kappa} \ag^{\!\lambda}_{\,\sigma}\, .\eqn{F.35}\fe
It is implicit in the separation expressed by (\ref{F.33}) that the mixed
tangential and orthogonal projection of the adapted curvature must vanish:
\be \Ff_{\mu\nu\ \lambda}^{\,\ \ \kappa}\,\og^\rho_{\ \kappa}
\!\ag^{\!\lambda}_{\, \sigma}=0\, .\eqn{F.36}\fe 

To get back, from the extended foliation curvature tensors that have just been
introduced, to their antecedent analogues \citeA{Carter92a} for an individual
embedded surface, the first step is to construct the {\it amalgamated
embedding curvature} tensor, $\Fe_{\mu\nu\ \sigma}^{\,\ \ \rho}$ say, which will
be obtainable from the corresponding amalgamated foliation curvature
$\Ff_{\mu\nu\ \sigma}^{\,\ \ \rho}$ by a doubly tangential projection having
the form
\be \Fe{_{\mu\nu\ \sigma}^{\,\ \ \rho}}=\og^\alpha_{\ \mu}\,\og^\beta_{\ \nu}                                
\Ff_{\alpha\beta\ \sigma}^{\,\ \ \rho}\, .
\eqn{F.40}\fe
As did the extended foliation curvature, so also  this amalgated embedding
curvature will separate as the sum of ``inner" and ``outer" parts in the 
form
\be \Fe{_{\mu\nu\ \sigma}^{\,\ \ \rho}}=\Re{_{\mu\nu\  \sigma}^{\,\ \ \rho}}
+\Ome{_{\mu\nu\ \sigma}^{\,\ \ \rho}}\, ,
\eqn{F.41}\fe
in which the ``inner" embedding curvature is given, consistently
wit (\ref{1.26}) by another doubly tangential projection as
\be \Re{_{\mu\nu\ \sigma}^{\,\ \ \rho}}=\Fe{_{\mu\nu\ \lambda}^{\,\ \ \kappa}}\,
\og^\rho_{\ \kappa}\,\og^\lambda_{\ \sigma} 
=\og^\alpha_{\ \mu}\og^\beta_{\ \nu}\Pf_{\alpha\beta\ \sigma}^{\,\ \ \rho}
\, ,\eqn{F.42}\fe
while  the ``outer" embedding curvature is given, consistently with
(\ref{1.27}) by the corresponding doubly lateral projection as
\be \Ome{_{\mu\nu\ \sigma}^{\,\ \ \rho}}=\Fe{_{\mu\nu\ \lambda}^{\,\ \ \kappa}
\!\ag^{\!\rho}_{\,\kappa}} \ag^{\!\lambda}_{\,\sigma}
=\og^\alpha_{\ \mu}\og^\beta_{\ \nu}\Qf_{\alpha\beta\ \sigma}^{\,\ \ \rho}
\, . \eqn{F.43}\fe

The non-trival separation identity (\ref{F.36}) can be considered as a 
generalisation to the case of foliations of the relation that is itself 
interpretable as an extended generalisation to higher dimensions of the 
historic Codazzi equation that was originally formulated in the restricted
context of 3-dimensional flat space. It can be seen from (\ref{F.32}) that this 
extended Codazzi identity is expressible as
\be 2\Xf_{[\mu\nu]}{^\rho_{\ \sigma} }+\calR_{\mu\nu\ \lambda}^{\,\ \ \kappa}\,
 \og^\rho_{\ \kappa}\!\ag^{\!\lambda}_{\,\sigma} = 0\, ,\eqn{F.47}\fe
which shows that the relevant higher derivatives are all determined entirely
by the Riemannian background curvature so that no specific knowledge of the
second deformation tensor is needed. The formula (\ref{F.32}) can also be used
to evaluate the ``inner" tangential part of the foliation curvature tensor as 
\be \Pf_{\mu\nu\ \sigma}^{\,\ \ \rho} = 2\Kf_{[\nu}{^{\rho\tau}}
\Kf_{\mu]\sigma\tau} +\calR_{\mu\nu\ \lambda}^{\,\ \ \kappa}\,
 \og^\rho_{\ \kappa}\,\og^{\lambda}_{\ \sigma}\eqn{F.48}\fe
and to evaluate the ``outer" orthogonal part of the foliation curvature
tensor as
\be \Qf_{\mu\nu\ \sigma}^{\,\ \ \rho} = 2\Kf_{[\mu}{^{\tau\rho}}
\Kf_{\nu]\tau\sigma}\calR_{\mu\nu\ \lambda}^{\,\ \ \kappa}\,
 \ag^{\!\rho}_{\,\kappa}\!\ag^{\!\lambda}_{\,\sigma} \, .\eqn{F.49}\fe

The formula (\ref{F.48}) for  the ``inner" foliation curvature is evidently
classifiable as an extension of the preceeding generalisation (\ref{1.26}) of
the historic Gauss equation, while the formula (\ref{F.49}) for the ``outer"
foliation curvature is an extension the more recently derived \citeA{Carter92a}
generalisation (\ref{1.27}) of what has sometimes been referred to as the
``Ricci equation" but what would seem more appropriately describable  as the
{\it Schouten equation}, with reference to the earliest relevant source with
which I am familiar \citeA{Schouten54}, since long after the time of Ricci it was
not yet understood even by such a leading geometer as Eisenhart  \citeA{Eisenhart26}.
In much the same way, the non-trivial separation identity (\ref{F.36})
provides the relation (\ref{F.47}) that can be considered as a generalisation
to the case of foliations of the identity (\ref{1.23}) that is itself
interpretable as an extended generalisation to higher dimensions of the
historic Codazzi equation that was originally formulated in the restricted
context of 3-dimensional flat space. The corresponding doubly lateral
projection of the foliation curvature would provide the analogous result for
the orthogonal foliation by (n-d)-surfaces that would exist in the
irrotational case for which the lateral turning tensor given by (\ref{F.10})
is symmetric. Finally the corresponding mixed tangential and lateral
projection of (\ref{F.32}) gives an identity that is expressible in terms of
foliation adapted differentiation (\ref{F.17}) as \be
\ag^{\!\tau}_{\,\mu}\!\Df^{\,}_{\!\tau} \Ke_{\nu\ \sigma}^{\ \rho} +\og^\tau_{\
\nu}\Df^{\,}_{\!\tau} L_{\mu\sigma}^{\,\ \ \rho} =\Ke_{\nu\ \mu}^{\
\lambda}\Ke_{\lambda\ \sigma}^{\ \rho}+L_{\mu\ \nu}^{\ \lambda}
L_{\lambda\sigma}^{\ \ \rho}+\og^\alpha_{\ \nu}\!\ag^{\!\beta}_{\,\mu}\!
\calR_{\alpha\beta\ \lambda}^{\,\ \ \kappa}\, \og^\rho_{\
\kappa}\!\ag^{\!\lambda}_{\,\sigma}\, . \eqn{F.56}\fe This last result is
interpretable as the translation into the pure background tensorial formalism
used here of the recently derived generalisation \citeA{CapovillaGuven95} to higher
dimensional foliations of the well known Raychaudhuri equation (whose original
scalar version, and its tensorial extension \citeA{HawkingEllis73}, were formulated just
for the special case of a foliation by 1-dimensional curves). The complete
identity (\ref{F.47}) is therefore interpretable  as an amalgamated
Raychaudhuri-Codazzi identity.

\subsection{*The special case of a string worldsheet in 
4-dimensions} \label{1-8}

The application with which we shall mainly be concerned in the following work
will be the case d=2 of a string. An orthonormal tangent frame will consist in
this case just of a timelike unit vector, $\iote\nauti{^\mu}$, and a spacelike
unit vector, $\iote\wuni{^\mu}$, whose exterior product vector is the frame
independent antisymmetric unit surface element tensor 
\be \calE^{\mu\nu}=2\iote\nauti{^{[\mu}}\iote\wuni{^{\nu]}}
=2\big(\!-\!\vert\hg\vert\big)^{-1/2}\,x^{[\mu}_{\ \, ,\nauti}x^{\nu]}_{\ ,\wuni}
\, ,\eqn{1.28}\fe
whose tangential gradient satisfies
\be \onab_\lambda\calE^{\mu\nu}=-2\Ke_{\lambda\rho}{^{[\mu}}\calE^{\nu]\rho}\, .
\eqn{1.29}\fe
(This is the special d=2 case of what is, as far as I am aware, the only
wrongly printed formula in the more complete analysis \citeA{Carter92a} on which this
section is based: the relevant general formula (B9) is valid as printed only
for odd d, but needs insertion of a missing sign adjustment factor $(-1)^{\rm
d-1}$ in order to hold for all d.) In this case the inner rotation pseudo
tensor (\ref{1.8}) is determined just by a corresponding rotation covector
$\rhoe_\mu$ according to the specification
\be \rhoe_{\lambda\ \nu}^{\ \mu}={_1\over^2}\,\calE^\mu_{\ \nu}\rhoe_\lambda\, ,
\hskip 1 cm \rhoe_\lambda=\rhoe_{\lambda\ \nu}^{\ \mu}\calE^\nu_{\ \mu}  
\, .\eqn{1.30}\fe
This can be used to see from (\ref{1.11}) that the Ricci scalar,
\be \Re = \Re{^\nu_{\ \nu}}\, 
\hskip 1 cm \Re{_{\mu\nu}}=\Re{_{\rho\mu}}{^\rho}{_\nu}
\, ,\eqn{1.31a}\fe
of the 2-dimensional worldsheet will have the well known property
of being a pure surface divergence, albeit of a frame gauge dependent 
quantity:
\be \Re=\onab_\mu\big(\calE^{\mu\nu} \rhoe_\nu\big) \, .\eqn{1.31b}\fe
In the specially important case of a string in ordinary 4-dimensional
spacetime, i.e. when we have not only d=2 but also n=4, the antisymmetric
background measure tensor $\bepsilon^{\lambda\mu\nu\rho}$ can be used to
determine a scalar (or more strictly, since its sign is orientation
dependent, a pseudo scalar) magnitude $\Ome$ for the outer curvature
tensor (\ref{1.12}) (despite the fact that its traces are identically zero)
according to the specification
\be \Ome={_1\over^2}\,\Ome_{\lambda\mu\nu\rho}\,\bepsilon^
{\lambda\mu\nu\rho}\, .\eqn{1.32}\fe
Under these circumstances one can also define a ``twist" covector
$\pome_\mu$, that is the outer analogue of $\rhoe_\mu$, according to the
specification
\be \pome_\nu={_1\over^2}\,\pome_\nu^{\ \mu\lambda}\,
\bepsilon_{\lambda\mu\rho\sigma}\,\calE^{\rho\sigma}\, .\eqn{1.33}\fe
This can be used to deduce from (\ref{1.12}) that the outer curvature (pseudo)
scalar $\Omega$ of a string worldsheet in 4-dimensions has a divergence
property of the same kind as that of its more widely known Ricci analogue
(\ref{1.31b}), the corresponding formula being given by
\be \Ome=\onab_\mu\big(\calE^{\mu\nu} \pome_\nu\big) \ .\eqn{1.34}\fe
It is to be remarked that for a compact spacelike 2-surface the integral of
(\ref{1.29}) gives the well known Gauss Bonnet invariant, but that the
timelike string worldsheets under consideration here will not be characterised
by any such global invariant since they will not be compact (being open in the
time direction even for a loop that is closed in the spacial sense). The outer
analogue of the Gauss Bonnet invariant that arises from (\ref{1.32}) for a
spacelike 2-surface has been discussed by Penrose and Rindler \citeA{PenroseRindler84} but
again there is no corresponding global invariant in the necessarily
non-compact timelike case of a string worldsheet.

\section{ Laws of motion for a regular pure brane complex}
\label{Section2}

\subsection{Regular and distributional formulations of a 
brane action} \label{2-1}

The term p-brane has come into use \citeA{Achucarroetal87}, \citeA{BarsPope88} to describe a dynamic
system localised on a timelike support surface of dimension d=p+1, 
imbedded in a spacetime background of dimension n$>$p. Thus at the
low dimensional extreme one has the example of a zero - brane, meaning what
is commonly referred to as a ``point particle", and of a 1-brane meaning
what is commonly referred to as a ``string". At the high dimensional extreme
one has the ``improper" case of an (n--1)-brane, meaning what is commonly
referred to as a ``medium" (as exemplified by a simple fluid), and of an
(n--2)-brane, meaning what is commonly referred to as a ``membrane" (from which 
the generic term ``brane" is derived). A membrane (as exemplified by a
cosmological domain wall) has the special feature of being supported by a
hypersurface, and so being able to form a boundary between separate
background space time regions; this means that a 2-brane has the status of
being a membrane in ordinary 4-dimensional spacetime (with $n=4$) but not in
a higher dimensional (e.g. Kaluza Klein type) background.

The purpose of the present section is to consider the dynamics not just of
an individual brane but of a {\it brane complex} or ``rigging model" \citeA{Carter90}
such as is illustrated by the nautical archetype in which the wind -- a
3-brane -- acts on a boat's sail -- a 2-brane -- that is held in place by
cords -- 1-branes -- which meet at knots, shackles and pulley blocks that
are macroscopically describable as point particles -- i.e. 0-branes. In
order for a a set of branes of diverse dimensions to qualify as
a``geometrically regular'' brane complex or ``rigging system" it is required
not only that the support surface of each (d--1)-brane should be a smoothly
imbedded d-dimensional timelike hypersurface but also that its boundary, if
any, should consist of a disjoint union of support surfaces of an attatched
subset of lower dimensional branes of the complex. (For example in order
qualify as part of a regular brane complex the edge of a boat's sail can not
be allowed to flap freely but must be attatched to a hem cord belonging to
the complex.) For the brane complex to qualify as regular in the strong
dynamic sense that will be postulated in the present work, it is also
required that a member p-brane can exert a direct force only on an an
attached (p--1)-brane on its boundary or on an attached (p+1)-brane on whose
boundary it is itself located, though it may be passively subject to forces
exerted by a higher dimensional background field. For instance the
Peccei-Quin axion model gives rise to field configurations representable as
regular complexes of domain walls attached to
strings \citeA{VilenkinEverett82}, \citeA{Sikivie82}, \citeA{Shellard90}, and a bounded (topological or
other) Higgs vortex defect terminated by a pair of pole
defects \citeA{Nambu77}, \citeA{Manton83}, \citeA{Copelandetal88a}, \citeA{VachaspatiAchucarro91}, \citeA{VachaspatiBarriola92}, \citeA{MartinVilenkin96} 
may be represented as a regular brane complex consisting of a finite cosmic
string with a pair of point particles at its ends, in an approximation
neglecting Higgs field radiation. (However allowance for radiation
would require the use of an extended complex including the Higgs medium
whose interaction with the string -- and a fortiori with the
terminating particles -- would violate the regularity condition: the
ensuing singularities in the back reaction would need to be treated by
a renormalisation procedure of a
kind \citeA{Shellard90}, \citeA{DabholkarQuashnock90}, \citeA{BattyeShellard95}, \citeA{BattyeShellard96} whose development so
far has been beset with difficulties in preserving exact local Lorentz
invariance, an awkward problem that is beyond the scope of the present
article.)

The present section will be restricted to the case of a brane complex that
is not only regular in the sense of the preceeding paragraph but that is
also {\it pure} (or ``fine") in the sense that the lengthscales
characterising the internal structure of the (defect or other) localised
phenomenon represented by the brane models are short compared with those
characterising the macroscopic variations under consideration so that
polarisation effects play no role. For instance in the case of  a point
particle, the restriction that it should be describable as a ``pure" zero
brane simply means that it can be represented as a simple monopole without
any dipole or higher multipole effects. In the case of a cosmic string the
use of a ``pure" 1-brane description requires that the underlying vortex
defect be sufficiently thin compared not only compared with its total length
but also compared with the lengthscales characterising its curvature
and the gradients of any currents it may be carrying. The effect of the
simplest kind of curvature corrections beyond this ``pure brane" limit will
be discussed in Subsection \ref{3-5}, but in the rest of these lectures,
as in the present section, it will be assumed that the ratio of microscopic to
macroscopic lengthscales is sufficiently small for description in terms of
``pure" p-branes to be adequate.

The present section will not be concerned with the specific details of
particular cases but with the generally valid laws that can be derived as
Noether identities from the postulate that the model is governed by dynamical
laws derivable from a variational principle specified in terms of an action
function ${\Ac}$. It is however to be emphasised that the validity at a
macroscopic level of the laws given here is not restricted to cases
represented by macroscopic models of the strictly conservative type directly
governed by a macroscopic variational principle. The laws obtained here will
also be applicable to classical models of dissipative type (e.g. allowing for
resistivity to relative flow by internal currents) as necessary conditions for
the existence of an underlying variational description  of the microscopic
(quantum) degrees of freedom that are allowed for merely as entropy in the
macroscopically averaged classical description.

In the case of a brane complex, the total action $\Ac$ will be given as a sum
of distinct d-surface integrals respectively contributed by the various
(d--1)-branes of the complex, of which each is supposed to have its own
corresponding Lagrangian surface density scalar $\ud \olag$ say. Each supporting
d-surface will be specified by a mapping $\sigme\mapsto x\{\sigme\}$ giving
the local background coordinates $x^\mu$ ($\mu$=0, .... , n--1) as functions
of local internal coordinates $\sigme^\ii\ $ ( i=0, ... , d--1). The
corresponding d-dimensional surface metric tensor $\ud \hg_{\ii\ji}$ that is
induced (in the manner described in Subsection \ref{1-1}) as the pull back
of the n-dimensional background spacetime metric $\bg_{\mu\nu}$, will
determine the natural surface measure, $\ud d\!\oS$, in terms of 
which the total action will be expressible in the form 
\be \Ac= \sum_{\rm d} \int\!\! \ud d\!\oS\, \ud \olag \ ,
 \hskip 1 cm \ud d\!\oS=\sqrt{\Vert^{_{\rm (d)}} \hg \Vert}\,
 d^{\rm d}\!\sigme \, . \eqn{2.1}\fe
As a formal artifice whose use is an unnecessary complication in ordinary
dynamical calculations but that can be useful for  purposes such as the
calculation of radiation, the {\it confined} (d-surface supported) but
locally {\it regular} Lagrangian scalar fields $\ud \olag$ can be
replaced by corresponding unconfined, so no longer regular but {\it
distributional} fields $\ud \hlag$, in order to allow the the basic
multidimensional action (\ref{2.1}) to be represented as a single integral, 
\be \Ac =\int\!\! d\!{\calS}\, \sum_{\rm d} \ud \hlag \ ,
 \hskip 1 cm d\!{\calS}=\sqrt{\Vert \bg \Vert}\, d^{\rm n}x \, .\eqn{2.2}\fe
over the n-dimensional {\it background} spacetime. In order to do this, it is
evident that for each (d--1)-brane of the complex the required distributional
action contribution $\ud\hlag$  must be constructed in terms of the
corresponding regular d-surface density scalar $\ud\olag$ according to the
prescription that is expressible in standard Dirac notation as
\be  \ud\hlag=\Vert \bg\Vert^{-1/2}\int\!\! \ud d\!\oS\,
\ud\olag\, \delta^{\rm n}[x-x\{\sigme\}]\, .\eqn{2.3}\fe

\subsection{Current, vorticity, and stress-energy tensor}
\label{2-2}

In the kind of model under consideration, each supporting d-surface is
supposed to be endowed with its own independent internal field variables which
are allowed to couple with each other and with their derivatives in the
corresponding d-surface Lagrangian contribution $\ud\olag$, and which are also
allowed to couple into the Lagrangian contribution $\udi\olag$ on any of its
attached boundary (d--1) surfaces, though -- in order not to violate the
strong dynamic regularity condition -- they are not allowed to couple into
contributions of dimension (d--2) or lower. As well as involving its own
d-brane surface fields and those of any (d+1) brane to whose boundary it may
belong, each contribution $\ud\olag$ may also depend passively on the fields
of a fixed higher dimensional background. Such fields will of course always
include the background spacetime metric $\bg_{\mu\nu}$ itself. Apart from that,
the most commonly relevant kind of backround field (the only one allowed for
in the earlier analysis, \citeA{Carter90}) is a Maxwellian gauge potential $\bA_\mu$ whose
exterior derivative is the automatically ``closed" electromagnetic field,
\be \bF_{\mu\nu}=2\nabl_{[\mu}\bA_{\nu]} \ , \hskip 1 cm
\nabl_{[\mu}\bF_{\nu\rho]}= 0\, .\eqn{2.4}\fe
Although many other possibilities can in principle be enviseaged, for the
sake of simplicity the  following analysis will not go beyond allowance for
the only one that is important in applications to the kind of cosmic or
superfluid defects that are the subject of discussion in the present volume,
namely an antisymmetric Kalb-Ramond gauge field $\bB_{\mu\nu}=-\bB_{\nu\mu}$
whose exterior derivative is an automativally closed physical current
3-form,
\be \bN_{\mu\nu\rho}=3\nabl_{[\mu}\bB_{\nu\rho]} \, ,\hskip 1 cm
\nabl_{[\mu}\bN_{\nu\rho\sigma]}=0\, .\eqn{2.5}\fe
Just as a Maxwellian gauge transformation of the form $\bA_\mu\mapsto
\bA_\mu+\nabl_\mu\alpha$ for an arbitrary scalar $\alpha$ leaves the
electromagnetic field (\ref{2.4}) invariant, so analogously a Kalb-Ramond
gauge transformation $\bB_{\mu\nu}\mapsto \bB_{\mu\nu}+2\nabl_{[\mu}\chi_{\nu]}$
for an arbitrary covector $\chi_\mu$ leaves the corresponding current 3-form
(\ref{2.5}) invariant. In applications to ordinary 4-dimensional spacetime,
the current 3-form will just be the dual
$\bN_{\mu\nu\rho}=\bepsilon_{\mu\nu\rho\sigma} \bN^\sigma$ of an ordinary current
vector  $\bN^\mu$ satisfying a conservation law of the usual type, $\nabl_\mu
\bN^\mu=0$. Such a Kalb-Ramond representation can be used to provide an elegant
variational formulation for ordinary perfect fluid theory \citeA{Carter94} and is
particularly convenient for setting up ``global" string models of vortices
both in a simple cosmic axion or Higgs field \citeA{VilenkinVachaspati87}, \citeA{DavisShellard89}, \citeA{Sakellariadou91}
and in a superfluid \citeA{BenYaacov92} such as liquid Helium-4.

In accordance with the preceeding considerations, the analysis that follows
will be based on the postulate that the action is covariantly and gauge
invariantly determined by specifying each scalar Lagrangian contribution
$\ud\olag$ as a function just of the background fields, $\bA_\mu$,
$\bB_{\mu\nu}$ and of course $\bg_{\mu\nu}$, and of any relevant internal fields
(which in the simplest non-trivial case -- exemplified by  string
models \citeA{Carter89}, \citeA{Larsen93} of the category needed for the macroscopic
description of Witten type \citeA{Witten85} superconducting vortices -- consist just
of a phase scalar $\phie$). In accordance with the restriction that the
branes be ``pure" or ``fine" in the sense explained above, it is postulated
that polarisation effects are excluded by ruling out couplings involving
gradients of the background fields. This means that the effect of making
arbitrary infinitesimal ``Lagrangian" variations $\dL \bA_\mu$,
$\dL \bB_{\mu\nu}$, $\dL \bg_{\mu\nu}$ of the background fields
will be to induce a corresponding variation $\d\Ac$ of the action that simply
has the form
\be \d \Ac=\sum_{\rm d}\int\!\! \ud d\!\oS\left\{
\ud\oj{^\mu} \dL \bA_\mu +{_1\over ^2}\ud\oW{^{\mu\nu}}
\dL \bB_{\mu\nu}+ {_1\over^2} \ud\oT{^{\mu\nu}} 
\dL  \bg_{\mu\nu} \right\} \, , \eqn{2.6}\fe
provided either that that the relevant independent internal field components
are fixed or else that the internal dynamic equations of motion are
satisfied in accordance with the variational principle stipulating that
variations of the relevant independent field variables should make no
difference. For each d-brane of the complex, this partial differentiation
formula implicitly specifies the corresponding {\it electromagnetic surface
current density} vector $\ud\oj{^{\mu}}$, the {\it surface vorticity flux} 
bivector $\ud \oW{^{\mu\nu}}=-\ud\oW{^{\nu\mu}}$, and the {\it surface stress 
momentum energy density} tensor $\ud\oT{^{\mu\nu}}=\ud\oT{^{\nu\mu}}$, 
which are formally expressible more explicitly as 
\be \ud\oj{^{\mu}} = {\partial\!\ud \olag\over\partial \bA_\mu }
\, , \hskip 0.6 cm 
\ud\oW{^{\mu\nu}} = 2{\partial\! \ud\olag\over\partial \bB_{\mu\nu}}
\, ,\fe 
and 
\be \ud\oT{^{\mu\nu}}=2{\partial\!\ud \olag\over\partial 
\bg_{\mu\nu}}+\ud\olag\ud\eta^{\mu\nu}  
\, , \eqn{2.7}\fe
of which the latter is obtained using the formula
\be \dL (\ud d\!\oS)={_1\over^2}\ud\og^{\mu\nu}
(\dL \bg_{\mu\nu}) \ud d\!\oS\, ,\eqn{varmes}\fe
where $\ud\og^{\mu\nu}$ is the rank-d {\it fundamental tensor} of the
d-dimensional imbedding, as defined in the manner described in the
preceeding section.

\subsection{Conservation of current and vorticity}
\label{2-3}

The condition that the action be gauge invariant means that if one simply
sets $\dL \bA_\mu=\nabl_\mu \alpha$, $\dL \bB_{\mu\nu}
=2\nabl_{[\mu}\chi_{\nu]}$, $d_{_{\rm L}}\bg_{\mu\nu}=0$, for arbitrarily 
chosen $\alpha$ and $\chi_\mu$ then $\d\Ac$ should simply vanish, i.e.
\be \sum_{\rm d}\int\!\! d\!\ud\oS\left\{
\ud\oj{^\mu} \nabl_\mu\alpha+\ud\oW{^{\mu\nu}}\nabl_\mu\chi_\nu
 \right\}=0 \, . \eqn{2.8}\fe
In order for this to be able to hold for all possible fields $\alpha$ and
$\chi_\mu$ it is evident that the surface current $\ud\oj{^\mu}$ and the
vorticity flux bivector $\ud\oW{^{\mu\nu}}$ must (as one would anyway expect
from the consideration that they depend just on the relevant internal
d-surface fields) be purely d-surface tangential, i.e. their contractions
with the relevant rank (n--d) orthogonal projector $\ud\!\!\ag^\mu_{\
\nu}=\bg^\mu_{\ \nu}- \ud\og^\mu_{\ \nu}$ must vanish:
\be \ud\!\!\ag^\mu_{\ \nu}\ud\oj{^\nu}=0\ , \hskip 1 cm \ud\!\!
\ag^\mu_{\ \nu}\ud\oW{^{\nu\rho}}=0\, .\eqn{2.9}\fe
Hence, decomposing the full gradient operator $\nabl_\mu$ as the sum of
its tangentially projected part $\ud\onab_\mu=\ud\og^\nu_{\ \mu}\nabl_\nu$
and of its orthogonally projected part $\ud\!\!\ag^\nu_{\ \mu}\nabl_\mu$, 
and noting that by (\ref{2.9}) the latter will give no contribution, one sees 
that (\ref{2.8}) will take the form
\be \sum_{\rm d}\int\!\!\ud d\!\oS\left\{ \ud\onab_\mu
\Big(\ud\oj{^\mu} \alpha+\ud\oW{^{\mu\nu}}\chi_\nu \Big)
-\alpha\ud\onab_\mu\oj{^\mu} -\chi_\nu\ud\onab_\mu
\ud\oW{^{\mu\nu}}\right\}=0  \, , \eqn{2.10}\fe
in which first term of each integrand is a pure surface divergence.
Such a divergence can be dealt with using Green's theorem, according to which,
for any d-dimensional support surface $\ud\oS$ of a (d--1)-brane,
one has  the identity
\be \int\!\! \ud d\!\oS\ud\onab_\mu\ud\oj{^\mu}=
\oint\! \udi d\!\oS\udp\lamb_\mu\ud\oj{^\mu}\, ,\eqn{2.11}\fe
where the integral on the right is taken over the boundary (d--1)-surface of
$\partial\ud\oS$ of $\ud\oS$, and $\udp\lamb_\mu$ is the
(uniquely defined) unit tangent vector on the d-surface that is directed
normally outwards at its (d--1)-dimensional boundary. Bearing in mind that a
membrane support hypersurface can belong to the boundary of two distinct
media, and that for d$\leq$ n--3 a d-brane may belong to a common boundary
joining three or more distinct (d+1)-branes of the complex under
consideration, one sees that (\ref{2.10}) is equivalent to the condition
\begin{eqnarray} 
\sum_{\rm p}\int\!\! \up d\!\oS\,\Big\{ \alpha\Big(\!
\up\onab_\mu\up\oj{^\mu}\! -\sum_{\rm d=p+1}\!\!\udp\lamb_\mu
\ud\oj{^\mu}\Big) \qquad & & \nonumber\\ 
+ \chi_\nu\Big(\!\up\onab_\mu\up\oW{^{\mu\nu}}\! 
-\sum_{\rm d=p+1}\!\!\udp\lamb_\mu\ud\oW{^{\mu\nu}}\Big)\Big\} & = & 0
 \, ,\eqn{2.12}
\end{eqnarray}
where, for a particular p-dimensionally supported (p--1)-brane, the summation
``over d=p+1" is to be understood as consisting of a contribution from each
(p+1)-dimensionally supported p-brane attached to it, where for each such
p-brane, $\udp\lamb_\mu$ denotes the (uniquely defined) unit tangent vector
on its (p+1)-dimensional support surface that is directed normally towards the
p-dimensional support surface of the boundary (p--1)-brane. The Maxwell gauge
invariance requirement to the effect that (\ref{2.12}) should hold for
arbitrary $\alpha$ can be seen to entail an electromagnetic charge
conservation law of the form
\be \up\onab_\mu\up\oj{^\mu}=\sum_{\rm d=p+1}\!\!\udp\lamb_\mu
\ud\oj{^\mu}\, . \eqn{2.13}\fe
This can be seen from (\ref{2.11}) to be be interpretable as meaning
that the total charge flowing out of particular (d--1)-brane from
its boundary is balanced by the total charge  flowing  into it from
any d-branes to which it may be attached. The analogous Kalb-Ramond
gauge invariance requirement that (\ref{2.12}) should also hold for arbitrary
$\chi_\mu$ can be seen to entail a corresponding vorticity conservation
law of the form
\be \up\onab_\mu\up\oW{^{\mu\nu}}=\sum_{\rm d=p+1}\!\!
\udp\lamb_\mu\ud\oW{^{\mu\nu}}\, .\eqn{2.14}\fe
A more sophisticated but less practical way of deriving the foregoing
conservation laws would be to work not from the expression (\ref{2.1}) in
terms  of ordinary surface integrals but instead to use the superficially
simpler expression (\ref{2.2}) in terms of distributions, which leads to the
replacement of (\ref{2.13}) by the ultimately equivalent (more formally
obvious but less directly meaningful) expression
\be \nabl_\mu\Big(\sum_{\rm d}\!\!\ud\hj^{\mu}\Big)=0 \eqn{2.15}\fe
involving the no longer regular but Dirac
distributional current $\ud\hj^\mu$ that is given in terms of the
corresponding regular surface current $\ud\oj^\mu$ by
\be \ud \hj^{\mu}=\Vert\bg \Vert^{-1/2}\int\!\!\ud d\!\oS\,
\ud \oj{^{\mu}} \delta^{\rm n}[x-x\{\sigme\}]\, .\eqn{2.16}\fe
Similarly one can if one wishes rewrite the vorticity flux conservation 
law (\ref{2.14}) in the distributional form
\be \nabl_\mu\Big(\sum\!\! \ud \hW^{\mu\nu}\Big)=0\, , \eqn{2.17}\fe
where the distributional vorticity flux $\ud \hW^{\mu\nu}$ is given in
terms of the corresponding regular surface flux $\ud\oW^{\mu\nu}$ by
\be \ud \hW^{\mu\nu}=\Vert\bg \Vert^{-1/2}\int\!\!\ud d\!\oS\,
\ud \oW{^{\mu\nu}} \delta^{\rm n}[x-x\{\sigme\}]\, .\eqn{2.18}\fe
It is left as an entirely optional exercise for any  readers who may be adept
in distribution theory to show how the ordinary functional relationships
(\ref{2.13}) and (\ref{2.14}) can be recovered by by integrating out the Dirac
distributions in (\ref{2.15}) and (\ref{2.17}).

\subsection{Force and the stress balance equation}
\label{2-4}

The condition that the hypothetical variations introduced in (\ref{2.6})
should be ``Lagrangian" simply means that they are to be understood to be
measured with respect to a reference system that is comoving with the various
branes under consideration, so that their localisation with respect to it
remains fixed. This condition is necessary for the variation to be meaningly
definable at all for a field whose support is confined to a particular brane
locus, but in the case of an unrestricted background field one can enviseage
the alternative possibility of an ``Eulerian" variation, meaning one defined
with respect to a reference system that is fixed in advance, independently of
the localisation of the brane complex, the standard example being that of a
Minkowski reference system in the case of a background that is flat. In such a
case the relation between the  more generally meaningfull Lagrangian
(comoving) variation, denoted by $\dL $, and the corresponding
Eulerian (fixed point) variation denoted by $\dE $ say will be given
by Lie differentiation with respect to the vector field $\xi^\mu$ say that
specifies the infinitesimal of the comoving reference system with respect to
the fixed background, i.e. one has
\be \dL -\dE=\vec{\ \xi\Libra}\, ,\eqn{2.19}\fe
where the Lie differentiation operator $\vec{\ \xi\Libra}$ is given for the
background fields under consideration here by
\be \vec{\ \xi\Libra} \bA_\mu=\xi^\nu\nabl_\nu 
\bA_\mu+\bA_\nu\nabl_\mu\xi^\nu \, , \fe \be
\vec{\ \xi\Libra} \bB_{\mu\nu}=\xi^\nu\nabl_\nu \bB_{\mu\nu}
+2\bB_{\rho[\nu}\nabl_{\mu]}\xi^\rho \, , \fe  \be
\vec{\ \xi\Libra} \bg_{\mu\nu}= 2\nabl_{(\mu}\xi_{\nu)}\, .\eqn{2.20}\fe
 
This brings us to the main point of this section which is the derivation of
the dynamic equations governing the extrinsic motion of the branes of the
complex, which are obtained from the variational principle to the effect that
the action $\Ac$ is left invariant not only by infinitesimal variations of the
relevant independent intrinsic fields on the support surfaces but also by
infinitesimal displacements of the support surfaces themselves. Since the
background fields $\bA_\mu$, $\bB_{\mu\nu}$, and $\bg_{\mu\nu}$ are to be considered
as fixed, the relevant Eulerian variations simply vanish, and so the resulting
Lagrangian variations will be directly identifiable with the corresponding Lie
derivatives -- as given by (\ref{2.20}) -- with respect to the generating
vector field $\xi^\mu$ of the infinitesimal displacement under consideration.
The variational principle governing the equations of extrinsic motion is thus
obtained by setting to zero the result of substituting these Lie derivatives
in place of the corresponding Lagrangian variations in the more general
variation formula (\ref{2.6}), which gives
\be \sum_{\rm d}\int\!\! \ud d\!\oS\left\{
\ud\oj{^\mu} \vec{\ \xi\Libra} \bA_\mu +{_1\over ^2}\ud\oW{^{\mu\nu}}
\vec{\ \xi\Libra} \bB_{\mu\nu}+ {_1\over^2} \ud\oT{^{\mu\nu}}\vec{\ \xi\Libra}
\bg_{\mu\nu} \right\}=0 \, . \eqn{2.21}\fe
The requirement that this should hold for any choice of $\xi^\mu$ evidently
implies that the tangentiality conditions (\ref{2.9}) for the surface fluxes
$\ud\oj{^\mu}$ and $\ud\oW{^{\mu\nu}}$ must be supplemented by an
analogous d-surface tangentiality condition for the surface stress momentum
energy tensor $\ud\oT{^{\mu\nu}}$, which must satisfy
\be \ud\!\!\ag^\mu_{\ \nu}\ud\oT{^{\nu\rho}}=0\, .\eqn{2.22}\fe
(as again one would  expect anyway from the consideration that it depends just
on the relevant internal d-surface fields). This allows  (\ref{2.20}) to be
written out in the form
\begin{eqnarray} 
\sum_{\rm d}\int\!\!\ud d\!\oS\Big\{
\xi^\rho\Big(\bF_{\rho\mu}\ud\oj{^\mu}\!+{_1\over^2}\bN_{\rho\mu\nu}\ud\oW
^{\mu\nu}\! \qquad\qquad & & \nonumber\\ 
-\ud\onab_\mu\ud\oT{^\mu_\rho}-\bA_\rho\ud\onab_\mu
\ud\oj{^\mu}\! -\bB_{\rho\nu}\ud\onab_\mu\ud\oW{^{\mu\nu}}\Big) & & \nonumber\\  
+ \ud\onab_\mu\Big(\xi^\rho(
\bA_\rho \ud\oj{^\mu}\! + \bB_{\rho\nu} \ud\oW{^{\mu\nu}}\!+
\ud\oT{^\mu}_\rho)\Big) \Big\} & = & 0 \, , \eqn{2.23}
\end{eqnarray}
in which the final contribution is a pure surface divergence that can be dealt
with using Green's theorem as before. Using the results (\ref{2.13}) and
(\ref{2.14}) of the analysis of the consequences of gauge invariance and
proceeding as in their derivation above, one sees that the condition for
(\ref{2.23}) to hold for an arbitrary field $\xi^\mu$ is that, on each
(p--1)-brane of the complex, the dynamical equations
\be \up\onab_\mu\up\oT{^\mu}_\rho= \up \tf_\rho \, , \eqn{2.24}\fe
should be satisfied for a total force density $\up \tf_\rho$ given by
\be \up \tf_\rho=\up\of_\rho+\up\cf_\rho \, , \eqn{2.24a}\fe
where $\up\cf_\rho$ is the contribution of the contact force exerted
on the p-surface by other members of the brane complex, which takes the
form
\be \up\cf_\rho= \sum_{\rm d=p+1}\!\!
\udp\lamb_\mu\ud\oT{^\mu}_\rho  \, ,\eqn{2.25}\fe
while the other force density contribution $\up\of_\rho$ represents
the effect of the external background fields, which  is given by
\be \up\of_\rho=\bF_{\rho\mu}\up\oj{^\mu}+{_1\over^2}
\bN_{\rho\mu\nu}\up\oW{^{\mu\nu}}\, .\eqn{2.26}\fe
As before, the summation ``over d=p+1" in (\ref{2.25}) is to be understood as
consisting of a contribution from each of the p-branes attached to the (p--1)-
brane under consideration, where for each such attached p-brane,
$\udp\lamb_\mu$ denotes the (uniquely defined) unit tangent vector on its
(p+1)-dimensional support surface that is directed normally towards the
p-dimensional support surface of the boundary (p--1)-brane. The first of the
background force contibutions in (\ref{2.26}) is of course the Lorentz type
force density resulting from the effect of the electromagnetic field on the
surface current, while the other contribution in (\ref{2.26}) is a Joukowsky
type force density (of the kind responsible for the lift on an aerofoil)
resulting from the Magnus effect, which acts in the case of a ``global"
string \citeA{VilenkinVachaspati87}, \citeA{DavisShellard89} though not in the case of a string of the
``local" type for which the relevant vorticity flux $\up\oW{^{\mu\nu}}$ will
be zero. As with the conservation laws (\ref{2.13}) and (\ref{2.14}), so also
the explicit force density balance law expressed by (\ref{2.24})  can
alternatively be expressed in terms of the corresponding Dirac distributional
stress momentum energy and background force density tensors, 
$\ud\hT^{\mu\nu}$ and $\ud\hf_\mu$, which are given for each (d--1)-brane in
terms of the corresponding regular surface stress momentum energy and
background force density tensors $\ud\oT{^{\mu\nu}}$ and $\ud\of_\mu$ by
\be  \ud \hT^{\mu\nu}=\Vert\bg \Vert^{-1/2}\int\!\!\ud d\!\oS\,
\ud \oT{^{\mu\nu}} \delta^{\rm n}[x-x\{\sigme\}]\ \eqn{2.27}\fe
and
\be \ud \hf_\mu=\Vert\bg \Vert^{-1/2}\int\!\!\ud d\!\oS\,
\ud \of_\mu \delta^{\rm n}[x-x\{\sigme\}]\, . \eqn{2.28}\fe
The equivalent -- more formally obvious but less explicitly meaningful --
distributional versional version of the force balance law (\ref{2.24}) takes
the form
\be \nabl_\mu\Big(\sum_{\rm d}\!\! \ud \hT^{\mu\nu}\Big)
=\hf_\mu\, , \eqn{2.29}\fe
where the total Dirac distributional force density is given in terms of the 
electromagnetic current distributions (\ref{2.16}) and the vorticity flux 
distributions (\ref{2.18}) by
\be \hf_\mu=\bF_{\rho\mu}\sum_{\rm d}\!\!\ud\hj^\mu+{_1\over^2}
\bN_{\rho\mu\nu}\sum_{\rm d}\!\!\ud\hW^{\mu\nu}\, .\eqn{2.30}\fe
It is again left as an optional exercise for readers who are adept in the use
of Dirac distributions to show that the system (\ref{2.24}), (\ref{2.25}),
(\ref{2.26}) is obtainable from (\ref{2.29}) and (\ref{2.30}) by substituting
(\ref{2.16}), (\ref{2.18}), (\ref{2.27}), (\ref{2.28}).

As an immediate corollary of (\ref{2.24}), it is to be noted that for any 
vector field $\el^\mu$ that generates a continuous symmetry of the background
spacetime metric, i.e. for any solution of the Killing equations
\be \nabl_{(\mu}\el_{\nu)}=0\, ,\eqn{2.31}\fe
one can construct a corresponding surface momentum or energy density current
\be \up\oP{^\mu}=\up\oT{^{\mu\nu}}\el_\nu\ ,\eqn{2.32}\fe
that will satisfy
\be \up\onab_\mu\up\oP{^\mu}= \sum_{\rm d=p+1}\!\!
\udp\lamb_\mu\ud\oP{^\mu}+\up\of_\mu k^\mu  \, .\eqn{2.33}\fe
In typical applications for which the n-dimensional background spacetime can
be taken to be flat there will be n independent translation Killing vectors
which alone (without recourse to the further n(n--1)/2 rotation and boost
Killing vectors of the Lorentz algebra) will provide a set of relations of the
form (\ref{2.33}) that together provide the same information as that in the
full force balance equation (\ref{2.24}) or (\ref{2.29}).

\subsection{The equation of extrinsic motion}
\label{2-5}

Rather than the distributional version (\ref{2.29}), it is the explicit
version (\ref{2.24}) of the force balance law that is directly useful for
calculating the dynamic evolution of the brane support surfaces. Since the
relation (\ref{2.29}) involves n independent components whereas the support
surface involved is only p-dimensional, there is a certain redundancy, which
results from the fact that if the virtual displacement field $\xi^\mu$ is
tangential to the surface in question it cannot affect the action. Thus if
$\up\!\!\ag^{\!\mu}_{\,\nu}\xi^\nu=0$, the condition (\ref{2.21}) will be
satisfied as a mere identity -- provided of course that the field equations
governing the internal fields of the system are satisfied. It follows that the
non-redundent information governing the extrinsic motion of the p-dimensional
support surface will be given just by the orthogonally projected part of
(\ref{2.24}). Integrating by parts, using the fact that, by (\ref{1.6}) and
(\ref{1.15}), the surface gradient of the rank-(n--p) orthogonal projector
$\up\!\!\ag^{\!\mu}_{\,\nu}$ will be given in terms of the second
fundamental tensor $\up \Ke_{\mu\nu}^{\ \ \,\rho}$ of the p-surface by
\be \up\onab_\mu\up\!\!\ag^{\!\nu}_{\,\rho}=-\up \Ke_{\mu\nu}^{\ \ \,\rho}
-\up \Ke_{\mu\ \nu}^{\ \rho} \, ,\eqn{2.34}\fe
it can be seen that the extrinsic equations of motion obtained as the
orthogonally projected part of (\ref{2.24}) will finally be expressible by
\be \up\oT{^{\mu\nu}}\up \Ke_{\mu\nu}^{\ \ \,\rho}=
\up\!\!\ag^{\!\rho}_{\,\mu}\up \tf^\mu\, .\eqn{2.35}\fe

It is to be emphasised that the formal validity of the formula that has just
been derived is not confined to the variational models on which the above
derivation is based, but also extends to dissipative models (involving effects
such as external drag by the background
medium \citeA{Carteretal94}, \citeA{Vilenkin91}, \citeA{GarrigaSakellariadou93} or mutual resistance between
independent internal currents). The condition that even a non-conservative
macroscopic model should be compatible with an underlying microscopic model of
conservative type  requires the existence (representing to averages of
corresponding microscopic quantities) of appropriate stress momentum energy
density and force density fields satisfying (\ref{2.35}). 

The ubiquitously applicable formula (\ref{2.35})is interpretable as
being just the natural higher generalisation of ``Newton's law"
(equating the product of mass with acceleration to the applied force)
in the case of a particle. The surface stress momentum energy tensor,
$\up\oT{^{\mu\nu}}$, generalises the mass, and the second fundamental
tensor, $\up \Ke_{\mu\nu}^{\ \ \,\rho}$, generalises the acceleration.

The way this works out in the 1-dimensional case of a ``pure" point
particle (i.e. a monopole) of mass $\mag$, for which the Lagrangian is
given simply by $^{_{(1)}}\!\olag=-\mag$, is as follows. The 1-dimensional
energy tensor will be obtained in terms of the unit tangent vector
$\ue^\mu$ ($\ue^\mu \ue_\mu=-1$) as $^{_{(1)}}\! \oT{^{\mu\nu}}=\mag\, \ue^\mu
\ue^\nu$, and in this zero-brane case, the first fundamental tensor will
simply be given by $^{_{(1)}}\!\og^{\mu\nu}=-\ue^\mu \ue^\nu$, so that the
second fundamental tensor will be obtained in terms of the acceleration
$\dot \ue^\mu =\ue^\nu\nabl_\nu \ue^\mu$ as
$^{_{(1)}}\!\Ke_{\mu\nu}^{\ \ \,\rho} =\ue_\mu \ue_\nu \dot \ue^\rho$. Thus
(\ref{2.35}) can be seen to reduce in the case of a particle simply to
the usual familiar form $\mag\, \dot \ue^\rho=
^{_{(1)}}\!\ag^{\!\rho}_{\,\mu}{^{_{(1)}}}\! \tf^\mu$.

\section{Perturbations and curvature effects beyond the pure brane limit}
\label{Section3}

\subsection{First order perturbations and the extrinsic 
characteristic equation} \label{3-1}
 
Two of the most useful formulae for the analysis of small perturbations of a
string or higher brane worldsheet are the expressions for the infinitesimal
Lagrangian (comoving) variation of the first and second fundamental tensors in
terms of the corresponding comoving variation $\dL \bg_{\mu\nu}$ of the metric
(with respect to the comoving reference system). For the first fundamental
tensor one easily obtains 
\be\dL\og^{\mu\nu}= -\og^{\mu\rho}\og^{\mu\sigma}\dL \bg_{\rho\sigma}\ ,
\hskip 1 cm\dL \og^\mu_{\ \nu}=\og^{\mu\rho}\!\ag^{\!\sigma}_{\,\nu}
\dL \bg_{\rho\sigma} 
\eqn{3.1}\fe
and, by substituting this in the defining relation (\ref{1.10}),  the
corresponding Lagrangian variation of the second fundamental tensor is
obtained \citeA{Carter93} as 

\be \dL \Ke_{\mu\nu}^{\ \ \,\rho}=\ag^{\!\rho}_{\,\lambda}\og^\sigma_{\,\mu}
\og^\tau_{\,\nu}\,\dL\Gamma_{\sigma\ \tau}^{\ \lambda} +\big(2\ag^{\!\sigma}
_{\,(\mu} \Ke_{\nu)}^{\ \ \tau\rho}-\Ke_{\mu\nu}^{\ \ \,\sigma}\og^{\tau\rho}
\big)\dL \bg_{\sigma\tau} \, ,\eqn{3.2}\fe
where the Lagrangian variation of the connection (\ref{1.18}) is given 
by the well known formula
\be \dL\Gamma_{\sigma\ \tau}^{\ \lambda} =g^{\lambda\rho}\big(
\nabl_{(\sigma\,}\dL \bg_{\tau)\rho}-{_1\over^2}\nabl_{\rho\,}\dL \bg_{\sigma\tau}
\big)\, .\eqn{3.3}\fe
Since we are concerned here only with cases for which the background is fixed
in advance so that the Eulerian variation $d_{\rm_E}$ will vanish in
(\ref{2.19}), the Lagrangian variation of the metric will be given just by its
Lie derivative with respect to the infinitesimal displacement vector field
$\xi^\mu$ that generates the displacement of the worldsheet under
consideration, i.e. we shall simply have
\be \dL \bg_{\sigma\tau}=2\nabl_{(\sigma}\xi_{\tau)} \, .\eqn{3.4}\fe
It then follows from (\ref{3.3}) that the Lagrangian variation of the
connection will be given by
\be \dL\Gamma_{\sigma\ \tau}^{\ \lambda}=\nabl_{(\sigma}\nabl_{\tau)}
\xi^\lambda-\calR^\lambda_{\ (\sigma\tau)\rho}\xi^\rho\, ,\eqn{3.5}\fe
where $\calR^\lambda_{\ \sigma\tau\rho}$ is the background Riemann curvature 
(which will be negligible in typical applications for which the lengthscales
characterising the geometric features of interest will be small compared
with those characterising any background spacetime curvature).
The Lagrangian variation of the first fundamental tensor is thus
finally obtained in the form
\be \dL\og^{\mu\nu}=-2\og_\sigma^{\,(\mu}\onab{^{\,\nu)}}\xi^\sigma\, ,
\eqn{3.6}\fe
while that of the second fundamental tensor is found to be given by
\be \dL \Ke_{\mu\nu}^{\ \ \,\rho}=\ag^{\!\rho}_{\,\lambda}\big(
\onab_{(\mu}\onab_{\nu)}\xi^\lambda-\og^\sigma_{\,(\mu}
\og^\tau_{\ \nu)}\calR^\lambda_{\ \sigma\tau\rho}\xi^\rho-\Ke^\sigma_{\ (\mu\nu)}
\onab_\sigma\xi^\lambda\big)+\fe \be \hskip 2cm
\big(2\ag^{\!\sigma}_{\,(\mu}\Ke_{\nu)\tau}^{\ \ \ \rho}-g^\rho_{\,\tau}
\Ke_{\mu\nu}^{\ \ \,\sigma}\big)\big(\nabl_\sigma\xi^\tau +
\onab{^{\,\tau}}\xi_\sigma\big)\,  .\eqn{3.7}\fe

It is instructive to apply the forgoing formulae to the case of a
{\it free} pure brane worldsheet, meaning one for which there is no external
force contribution so that the equation of extinsic motion reduces to the
form
\be \oT{^{\mu\nu}}\Ke_{\mu\nu}^{\ \ \,\rho}=0\, .\eqn{3.8}\fe
On varying the relation (\ref{3.8}) using (\ref{3.7}) in conjunction with the
orthogonality property (\ref{2.22}) and the unperturbed equation (\ref{3.8})
itself, the equation governing the propagation of the infinitesimal
displacement vector is obtained in the form
\be \ag^{\!\rho}_{\, \lambda}\oT{^{\mu\nu}}\big(\onab_\mu\onab_\nu
\xi^\lambda-\calR^\lambda_{\ \mu\nu\sigma}\xi^\sigma\big)
=-\Ke_{\mu\nu}^{\ \ \,\rho\,}\dL \oT{^{\mu\nu}}\ .\eqn{3.9}\fe

The extrinsic perturbation equation (\ref{3.9}) is by itself only part of the
complete system of perturbation equations governing the evolution of the
brane, the remaining equations of the system being those governing the
evolution of whatever surface current \citeA{Carter89a} and other relevant internal
fields on the supporting worldsheet may be relevant. The perturbations of such
fields are involved in the source term on the right of (\ref{3.9}), whose
explicit evaluation depends on the specific form of the relevant currents or
other internal fields. However it is not necessary to know the specific form
of such internal fields for the purpose just of deriving the characteristic
velocities of propagation of the extrinsic propagations represented by the
displacement vector $\xi^\mu$, so long as they contribute to the source term
on the right of the linearised perturbation equation (\ref{3.9}) only at first
differential order, so that the characteristic velocities will be completely
determined by the first term on the left of (\ref{3.9}) which will be the only
second differential order contribution. It is apparent from (\ref{3.9}) that
under these conditions the equation for the characteristic tangent covector
$\chie_\mu$ say will be given independently of any details of the surface
currents or other internal fields simply \citeA{Carter90} by
\be \oT{^{\mu\nu}}\chie_\mu\chie_\nu=0\, .\eqn{3.10}\fe
(It can be seen that the unperturbed surface stress momentum energy density
tensor $\oT{^{\mu\nu}}$ plays the same role here as that of the unperturbed
metric tensor $\bg^{\mu\nu}$ in the analogous characteristic equation for the
familiar case of a massless background spacetime field, as exemplified by
electromagnetic or gravitational radiation.)

\subsection{*Higher order displacements by ``straight'' 
transportation} \label{3-2}

Mere linear perturbation theory, such as described in the preceeding
subsection, is of course insufficient for many physical purposes (such as the
treatment of gravitational radiation reaction for which it is well known to be
necessary to go even beyond second order in the post Newtonian approximation).
Allowance for actual  physical effects involving perturbations at quadratic or
higher order is beyond the scope of the present notes. Nevertheless even when
one's attention is restricted to first order in so far as ``physical''
(on shell) perturbations are concerned, it is still necessary to go to quadratic
order in the treatment of  ``virtual'' (off shell) perturbations if one wants a
variational formulation of the linearised dynamical equations.

The treatment of higher order perturbations -- whether ``physical'' or
``virtual'' -- is intrinsically straightforeward (though it may of course
involve very heavy algebra) when it only involves continuous background fields
for which a purely ``Eulerian'' (effectively fixed) reference system can be
used. However when point particles, strings, or higher branes are involved one
needs to use a comoving ``Lagrangian'' type reference system. in the manner
developed in Section \ref{Section2}. In so far as only first order
perturbation theory is concerned, the necessary adjustment between Eulerian
and Lagrangian systems  can be treated in accordance with (\ref{2.19}) by the
well known procedure whereby the corresponding differences are obtained as Lie
derivatives with respect to the relevant displacement vector field. However a
new issue arises when one comes to generalise such a procedure to quadratic
and higher order perturbation theory because in a general mathematical context
it is only to first order that a small coordinate displacement will
unambiguously characterise and be characterised by a corresponding vector. 

Fortunately for what follows, the physical context with which the present
notes are concerned is not of the most general mathematical kind, but one in
which the background under consideration is endowed with a preferred
connection, $\Gamma_{\nu\ \rho}^{\ \mu}$ say, namely the one derived according
to the standard Riemannian prescription (\ref{1.18}) from the spacetime metric
$\bg_{\mu\nu}$. So long as one has a preferred connection (whether Riemannian or
not) to work with, there will after all be a corresponding precription, namely
the one naturally specified by the corresponding affinely parametrised
geodesic, which we shall refer to simply as ``straight'' transportation,
whereby even to second and higher order, a given vector will unambiguously
determine a corresponding displacement and also (at least provided the
displacement is not unduly large) vice versa. 

To be explicit, let $\xi^\mu$ be the components of a vector at a given
position with coordinates $x^\mu$, and let $\epsilon$ be the corresponding
affine parameter along the geodesic starting from $x^\mu$ for which the
initial affinely normalised tangent vector is $\xi^\mu$. This will
characterise a well defined displacement $x^\mu\mapsto x_{\{1\}}^{\,\mu}$ that
is specified by setting $\epsilon=1$ in the solution
$x_{\{\epsilon\}}^{\,\mu}$ with initial conditions $x_{\{0\}}^{\,\mu}=x^\mu$,
$\ \dot x_{\{0\}}^{\,\mu}=\xi^\mu$ of the affinely parameterised geodesic
equation
\be \ddot x^{\mu}_{\{\epsilon\}}+
\Gamma_{\!\!\{\epsilon\}\nu\ \rho}^{\,\ \ \ \mu}\ \dot x^{\mu}_{\{\epsilon\}}
\dot x^{\rho}_{\{\epsilon\}} =0 \, ,\eqn{afgeod}\fe
where a dot denotes the derivative with respect to the affine parameter. The
required solution of this geodesic equation is conveniently expressible by a
Taylor expansion whose expression to third order will have the form
\be x_{\{\epsilon\}}^{\,\mu}-x^\mu= \epsilon\, \$ x^\mu
+{\epsilon^2\over 2!} \$^2 x^\mu+{\epsilon^3\over 3!} \$^3 x^\mu
 + {\cal O}\{\epsilon^4\} \, ,\eqn{straexp}\fe
using a systematic notation scheme whereby the symbol $ \$ $ is introduced for
the purpose of denoting {\it differentiation with respect to the affine
parameter} at its initial value, $\epsilon=0$, so that for example we shall
have the abbreviation $ \$ x^\mu=\dot x_{\{0\}}^{\,\mu} $. Proceeding by
successive approximations, it is a straightforward exercise to verify that, up
to third order, the required coefficients will be given by
\be \$ x^\mu =\xi^\mu\, ,\hskip 1 cm  \$^2  x^\mu 
= - \Gamma_{\lambda\ \nu}^{\ \mu} \xi^\lambda\xi^\lambda \, ,\fe
\be \$^3  x^\mu =\big( \Gamma_{(\lambda\ \nu)}^{\ \ \mu}
\Gamma_{\rho\ \sigma}^{\ \lambda} -\Gamma_{\nu\ \rho,\sigma}^{\ \mu}\big)
\xi^\nu \xi^\rho \xi^\sigma \, .\eqn{stracoef}\fe

The purpose of this scheme is to obtain a systematic calculus that can be used
to obtain the generalisation to higher orders of the standard formula
(\ref{2.19}) whereby starting from a first order Eulerian variation operator
$\dE$ the corresponding Lagrangian variation operator $\dL$
as specified with respect to a displaced reference system will be given by
$\dL-\dE=\Libra$ where $\Libra$ is the operator of Lie
differentiation with respect to the vector $\xi^\mu$ that characterises the
relative displacement. Starting from a finite Eulerian variation operator
$\DE$ allowing for higher order corrections obtained by
successive applications of the corresponding infinitesimal differential
operator $d_{_{\rm L}}$ via an Taylor expansion of the form $\DE=$ 
$\dE+{^1/_2} \dE^{\ 2}+ ...$ we want to be able to
construct the corresponding finite Lagrangian variation operator
$\DL=$ $\dL+{^1/_2} \dL^{\ 2}+ ...$ as specified
with respect to a reference system that has been subject to a relative
transportation operation representable in terms of local coordinates as
the mapping $x^\mu\mapsto x_{\{1\}}^{\,\mu}$. By expressing this
mapping in terms of the displacement vector $\xi^\mu$ in the manner that
has just beeen described, the required difference will be obtainable
as a Taylor expansion in powers of the vector $\xi^\mu$, and of its first
(but not higher) order gradient tensor $\nabl_\nu\xi^\mu$, in the form
\be \DL-\DE = \$+{_1\over^2}\$^2+{_1\over ^6}\$^3+ ...
 \eqn{Lagdiff}\fe
where $ \$ $, $ \$^2 $, and so on are the relevant operators of first, second,
and higher order ``straight'' differentiation with respect to the
$\xi^\mu$.  The way to work out the explicit form of the action of
such ``straight'' differentiation operators on the fields of the kind
that is of interest -- of which the most important is of course the
metric $\bg_{\mu\nu}$ -- will be described in the following subsection.

Before proceeding, it is to be observed that the  geodesic or ``straight''
transportation operation, based in the manner just described on the solution
of the second order equation (\ref{afgeod}), needs to be distinguished from
the corresponding Lie transportation operation, which is instead based on the
solution of a first order equation of the form $\dot x_{\{\epsilon\}}^{\,\mu}
=\xi^{\,\mu}_{\{\epsilon\}}$. It would in principle be possible to develop
higher order displacement perturbation theory in terms of Lie transportation,
but the disadvantage of such an aproach is is that (as the price of avoiding
dependence on the preferred connection) it would introduce an inordinately
high degree of gauge dependence, since it would depend on the specification of
the vector $\xi^\mu$ not just at the initial point but all the way along the
trajectory of the displacement. However the distinction arises only at
quadratic and higher order: at first order there is no difference between the
``straight'' differentiation operator denoted by $ \$ $, that has just been
defined and the corresponding Lie differentiation operation as customarily
denoted by the symbol $\Libra$. 

It is to be remarked by the way that, just as is customary to use the more
explicit notation  $\vec \xi \Libra$ or alternatively $\Libra_\xi$ whenever it
is necessary to indicate the particular vector field $\xi^\mu$ with respect to
which the Lie differentiation operation $\Libra$ is defined, so analogously we
can use the more explicit notation $\vec\xi \$ $ or alternatively $ \$_\xi$ to
indicate the particular vector field  $\xi^\mu$ with respect to which the
``straight'' differentiation operation $ \$ $ is defined.

\subsection{*``Straight'' differentiation of covariant fields}
\label{3-3}

The ``straight'' displacement scheme set up in the preceeding subsection
will evidently determine corresponding transformations of any continuous 
fields that may be given on the background space. The simplest case is that 
of an ordinary scalar field $\bPhi$ say, for which the (affinely parametrised) 
displacement $x^\mu \mapsto x_{\{\epsilon\}}^{\,\mu}$ 
will determine a corresponding pullback mapping
$\bPhi \mapsto \bPhi_{\{\epsilon\}}$ with $\bPhi_{\{\epsilon\}}\{x\}=\bPhi\{
x_{\{\epsilon\}} \}$. Evaluating this using an expansion of the form
\be \bPhi_{\{\epsilon\}}=\bPhi+\bPhi_{,\mu}(x_{\{\epsilon\}}^{\,\mu}-x^\mu)
+{1\over 2}\bPhi_{,\mu\nu}(x_{\{\epsilon\}}^{\,\mu}-x^\mu)
(x_{\{\epsilon\}}^{\,\nu}-x^\nu)+{\cal O}\{\epsilon^3\}\, ,\eqn{standexp}\fe
it can be seen from (\ref{straexp}) and (\ref{stracoef}) that up to second
order the required result will be given  by an expansion of the standard form
\be \bPhi_{\{\epsilon\}}=\bPhi+\epsilon\, \$ \bPhi +{\epsilon^2\over 2}
\$^2\bPhi +{\cal O}\{\epsilon^3\} \fe
in which the relevant ``straight'' derivative coefficients are given by 
\be \$ \bPhi=\xi^\mu\nabl_\mu\bPhi\, ,\hskip 1cm  \$^2\bPhi
=\xi^\mu\xi^\nu\nabl_\mu\nabl_\nu\bPhi\, .\eqn{strascal}\fe
It is to be noticed that the formulae obtained in this way are of a strictly
tensorial nature, unlike those in (\ref{stracoef}), whose non tensorially
coordinate dependent nature is due to that of the coordinates to which they
apply. It is also to be noticed that the formula for the second order
``straight'' derivative $ \$^2\bPhi $ does not involve derivation of the
displacement vector field $\xi^\mu$, unlike the formula for the corresponding
second order Lie derivative, which is given by the expression $\Libra^2\bPhi$
$=\xi^\mu\nabl_\mu\big(\xi^\nu\nabl_\nu\bPhi\big)$ $=\$^2
\bPhi+(\xi^\mu\nabl_\mu \xi^\nu) \nabl_\nu\bPhi $.

After having thus dealt with the case of a scalar, the next simplest case to
be considered is that of what Cartan would call a one-form, meaning just
a covector, $\bA_\mu$ say, such as the electromagnetic gauge potential, for which
the displacement $x^\mu\mapsto x_{\{\epsilon\}}^{\,\mu}$ will determine a
corresponding pullback mapping $\bA_\mu\mapsto \bA_{\{\epsilon\}\mu}$ with
$\bA_{\{\epsilon\}\mu}\{x\}=\bA_\nu\{ x_{\{\epsilon\}} \} x^{\
\nu}_{\{\epsilon\},\mu} $. In order to work this out, we need to evaluate the
relevant tensor transformation matrix $x^{\ \nu}_{\{\epsilon\},\mu} $ whose
components are the partial derivatives of the displaced coordinates with
respect to the initial coordinates, so that they will be given by an expansion
that is obtainable from (\ref{afgeod}) by partial differentiation of the
coefficients (\ref{stracoef}) in the form
\be x^{\ \mu}_{\{\epsilon\},\lambda}=\delta^\mu_\lambda+
\epsilon\, \xi^\mu_{\, ,\lambda} - {\epsilon^2\over
2}\big(\Gamma_\nu{^\mu}{_{\rho,\lambda}}\xi^\nu\xi^\rho
+2\Gamma_\nu{^\mu}{_\rho}\xi^{(\nu}\xi^{\rho)}{_{,\sigma}}\big)+ {\cal
O}(\epsilon^3) \, , \eqn{tenstran}\fe 
in which it is of course to be understood that the fields and their partial
coordinate derivatives indicated by the use of a comma, are all to be
evaluated as the corresponding functions of the initial (undisplaced)
coordinates $x^\mu$.  The covectorial analogue of the scalar ``straight''
transportation formula (\ref{standexp}) is thus obtainable in the standard
form
\be \bA_{\{\epsilon\}\mu}=\bA_\mu+\epsilon\, \$ \bA_\mu +{\epsilon^2\over 2}
\$^2 \bA_\mu +{\cal O}\{\epsilon^3\}  \fe
with the relevant ``straight'' derivative coefficients  given by 
\be \$\bA_\mu =\xi^\nu\nabl_\nu\ \bA_\mu+\bA_\nu\nabl_\mu\xi^\nu\, ,\fe
\be \$^2 \bA_\mu=\xi^\nu\xi^\rho(\nabl_\rho\nabl_\nu \bA_\mu
-{\calR}_{\mu\nu}{^\lambda}{_\rho}\bA_\lambda) + 2\xi^\nu(\nabl_\mu\xi^\rho)
 \nabl_\nu \bA_\rho \, ,\eqn{stracov}\fe
where ${\calR}_{\mu\nu}{^\lambda}{_\rho}$ is the Riemann tensor associated
with the chosen connection, which is definable, using the (MTW) sign
convention, by
\be 2\nabl_{[\mu}\nabl_{\nu]}\bA_\rho=
-{\calR}_{\mu\nu}{^\lambda}{_\rho}\bA_\lambda \, .\eqn{Rid}\fe
The cancelling out of all the non-tensorially coordinate dependent
contributions so as to give a strictly tensorial result is of course not an
accident but an automatic consequence of the covariant character of the
definition of the straight differentation procedure. 

The next case that is of interest in the context of the kind of variational
analysis that has been discussed in preceeding sections, is that of a
two-index covariant tensor, $\bB_{\mu\nu}$, which, if it were antisymmetric, as
in the example of the Kalb-Ramond potential that was considered in previous
work, would be interpretable as a two-form in the sense of Cartan.  Whether or
not it has such an antisymmetry property, the effect on any two-index
covariant tensor of the``straight'' transportation $x^\mu\mapsto
x_{\{\epsilon\}}^{\,\mu}$ will, as in the previous case, be to induce a
corresponding pullback mapping $\bB_{\mu\nu}\mapsto \bB_{\{\epsilon\}\mu\nu}$
which will be given by $\bB_{\{\epsilon\}\mu\nu}\{x\}=\bB_{\rho\sigma}\{
x_{\{\epsilon\}} \} x^{\ \rho}_{\{\epsilon\},\mu} x^{\
\sigma}_{\{\epsilon\},\nu} $. Working this out, using (\ref{afgeod}),
(\ref{stracoef}), and (\ref{tenstran}) as in the previous single index case,
one can obtain the corresponding expansion in the standard form
\be \bB_{\{\epsilon\}\mu\nu}=\bB_{\mu\nu}+\epsilon\, \$ \bB_{\mu\nu}
 +{\epsilon^2\over 2} \$^2 \bB_{\mu\nu} +{\cal O}\{\epsilon^3\} \fe
in which the required ``straight'' derivative coefficients work out as
\be \$\bB_{\mu\nu}=\xi^\rho\nabl_\rho+\bB_{\rho\nu}\nabl_\mu\xi^\rho
+\bB_{\mu\rho}\nabl_\nu\xi^\rho\hskip 4 cm\, ,\fe
\be \$^2\bB_{\mu\nu}=\xi^\rho\xi^\sigma\big(\nabl_\rho\nabl_\sigma \bB_{\mu\nu}
-{\calR}_{\mu\rho\ \sigma}^{\,\ \ \lambda} \bB_{\lambda\nu}
-{\calR}_{\nu\rho\ \sigma}^{\,\ \ \lambda} \bB_{\mu\lambda}\big)\hskip 2cm \fe  
\be	\hskip 2 cm +2\xi^\rho\big( (\nabl_\mu\xi^\lambda)\nabl_\rho 
\bB_{\lambda\nu} + (\nabl_\nu\xi^\lambda)\nabl_\rho \bB_{\mu\lambda}\big)
+2 \bB_{\rho\sigma}(\nabl_\mu\xi^\rho)\nabl_\nu\xi^\sigma\, . \fe

The most important application of the preceeding result is to the case in
which $\bB_{\mu\nu}$ is to taken to be the spacetime metric $\bg_{\mu\nu}$ itself,
for which one thus obtains
\be \bg_{\{\epsilon\}\mu\nu}=\bg_{\mu\nu}+\epsilon\, \$ \bg_{\mu\nu}
 +{\epsilon^2\over 2} \$^2 \bg_{\mu\nu} +{\cal O}\{\epsilon^3\} \fe
in which the relevant straight derivatives are given by the comparitively
simple looking formulae
\be \$ \bg_{\mu\nu}=2\nabl_{(\mu}\xi_{\nu)}\ ,\hskip 1 cm
\$^2 \bg_{\mu\nu}=2(\nabl_{(\mu}\xi^{\rho})\nabl_{\nu)}\xi_\rho 
-2\xi^\rho\xi^{\sigma} {\calR}_{\mu \rho\nu\sigma} \, , \eqn{stramet}\fe
of which the former is the same as the usual Lie differentiation
formula while the latter agrees with a result that was obtained in
a different context by Boisseau and Letelier \citeA{BoisseauLetelier92}.

\subsection{*Perturbed Dirac Goto Nambu action}
\label{3-4}

The calculus developed in the preceding sections is intended for application
to cases in which one is concerned with a perturbed configuration that is
obtainable from a given unpertubed perturbation by a small deviation, of order
$\d$ say, via a continuously differentiable (homotopic) variation, which may
be taken to be specified by a homotopy parameter $\lambda$ say ranging from
$0$ in the unperturbed configuration to $1$ in the particular perturbed
configuration under consideration. Using an overhead prime to denote
differentiation with respect to $\lambda$ the ``straight'' transport vector
$\xi^\mu$ characterising the relative displacement  of the final perturbed
configuration (for which $\lambda=1$) will be expressible by an expansion of
the form
\be \xi^\mu=\xiI^\mu+{_1\over ^2}\xiII^{\mu}+{\cal O}\{\d^3\}\, .
\eqn{dispexp}\fe 
 Specifying the differential operator $d$ with respect to the same parameter
$\lambda$, so that for example in the case of the action $\Ac$ we obtain the
identification $\d\Ac= \AcI$, the operator $\Delta$ giving the deviation between
the original and perturbed value of any additive quantity will be expressible
by a corresponding expansion
\be \Delta=\d+{_1\over^2}\d^2 +{\cal O}\{\delta^3\}\, .\eqn{deltexp}\fe

The reason why we are interested in going to second order in the present notes
is that -- according to a very general principle first drawn to my attention
many years ago by Taub \citeA{Taub69}, \citeA{FriedmanSchutz75} -- if a system of dynamical
equations is obtainable by a variational principle from an action $\Ac$
(meaning that the dynamical equations are the condition for the first order
locally perturbed action $\d\Ac$ to vanish)  then the corresponding first order
perturbed dynamical equations will be similarly obtainable from the {\it
second} order perturbed action $\d^2\Ac$.

For a free brane brane action of the kind governed by (\ref{3.8}), the first
order perturbed action with respect to a Lagrangian reference system (meaning
one that is comoving with the brane worldsheet) is directly obtainable (just
by setting the external field variables $\bA_\mu$ and $\bB_{\mu\nu}$ set to zero)
from (\ref{2.6}) in the form 
 \be \d\Ac={1\over 2}\int d\!\oS\ 
\oT{^{\mu\nu}} \dL \bg_{\mu\nu} \, , \eqn{firstact}\fe
and so the corresponding second order perturbed action will be
obtainable, using (\ref{varmes}), in the form
\be \d^2\Ac={1\over 2}\int  d\!\oS\, \Big( 
(\dL\oT{^{\mu\nu}}) \dL \bg_{\mu\nu} + \oT{^{\mu\nu}}\big(
{_1\over^2}\og^{\rho\sigma}(\dL \bg_{\rho\sigma}) \dL \bg_{\mu\nu}
+\dL^{\, 2}\bg_{\mu\nu}\big)\Big) \, .\eqn{secondact}\fe

In the case to which our attention will be restricted here,
the background metric will be taken to be held fixed in the sense
that its Eulerian variation vanishes, $\DE \bg_{\mu\nu}=0$. This means that
according to the general displacement rule (\ref{Lagdiff}) the Lagrangian
(comoving) metric variation in which we are interested will be given by
\be \DL=\vec{\xi} \$+{_1\over^2}\big(\vec{\xi}\$\big)^2+ {\cal O}\{\delta^3\}
\, . \fe
The concise abbreviation $ \$ $ for the straight differentation operator
involved in (\ref{Lagdiff}) has been replaced here by the more explicit
notation $\vec{\xi}\$ $ to indicate its dependence on the displacement vector
whose expansion is given by (\ref{dispexp}). It can thus be seen that the
first and second order Lagrangian variations of the metric will be given by
\be \dL \bg_{\mu\nu}=\vec{\xi^\prime} \$ \bg_{\mu\nu}\ , \hskip 1 cm 
\dL^2 \bg_{\mu\nu}= \big(\vec{\xi^\prime}\$\big)^2 \bg_{\mu\nu} +
\vec{\xi^{\prime\prime}}\$ \bg_{\mu\nu} \, . \fe

When the relevant unperturbed field equation, namely (\ref{3.8}), are
satisfied, the first order action variation due to any perturbation with
compact support will of course vanish by the variation principle, $\d^2\Ac=0$.
In these circumstances the contribution from the second order displacement
contribution $\xiII^\mu$ will automatically cancel out, so that
for the Dirac-Goto-Nambu model determined 
by a surface Lagrangian of the trivial constant form
\be \ud\olag=-\mag^{\rm d} \eqn{3.11}\fe
the second order action variation can be obtained from
(\ref{secondact}), using the formula (\ref{stramet}) in the form
\be \d^2\Ac=\int  {1\over 2}\int  d\!\oS\,
\oT^\mu_{\ \nu}\Big(\ag_{\rho\sigma}\big(
\onab_\mu\xiI^\rho\big)\onab^{\,\nu}\xiI^\sigma+2\big(\onab_{[\rho}
\xiI^\rho\big)\onab_{\mu]}\xiI^\nu-{\calR}_{\mu\rho\ \sigma}^{\ \ \,\nu}
\xiI^\rho\xiI^\sigma\Big)\, .\eqn{secondvaract} \fe
with
\be \ud\oT{^{\mu\nu}}=-\mag^{\rm d}\ud\og{^{\mu\nu}}\, .\eqn{3.12}\fe
where $\mag$ is a constant having the dimension of mass (which would be of the
order of magnitude of the relevant Higgs mass scale in the case of a vacuum
defect arising from the spontaneous symmetry breaking mechanism of the kind
most commonly considered \citeA{Kibble76}) and d is the dimension of the worldsheet
(i.e. d=1 for a simple point particle, d=2 for a string, and so on).

Substituting the expression (\ref{3.12}) for the surface stress energy
momentum tensor into (\ref{3.8}), the unperturbed Dirac Gotu Nambu equation of
motion is  obtained in a form that is given -- independently of the dimension
d indicated by the prefix $^{(\rm d)}$ which may therefore be dropped -- by
the well known harmonicity condition that is expressible as the vanishing of
the curvature vector, 
\be \Ke^\mu=0\, .\eqn{3.13}\fe
In this rather degenerate special case, since there are no internal
fields the linearised perturbation equation (\ref{3.9}) will of 
course completely determine the evolution of the displacements by itself.
By substituting (\ref{3.12}) in (\ref{3.9}) the
corresponding perturbation equation is obtained in the form
\be \ag^{\!\rho}_{\, \lambda}\big(\onab^{\,\mu}\onab_\mu
\xi^\lambda-\og^{\mu\nu}\calR^\lambda_{\ \mu\nu\sigma}\xi^\sigma\big)
=2\Ke_{\mu\nu}^{\ \ \,\rho\,}\onab^{\,\mu}\xi^\nu\, .\eqn{3.14}\fe
It is straightforeward to check that this does indeed agree -- subject to
(\ref{3.13}) -- with what is obtained \citeA{Carter93} by applying the variation
principle to the second order action (\ref{secondvaract}). As well as the compact
tensorial version (\ref{3.14}) given here, the literature includes other
equivalent but formally more complicated expressions \citeA{Guven93a}, \citeA{LarsenFrolov94}
(involving reference to internal coordinates or surface adapted frames of the
kind discussed in Subection \ref{1-1}) that generalise
earlier work which was restricted to a flat or De Sitter 
background \citeA{GarrigaVilenkin91}, \citeA{GarrigaVilenkin92}, \citeA{GarrigaVilenkin92} or specialised
to the hypersurface supported (``wall" or membrane)
special case \citeA{Guven93}.

\subsection{*Higher order geodynamic models}
\label{3-5}

For most practical physical purposes the most useful generalisations of the
Dirac-Goto-Nambu models governed by (\ref{3.13}) and  (\ref{3.14}) are those
of the very general category governed by (\ref{3.8}) and (\ref{3.9}) which
allow for internal fields such as the currents that can be used, not only to
represent the Witten type superconductivity effect in cosmic strings, but also
also to represent the effect of ordinary elasticity in terrestrial
applications, such as the strings of musical instruments which were already a
subject of scientific investigation, albeit at an empirical rather than
theoretical level, in the time of Pythagoras. However before proceeding to the
discussion of such internal field effects in the following sections, it is of
interest to consider how the simple Dirac-Goto-Nambu model can be generalised
in a way that goes beyond the ``pure" brane description characterised in the
free case by (\ref{3.8}) and in the presence of an external force by
(\ref{2.35}). 

The distinguishing property of a ``pure" brane model of the kind
considered in Section \ref{Section2} is the condition that the action depends
only on the undifferentiated background fields $\bg_{\mu\nu}$, $\bA_\mu$,
$\bB_{\mu\nu}$, but not on their gradients. One of the most familiar kinds of
example in which this condition fails to hold is that of an
electromagnetically polarised medium, whose action \citeA{Carter80} depends not just
on the gauge field $\bA_\mu$ but also directly on the associated field
$\bF_{\mu\nu}$ itself. For lack of time and space we shall not consider
such electromagnetic effects here, but will consider only the
simplest category of ``geodynamic" brane models, meaning those in which, as in
the ``pure" brane models of  the Dirac-Goto-Nambu category, the action depends
only on the imbedding geometry of the worldsheet and not on any other external
internal fields. The simplest such extension of the ``pure" Dirac-Goto-Nambu
model (whose action is proportional just to the surface measure which depends
only on $\bg_{\mu\nu}$ but not its derivatives) is based on a Lagrangian
consisting not only of a constant term but also of terms proportional to the
two independent scalars that can be constructed as quadratic functions of the
first derivatives of the metric, namely $\Ke_\mu \Ke^\mu$ and
$\Ke_{\mu\nu\rho}\Ke^{\mu\nu\rho}$. The inclusion of such ``stiffness" terms has
been suggested by Polyakov \citeA{Polyakov86} and others, one of the main reasons
being allowance for the deviations from the ``pure" Dirac-Goto-Nambu
description of cosmological string  \citeA{MaedaTurok88}, \citeA{Gregory88}, \citeA{Gregory93} or
domain wall  \citeA{GarfinkleGregory90}, \citeA{Gregoryetal91}, \citeA{SilveiraMaia93}, \citeA{Barrabesetal94}, \citeA{CarterGregory95}
defects that one would expect to arise if the curvature becomes too strong.
The kind of Lagrangian constructed in this way, namely \be \ud\olag=
-\mag^{\rm d}+\bag\,\ud \Ke_\rho\ud \Ke^\rho-\cag\,\ud \Ke_{\mu\nu\rho} \ud 
\Ke^{\mu\nu\rho}\, ,\eqn{3.15}\fe 
where $\mag$, $\bag$, and $\cag$ are constants, has recently been
the subject of investigations by a number of authors \citeA{Letelier90}, 
\citeA{HartleyTucker90}, \citeA{ArodzSitarzWegrzyn92}, 
\citeA{BoisseauLetelier92}, \citeA{Carter94a}, \citeA{CapovillaGuven95a}. 
It is convenient to use the abbreviation
\be \ud\kag^{\mu\nu\rho}=\bag\,\ud\og^{\mu\nu}\ud \Ke^\rho-\cag\,
\ud \Ke^{\mu\nu\rho}\, ,\eqn{3.16}\fe
which enables the Lagrangian (\ref{3.15}) to be expressed in condensed form as
\be \ud\olag=-\mag^{\rm d}+\ud\kag_{\mu\nu\rho}\ud \Ke^{\mu\nu\rho} \, .
\eqn{3.17}\fe
The variation needed for evaluating the change in such an ``impure"
brane action will thereby be obtainable from the formulae above in the 
corresponding form
\be \Vert\hg\Vert^{-1/2}\dL\big(\Vert\hg\Vert^{1/2}\olag\big)
=\dL\olag+{1\over 2}\olag\,\og^{\mu\nu}\dL \bg_{\mu\nu}\ \eqn{3.18}\fe
(again dropping the explicit reference to the brane dimension d) with
\be \dL\olag=\big(\kag^{\lambda\rho\mu} \Ke_{\lambda\rho}^{\ \ \nu}
-2\,\kag^{\lambda\mu}_{\ \ \rho} \Ke_\lambda^{\ \nu\rho}\big)\dL \bg_{\mu\nu}
+\big(2\,\kag^{\mu\lambda\nu}-\kag^{\mu\nu\lambda}\big)
\nabl_\lambda\dL \bg_{\mu\nu} \, .\eqn{3.19}\fe
Although it is still possible to construct a formally symmetric stress
momentum energy density tensor of the distributional type, the presence of the
gradient term on the right of (\ref{3.19}) will make it rather pathological,
with not just a contribution proportional to a Dirac distribution as in the
``pure" brane case described by (\ref{2.27}) but also with a contribution
proportional to the even more highly singular gradient of a Dirac
distribution \citeA{BoisseauLetelier92}. In order to be able to continue
working in terms of strictly regular surface supported field, 
it is necessary \citeA{Carter94a} to deal
with the gradient dependence of the action in such an ``impure" brane model by
having recourse to the use of a total stress momentum energy tensor ${\calT}
^\mu{_{\nu}}$ of the no longer no longer symmetric canonical type that can
be read out from the variation formula
\be \Vert\hg\Vert^{-1/2}\dL\big(\Vert\hg\Vert^{1/2}\olag\big)
={\calT}^\mu{_{\nu}}\onab_\mu\xi^\nu+2\kag_\mu^{\ 
\rho\sigma}\,{\calR}_{\rho\sigma\ \nu}^{\ \ \,\mu}\xi^\nu 
-\,\onab_\mu\big(2\,\kag_\nu^{\ 
\mu\rho}\nabl_\rho\xi^\nu\big)\eqn{3.20}\fe
that is obtained after substitution of (\ref{3.4}) in (\ref{3.18}) and
(\ref{3.19}). The regular but non-symmetric {\it canonical surface stress
momentum energy density} tensor ${\calT}^\mu{_{\nu}}$ obtained in this
way \citeA{Carter94a} is given by
\be {\calT}^\mu{_{\nu}}=\otT^\mu_{\ \,\nu}+\tT^\mu_{\ \nu}\, , \eqn{3.21}\fe
where $\otT^{\mu\nu}$ is symmetric and purely tangential to the worldsheet
(as a regular ``geometric" stress momentum energy density tensor would be)
with a merely algebraic dependence on the second fundamental tensor, being
given by
\be \otT^\mu{_{\nu}}=\otT_\nu^{\ \mu}={\calT}^\mu{_{\!\lambda}}
\og^\lambda_{\ \nu}= \olag\,\og^\mu_{\ \nu}-2\,\kag^\lambda_{\ \nu\rho}\,
\Ke_\lambda^{\ \mu\rho}\, ,\eqn{3.22}\fe
while the remainder, which is of higher differential order, is expressible in
terms of the third fundamental tensor (\ref{1.22}) as
\be \tT^\mu_{\ \nu}={\calT}^\mu{_{\!\lambda}}\!\ag^{\!\lambda}_{\,\nu}=
2\,\cag\,\Xe^{\lambda\ \mu}_{\ \,\lambda\ \nu}
- 2\,{\bag}\,\Xe^{\mu\ \lambda}_{\ \,\lambda\ \nu} \, , \eqn{3.23}\fe
which can conveniently be rewritten with the higher derivative
contributions regrouped in the form
\be \tT^\mu_{\ \nu}=2(\cag-\bag)\Xe^\mu_{\ \nu}+\cag\,\og^\mu_{\ \rho}
\og^{\sigma\tau}\calR^\rho{_{\sigma\tau\lambda}}\ag^{\!\lambda}_{\,\nu}
\, ,\eqn{3.24}\fe
where
\be \Xe^\mu_{\ \nu}=\Xe^{\mu\ \lambda}_{\ \lambda\ \nu}=
\ag^{\!\lambda}_{\,\nu} \onab^{\,\mu} \Ke_\lambda \, .\eqn{3.25}\fe
It is to be noticed that the total canonical surface stress momentum energy
density tensor obtained in this way is is still automatically tangential to
the worldsheet on its first (though no longer on its second) index, i.e.
\be \ag^{\!\lambda}_{\,\mu}\!{\calT}^\mu{_{\nu}}=0\, ,\eqn{3.26}\fe 
and that the higher derivative contribution proportional to the trace
$\Xe^\mu_{\ \nu}$ of the third fundamental tensor will drop out if the
coefficients have the same value, $\cag=\bag$.

The application of the variation principle to the effect that the surface
integral of the variation (\ref{3.20}) should vanish for any displacement
$\xi^\mu$ within a bounded neighbourhood can be seen  to lead
(via an application of Green's theorem as in the Section \ref{Section2}
to dynamical equations \citeA{Carter94a} of the form 
\be \onab_\mu{\calT}^\mu{_{\!\nu}}=2\,\kag_\mu^{\ \rho\sigma}\,
{\calR}_{\rho\sigma\ \nu}^{\ \ \,\mu} \, .\eqn{3.27}\fe
As in the pure Dirac-Gotu-Nambu case discussed above, the foregoing system 
of equations is partially redundant: although it involves n distinct 
spacetime vectorial equations, only n--d of them are dynamically 
independent, namely those projected orthogonally to the d-dimensional 
worldsheet. The others are merely Noether identities which follow 
independently of the variation principle from the fact that a 
displacement $\xi^\nu$ that is purely tangential to the worldsheet merely 
maps it onto itself and thus cannot affect the action, as can be verified 
directly using the generalised Codazzi identity (\ref{1.23}).  

The lack of symmetry of ${\calT}^\mu{_{\!\nu}}$ means that the construction
of the corresponding momentum current vector, ${\calP}^\mu$ say, associated
with a generic background spacetime Killing vector field $k^\mu$ will
not be quite as simple as in the ``pure" brane case for which an expression
of the simple form (\ref{2.32}) suffices. However using the well known
fact that the Killing equation (\ref{2.31}) entails the integrability 
condition
\be \nabl_\mu\nabl_\nu k^\rho=\calR^\rho_{\ \nu\rho\lambda}k^\lambda\, ,
\eqn{3.28}\fe
together with the observation that the antisymmetric part of the canonical 
stress momentum energy density tensor (\ref{3.22}) is given according to 
(\ref{3.23}) just by
\be {\calT}^{[\mu\nu]}=-2\onab_\lambda\Sigma^{\lambda[\mu\nu]}
\, ,\eqn{3.29}\fe
it can be seen that, for any solution of (\ref{2.31}), the ansatz \citeA{Carter94a}
\be {\calP}^\mu={\calT}^\mu{_{\!\nu}}k^\nu+ 2\kag^{\mu\nu}{_\rho}
\nabl_\nu k^\rho \eqn{3.30}\fe
provides a surface current, ${\calP}^\mu$, which satisfies the 
tangentiality condition
\be \ag^{\!\lambda}_{\,\mu}{\calP}^\mu=0\, ,\eqn{3.31}\fe
and for which the strict surface conservation law,
\be \onab_\mu{\calP}^\mu=0\, , \eqn{3.32}\fe
will hold whenever the equation of motion (\ref{3.27}) is satisfied. It is to
be remarked however that for a Killing vector of the irrotational kind for
which $\nabl_\mu k_\mu$ vanishes altogether the second term (interpretable as
a surface spin density contribution) in (\ref{3.30}) will not contribute, i.e.
an expression of the simpler form (\ref{2.32}) will suffice. This applies in
particular to the case of an ordinary translation generator in flat space, for
which the corresponding conserved surface current will represent ordinary
energy or linear momentum, whereas in the case of angular momentum the extra
(spin density) term in (\ref{3.30}) is indispensible.

The preceeding formulae all include allowance for arbitrary background 
curvature, but, to obtain the analogue of the non redundant
version (\ref{3.8}) of the equations of motion in a reasonably simple form,
the restriction that the background spacetime be flat, i.e.
$\calR^\mu{_{\nu\rho\sigma}}=0$, will now be imposed. This enables
the required system of dynamical equations to be expressed \citeA{Carter94a} in 
the form
\be \otT{^{\mu\nu}}\Ke_{\mu\nu\rho}=2(\bag-\cag) 
\ag^{\!\nu}_{\,\rho}\onab_\mu\big(\ag^{\!\sigma}_{\,\nu}\onab^{\,\mu}
\Ke_\sigma\big)\, ,\eqn{3.33}\fe
with the higher derivative terms grouped on the right hand side, which
vanishes if $\bag=\cag$.

In the particular case \citeA{CarterGregory95} of a membrane, meaning a brane supported by
a hypersurface of dimension d=n--1, the second fundamental tensor and its
trace will be given in terms of the unit normal $\lambde_\rho$ (which in this
case will be unique up to a choice of sign) by
$\Ke_{\mu\nu\rho}=\Kie_{\mu\nu}\lambde_\rho$ and $\Ke_\rho= \Kie\lambde_\rho$ with
$\Kie=\Kie^\nu_{\ \nu}$ where $\Kie_{\mu\nu}$ is the second fundamental form (whose
sign depends on that orientation chosen for the normal). In an ordinary
4-dimensional spacetime background, this membrane case corresponds to d=3, for
which the symmetric tangential part of the surface stress momentum energy
density tensor will be therefore be obtainable from (\ref{3.22}) in the form
\be \uth\otT{^{\mu\nu}}=-\big(\mag^3-\bag\uth \Kie^2+\cag\uth \Kie^\rho_{\ \sigma}
\uth \Kie^\sigma_{\ \rho}\big)\uth\og^{\mu\nu}-2\bag\uth \Kie\uth \Kie^{\mu\nu}
+2\cag \uth \Kie^{\lambda\mu}\uth \Kie_\lambda^{\ \nu}\, .\fe
In the case of a string with d=2 one can use the fact that the
trace free conformation tensor (\ref{1.20}) will satisfy
$2\utw \Ce^\lambda_{\  \,\mu\rho}\utw \Ce_{\lambda\nu}^{\ \ \,\rho}
=$ $\utw \Ce_{\kappa\lambda\rho}\utw \Ce^{\kappa\lambda\rho}\utw\og_{\mu\nu}$
to obtain a corresponding formula (which holds regardless of the background
spacetime dimension n) given \citeA{Carter94a} by
\be \utw\otT{^{\mu\nu}}=-\mag^2\utw\og^{\mu\nu}+2(\cag-\bag)\utw 
\Ce^{\mu\nu\rho} \utw \Ke_\rho \, .\fe

\section{Strings and other spacially isotropic brane models}
\label{Section4}

\subsection{The general category of ``perfect'' brane models}
\label{4-1}

It is reasonable to postulate that a ``weak" energy condition of the kind
formulated and justified by Hawking and Ellis \citeA{HawkingEllis73} should hold for any
pure p-brane model as a  condition for physical realism as a macroscopic
description of a (p+1)-surface supported physical system at a classical level,
meaning that the model's surface stress momentum energy density tensor (as
introduced in Section \ref{Section2}) should be such that the contraction 
$\oT{^{\mu\nu}}\bet_\mu\bet_\nu$ is non negative for any vector $\bet^\mu$
that is timelike. Furthermore the causality condition to the effect that there
should be no timelike characteristic covector (i.e. no superluminal
propagation) can be seen from (\ref{3.10}) to entail the further requirement
(going marginally beyond the ``weak" condition of Hawking and Ellis) that 
$\oT{^{\mu\nu}}\bet_\mu\bet_\nu$ should be strictly positive if $\bet^\mu$ is
timelike. This leads to the formulation of what may be called the ``minimal"
energy condition for a pure p-brane which is expressible  as
\be \bet^\mu\bet_\mu<0\hskip 0.6 cm \Rightarrow\hskip 0.6 cm
 \oT{^{\mu\nu}}\bet_\mu\bet_\nu>0 \, .\eqn{4.1}\fe

A (pure) p-brane model will consequently be characterised by a well defined
{\it surface energy density}, $\Ue$ say, that is specifiable by an eigenvalue
equation of the form 
\be \oT{^\mu}{_\nu}\bet^\nu=-\Ue \bet^\mu\, , \eqn{4.2}\fe 
where the corresponding eigenvector $\bet^\mu$ is distinguished by the
requirement that it be tangential and non-spacelike:
\be \ag^{\!\mu}_{\,\nu}\bet^\nu=0\, , \hskip 1 cm \bet^\mu\bet_\mu
\leq 0 \, .\eqn{4.3}\fe
It is apparent that the ``minimal''energy condition (\ref{4.1}) requires
that the eigenvalue $\Ue$ should be strictly positive unless $\bet^\mu$
is null in which case it may vanish:
\be \Ue\geq 0\, ,\hskip 1 cm \Ue=0\hskip 0.6 cm \Rightarrow\hskip 0.6 cm
 \bet^\mu\bet_\mu=0\, .\eqn{4.4}\fe

In the Dirac-Goto-Nambu model that is most familiar to present day
cosmologists, the eigenvector $\bet^\mu$ is indeterminate and the energy
density $\Ue$ is the same (in relativistic units such as are used here, with the
speed of light set to unity) as the corresponding (surface)  tension. However
in general it is essential to distinguish the concept of energy density $\Ue$
from the concept of the {\it tension} (as used in physics since the
formulation of Hooke's law at the time of Newton) from which term ``tensor''
is derived. The (surface) tension scalar, for which we shall use the
traditional symbol $\Te$, is defineable generically in a manner consistent with
traditional usage, for a ``pure" p-brane (i.e. a (p+1)-dimensionally supported
system) of the kind considered here by decomposing the trace of the surface
stress momentum energy density tensor in the form
\be \oT{_\nu^{\,\nu}}=-\Ue-{\rm p}\Te\ .\eqn{4.5}\fe

Apart from the degenerate Dirac-Goto-Nambu case for which $U$ and $T$ are
actually equal, the simplest possibility is that of a {\it perfect}
p-brane \citeA{Carter90}, meaning one whose surface stress momentum energy density
tensor is spacially (thus p-dimensionally) {\it isotropic} so that it is will
be expressible, for a suitable choice of the normalisation of the eigenvector
$\bet^\mu$ in (\ref{4.2}) by
\be \oT{^{\mu\nu}}=\bet^\mu\bet^\nu-\Te\og^{\mu\nu}\, ,\eqn{4.6}\fe
where the required normalisation is given by
\be \Ue-\Te=\bet^\mu\bet_\mu\leq 0\, .\eqn{4.7}\fe
This category includes the case of the Dirac-Goto-Nambu model (\ref{3.12}),
which is obtained (with $\Te=\mag^{\rm p+1}$) by normalising the (in this case
indeterminate) eigenvector to zero, i.e setting $\bet^\mu=0$. A more mundane
example is provided by the familiar ``improper" case p=n--1 (where n is the
background dimension) of an ordinary (relativistic) perfect fluid with
pressure $\Pe=-\Te$. Although the tension $\Te$ is negative in the ordinary fluid
case, it must be positive, as a condition for stability, for ``proper"
p-branes of lower dimension, p$<$n--1. Whereas the membrane models (with p=2
for n=4) that are appropriate for the description of nautical sails will not
in general have the isotropic form (\ref{4.6}), a familiar everyday example of
a membrane that will automatically be ``perfect" in this sense is provided by
the case of an ordinary soap bubble whose boundary hypersurface will be
characterised by $0<\Te\ll \Ue$. The kind of application on which we shall
concentrate in the following sections is that of a {\it string}, as given by
p=1, for which the ``perfection" property (\ref{4.6}) will always hold:
isotropy cannot fail to apply in this 1-brane case because only a single space
dimension is involved. 

Before specialising to the (automatically perfect) string case, it is to be
remarked that except when the $\bet^\mu$ is indeterminate (as in the
Dirac-Goto-Nambu case), or null (as can occur in special subspaces where the
current becomes null in a string model of the
kind \citeA{Carter89}, \citeA{Larsen93}, \citeA{Carter89a} appropriate for describing cosmic vortex
defects of the ``superconducting" kind proposed by Witten \citeA{Witten85}) this
eigenvector will determine a corresponding unit eigenvector,
\be  \ue^\mu=\big(\Ue-\Te\big)^{-1/2}\bet^\mu\, , 
\hskip 1cm \ue^\mu \ue_\mu=-1 \, ,\eqn{4.8}\fe
specifying a naturally preferred rest frame, in terms of which (\ref{4.6})
will be rewriteable in the standard form
\be \oT{^{\mu\nu}}=\big(\Ue-\Te\big)\ue^\mu \ue^\nu -\Te\og^{\mu\nu}\, .\eqn{4.9}\fe
With respect to the frame so defined, it can be seen that the velocity
$\cE$ say of propagation of extrinsic perturbations (what are
commonly referred to as ``wiggles") of the perfect brane worldsheet will be
given, according to the characteristic equation (\ref{3.10}), by
\be \cE{^2}={\Te\over \Ue}\, .\eqn{4.10}\fe
By substituting (\ref{4.8}) in the (\ref{3.3}), and using the expression
(\ref{1.16}) for the acceleration $\dot \ue^\mu$ of the preferred unit vector
(\ref{4.8}), the extrisic equation governing the free motion of the support
surface of any (proper) perfect brane is reducible to the standard
form \citeA{Carter90}, \citeA{Carter89a}
\be \cE{^2} \Ke^\mu=(1-\cE{^2})\ag^{\!\mu}_{\,\nu}\dot \ue^\nu
\, .\eqn{4.11}\fe
In the Dirac-Goto-Nambu case one has $\Ue=\Te$ and hence $\cE{^2}=1$,
(i.e. the ``wiggle" speed is that of light) so that the right hand side of
(\ref{4.11})  will drop out, leaving the harmonicity condition that is
expressed by the vanishing of the curvature vector given by (\ref{1.19}). At
the opposite extreme from this  relativistic limit case, one has applications
to such mundane examples as that of an ordinary soap bubble membrane or violin
string for which one has $\cE{^2} \ll 1$.

\subsection{The special case of ``barotropic''  brane models}
\label{4-2}

Except in the Dirac-Goto-Nambu limit case, the extrinsic equation of motion
(\ref{4.12}) will need to be supplemented by the dynamic equations governing
the evolution of the internal fields, and in particular of $\Te$, $\Ue$, and
$\ue^\mu$, on the brane surface. The simplest non trivial possibility is the
case of a perfect brane model that is ``barotropic", meaning that its tension
$\Te$ (or equivalently, in the case of a fluid, its pressure $\Pe$) is a function
only of the energy density $\Ue$ in the preferred rest frame. In this barotropic
or ``perfectly elastic" case, the complete set of dynamical equations
governing the internal evolution of the perfect brane will be provided just by
the tangential projection of the force balance law (\ref{2.24}) which, in the
free case to which the discussion of the present section is restricted, has
the simple form
\be \og^\lambda_{\ \nu}\onab_\mu\oT{^{\mu\nu}}=0\, .\eqn{4.12}\fe

In the barotropic case (which includes the kind of cosmic string 
models \citeA{Carter89}, \citeA{Larsen93}, \citeA{Carter89a} appropriate for describing the Witten
type superconducting vaccuum vortices \citeA{Witten85} whose investigation provided
the original motivation for developing the kind of analysis presented here)
experience of the ``improper" case of an (n--1)-brane, i.e.  the familiar
example of an ordinary barotropic perfect fluid, suggests the convenience of
introducing an idealised particle {\it number density}, $\nue$ say, and a
corresponding (relativistic) {\it chemical potential}, or effective mass per
idealised particle, $\mue$ say, that are specified modulo an arbitrary
normalisation factor by the equation of state according to a prescription of
the form 
\be {\rm ln\ }\nue=\int{d\Ue\over \Ue-\Te} \, ,\hskip 1 cm {\rm ln\ }\mue
=\int{d\Te\over \Te-\Ue}\,  , \eqn{4.13}\fe
which fixes them modulo a pair of constants of integration which are
not allowed to remain independent but are related by the imposition of
the restraint condition
\be \mue\nue=\Ue-\Te \, . \eqn{4.14}\fe
The quantities introduced in this way can be used for defining a surface
particle current density vector $\nue^\rho$ and a corresponding dynamically
conjugate energy - momentum (per particle) covector $\mue_\rho$ that are given
by
\be \nue^\rho=\nue \ue^\rho\ ,\hskip 1 cm \mue_\rho=\mue \ue_\rho\, ,\eqn{4.15}\fe
in terms of which the generic expression (\ref{4.9}) of the surface stress
momentum energy density tensor of a perfect brane can be rewritten in the form
\be \oT{^\rho}_{\!\sigma}=\nue^\rho\mue_\sigma-\Te\og^\rho_{\ \sigma}
\, ,\eqn{4.16}\fe
whose divergence can be seen to be given identically by 
\be \onab_\rho\oT{^\rho}_{\!\sigma}=\mue_\sigma\onab_\rho\nue^\rho
+2\nue^\rho\onab_{[\rho}\mue_{\sigma]}-\Te \Ke_\sigma \, .\eqn{4.17}\fe
In the free case for which this divergence vanishes, projection
orthogonal to the worldsheet gives back the extrinsic dynamical equation 
(\ref{4.11}), while by contraction with $\nue^\rho$ one obtains the simple 
surface current conservation law
\be \onab_\rho\nue^\rho=0\, . \eqn{4.18}\fe
The remainder of the system of equations of motion for the free perfect brane 
can hence be obtained from (\ref{4.12}), as a surface generalised version
of the standard perfect fluid momentum transport law \citeA{Carter76}, in the form
\be \ue^\mu \og^\rho_{\ \nu} \onab_{[\mu}\mue_{\rho]}=0\, ,\eqn{4.19}\fe
which is such as to entail a corresponding Kelvin type law of conservation
(under Lie transport by the flow) for the relevant surface vorticity
tensor $\og^\rho_{\ [\nu}\onab_{[\mu]}\mue_{\rho]}$ so that if it
vanishes initially it will remain zero throughout, which is the integrability
condition for the existence of a surface potential function, $\phie$
say, such that $\mue_\rho=\onab_\rho\phie$.

Appart from the extrinsic perturbations of the world sheet location itself,
which propagate with the ``wiggle" speed $\cE$ (relative to the
frame deterined by $u^\mu$) as already discused, the only other kind of
perturbation mode that can occur in a barytropic string are longitudinal
modes specified by the variation of $U$ or equivalently of $T$ within the
world sheet. As in the ``improper" special case of an ordinary perfect
fluid, for which they are interpretable just as ordinary sound waves, such
longitudinal ``woggle" or ``jiggle" perturbations can easily be seen 
from (\ref{4.16}) and (\ref{4.17}) to have a relative propagation velocity 
$\cL$ say that is given \citeA{Carter90}, \citeA{Carter89a} by
\be \cL{^2}={\nue\over\mue}{d\mue\over d\nue}=-{d\Te\over d\Ue}\, .
 \eqn{4.20}\fe
In order for a barotropic model to be well behaved, $\cE{^2}$ and
$\cL$ must of course both be positive, in order for the
velocities to be real (as a condition for local stability), and they must also
both be less than unity in order to avoid superluminal propagation (as a
condition for local causality). An important qualitative question that arises
for any {\it proper} barotropic p-brane, i.e. one with 1$\leq$ p$<$n--1
(excluding the trivial case p=0 of a point particle, and at the other extreme
the ``improper" case p=n--1 of a fluid) is whether the extrinsic propagation
speed is subsonic, meaning that ``wiggles" go slower than ``jiggles", i.e
$\cE < \cL$, transonic, meaning that the speeds are the
same, $\cE =\cL$ or supersonic, meaning that ``wiggles" go
faster than ``jiggles", $\cE >\cL$.  Much of the earliest
work \citeA{SpergelPiranGoodman87}, \citeA{VilenkinVachaspati87a}, \citeA{CopelandHindmarshTurok87}, \citeA{Copelandetal88}, \citeA{HawsHindmarshTurok88}, \citeA{SpergelPressScherrer89}, \citeA{Amsterdamski89} 
on the Witten type superconducting vaccuum vortex phenomena was implicitly
based on use of a string model of subsonic type, but a more careful analysis
of the internal structure of the Witten vortex by 
Peter \citeA{Peter92}, \citeA{Peter92a}, \citeA{Peter93} has since shown that models of
supersonic type are more appropriate. On the other hand such everyday examples
as that of an ordinary violin string are subsonic, a condition that is
sufficient \citeA{CarterMartin93} though not necessary \citeA{Martin94}, \citeA{MartinPeter95} for stability 
of the corresponding circular centrifugally supported ``vorton'' type loop 
configurations \citeA{DavisShellard88}, \citeA{DavisShellard89a}, \citeA{Carter90z}, \citeA{Peter93a} that will be 
discussed later on.
 
A particularly important special case is that of a barytropic p-brane of {\it
permanently transonic} type, meaning one characterised by an equation of state
of the non dispersive constant product form \be \Ue \Te=\mag^{\rm 2p+2} \,
,\eqn{4.21}\fe where $m$ is a constant having the dimensions of a mass, for
which application of the formulae (\ref{4.10}) and (\ref{4.18}) gives
$\cE{^2}=\cL{^2}$ not just for some critical transition
value of $\Ue$ or $\Te$ (as can occur for other kinds of equation of state) but
for all values. In the string case, p=1, it can easily be 
demonstrated \citeA{Carter90} that this non dispersive permanently transonic model 
represents the outcome of a (rather artificial) dimensional reduction process 
first suggested by Nielsen \citeA{Nielsen80}, \citeA{NielsenOlesen87}, \citeA{DavidsonWali88}, \citeA{DavidsonWali88a}. As
will be described below, this model can be shown \citeA{Carter90a} to be governed
by equations of motion that are explicitly integrable in a flat empty
background, and it can be used \citeA{Carter90a}, \citeA{Vilenkin90} to provide what
(contrary to a misleading claim \citeA{HongKimSikivie80} that has been made) can in 
fact \citeA{Carter95a}, \citeA{Martin95} be a highly accurate description of the macroscopic
effect of ``wiggles" in an underlying cosmic string of the simple Goto-Nambu
type.

\section{Essentials of elastic string dynamics}
\label{Section5}

\subsection{*Bicharacteristic formulation of extrinsic equations}
\label{5-1}

Between the hypersurface supported case of a membrane and the curve
supported case of a point particle, the only intermediate kind of brane that
can exist in 4-dimensions is that of 1-brane, i.e. a string model, which
(for any background dimension $n$) will have a first fundamental tensor that
is expressible as the square of the antisymmetric unit surface element
tensor ${\calE}^{\mu\nu}=-{\calE}^{\nu\mu}$ given by (\ref{1.28}): 
\be \og^\mu_{\ \nu}={\calE}^\mu_{\ \rho}{\calE}^\rho_{\ \nu} \, . 
 \eqn{5.1}\fe
In the case of string (to which the remainder of this article will be devoted)
the symmetric surface stress momentum energy tensor $\oT{^{\mu\nu}}$ that is
well defined for any ``pure" (i.e. unpolarised) model will not only be
expressible in the generic form (\ref{4.6}) that expresses the (in the string
case trivial) property of spacial anisotropy with respect to the preferred
timelike or null eigenvector $\bet^\mu$, but it will also be expressible
generically in {\it bicharacteristic form} as \be 
\oT{^{\mu\nu}}=\bet_{_+}^{[\mu}\bet_{_-}^{\nu]}\, ,\eqn{5.2}\fe in terms of
a pair of timelike or null tangent vectors $\bet_{_\pm}^{\,\mu}$ that can be
seen  to be extrinsic bicharacteric vectors, meaning that they lie
respectively along the two (``right moving" and ``left moving") directions of
propagation of extrinsic perturbations, and are thus respectively orthogonal
to the covectors $\chie_\mu$ given by the extrinsic characteristic equation
(\ref{3.10}) whose solutions can be seen from (\ref{5.2}) to be given by
$\bet_{_+}^{\,\mu}\chie_\mu=0$ or $\bet_{_-}^{\,\mu}\chie_\mu=0$. The
expression (\ref{5.2}) does not completely determine the bicharacteristic
vectors $\bet_{_\pm}^{\,\mu}$, but leaves open the possibility of a
reciprocal multiplicative rescaling of their magnitudes, subject to the
condition that their scalar product remains invariant, its value being
obtainable by comparison of (\ref{5.2}) with (\ref{4.6}) as
\be \bet_{_+\mu}\bet_{_-}^{\,\mu}=\oT{^\nu}_{\!\nu}=-\big (\Ue+\Te\big)\, .
\eqn{5.3}\fe
The geometric mean of the magnitudes of the bicharacteristic vectors
remains similarly invariant, with value given by the magnitude of the 
preferred eigenvector $\bet^\mu$ as given by (\ref{4.7}), i.e.
\be \big(\bet_{_+\mu}\bet_{_+}{^{\!\mu}}\big)\big(\bet_{_-\nu}
\bet_{_-}{^{\!\nu}}\big)=\big(\bet_\mu\bet^\mu\big)^2=(\Ue-\Te)^2\, . 
\eqn{5.4}\fe

To obtain the equations of motion of the string we need to use the
force balance equation (\ref{2.25}), which just takes the form
\be \onab_\mu\oT{^{\mu\nu}}=\of{^\nu}\, ,\eqn{5.5}\fe
for a string that is isolated, where $\of{^\mu}$ (which simply vanishes in
the free case) is the force exerted by any background fields, such as the
electromagnetic field and ambient (Higgs or other) fluid that are allowed for
in the explicit expression (\ref{2.26}). (The only difference if the string
were not isolated, i.e. if it belonged to the boundary of one or more attached
membranes, is that an additional contact force contribution $\cf{^\mu}$
of the form (\ref{2.25}) would also be needed as well as the background
contribution $\of{^{\mu}}$.) Using the Weingarten integrability condition
(\ref{1.13}) which is equivalent in the string case to the the projected Lie
commutativity condition
\be \ag^{\!\rho}_{\,\nu}\big(\bet_{_+}{^{\!\mu}}\onab_\mu
\bet_{_-}{^{\!\nu}}-\bet_{_-}{^{\!\mu}}\onab_\mu
\bet_{_+}{^{\!\nu}}\big)=0 \, ,\eqn{5.6}\fe
it can be seen from (\ref{5.2}) that the extrinsic equations of motion
governing the evolution of the string worldsheet will therefore \citeA{Carter89a} be
expressible in characteristic form as
\be \ag^{\!\rho}_{\,\nu}\bet_{_\pm}{^{\!\mu}}\onab_\mu
\bet_{_\mp}{^{\!\nu}}=\ag^{\!\rho}_{\,\mu}\of{^\mu}\, ,\eqn{5.7}\fe
for either choice of the sign $\pm$, one version being obtainable
from the other and vice versa by (\ref{5.6}).

Except where the string state is locally degenerate in the sense that
one of the bicharacteristic vectors $\bet_{_\pm}{^{\!\mu}}$ is null, which
by (\ref{5.5}) will only occur where $\Te=\Ue$ (as is everywhere the
case for the Goto-Nambu model, for which both bicharacteristic vectors are
null everywhere) it will be possible to fix their normalisation by requiring
that they have the same magnitude, $\bet_{_\pm\mu}\bet_{_\pm}{^{\,\mu}}
=-(\Te-\Ue)$ and hence to define a corresponding pair of {\it unit
bicharacteristic vectors}
\be  \ue_{_\pm}{^{\,\mu}}=\big (\Ue-\Te\big)^{-1/2}\bet_{_\pm}{^{\,\mu}}
\ ,\hskip 1 cm     \ue_{_\pm\mu}\ue_{_\pm}{^{\,\mu}}=-1\, .\eqn{5.8}\fe

\subsection{Preferred orthonormal diad for nondegenerate state}
\label{5-2}

In the generic case, $\Te<\Ue$, for which the eigenvectors of the
stress-energy tensor are non null, the corresponding unit vectors will
constitute a preferred diad consisting of the unit timelike
eigenvector $\ue^\mu$ introduced in (\ref{4.8}), together with an
orthogonal spacelike unit eigenvalue $\ve^\mu$.  These mutually
orthogonal unit vectors will be related to the corresponding unit
bicharacteristic vectors $\ue_{_+}{^{\,\mu}}$ and $\ue_{_-}{^{\,\mu}}$ (as
given by (\ref{5.8}) above) by the expressions
\be \ue^\mu={\sqrt{1-\cE{^2}}\over 2}\big(\ue_{_+}{^{\,\mu}} +
\ue_{_-}{^{\,\mu}}\big)\ ,\hskip 1 cm \ve^\mu={\sqrt{1-\cE{^2}}
\over 2 \cE }\big(\ue_{_+}{^{\,\mu}} -\ue_{_-}{^{\,\mu}}\big)\, .
\eqn{5.9}\fe

In terms of this preferred diad, the antisymmetric unit surface element
${\calE}^{\mu\nu}$ and the (first) fundamental tensor $\og^{\mu\nu}$
of the string worldsheet will be given by
\be {\calE}^{\mu\nu}=\ue^\mu \ve^\nu-\ve^\mu \ue^\nu\, ,\hskip 1 cm
\og^{\mu\nu}=-\ue^\mu \ue^\nu+\ve^\mu \ve^\nu \, ,\eqn{5.10}\fe
and  the standard form (\ref{4.9}) for the surface stress 
momentum energy density tensor will reduce to the simple form
\be \oT{^{\mu\nu}}=\Ue \ue^\mu \ue^\nu - \Te \ve^\mu \ve^\nu \, .\eqn{5.11}\fe
Proceeding directly from this last form, the extrinsic dynamical
equation (\ref{5.6}), i.e. the surface orthogonal projection of (\ref{5.5}),
can be rewritten as
\be \ag^{\!\mu}_{\,\nu}\big(\Ue\dot \ue^\nu-\Te \ve^{\prime\nu}\big)=\ag^{\!\mu}
_{\,\nu}\of{^\nu}\, ,\eqn{5.12}\fe
using the notation
\be \dot \ue^\mu=\ue^\nu\nabl_\nu \ue^\nu\, ,
\hskip 1 cm \ve^{\prime\mu}=\ve^\nu\nabl_\nu \ve^\nu \, .\eqn{5.13}\fe
in terms of which the worldsheet curvature vector is expressible as 
\be \Ke^\mu=\ag^{\!\mu}_{\,\nu}(\ve^{\prime\nu}-\dot \ue^\nu) \, .\eqn{5.14}\fe 

\subsection{Duality in elastic string models}
\label{5-3}

The formulae in  Subsection \ref{5-2} illustrate a special feature
distinguishing string models from point particle models on one hand and from
higher dimensional brane models on the other, namely the dual
symmetry \citeA{Carter89} that exists at a formal level between the timelike
eigenvector $\ue^\mu$  and the associated eigenvalue $\Ue$ on one hand, and the
the dynamically conjugate quantities that are the spacelike eigenvector and
$\ve^\mu$ and the associated eigenvalue $\Te$. This feature is particularly
informative in the barotropic or ``perfectly elastic" case discussed at the
end of the Section \ref{Section4}, i.e. in the case when the internal dynamics
is controlled just by an equation of state giving the energy density $\Ue$ as a
function of the tension $\Te$ only, since in there will be a corresponding
duality relation between the number density $\nue$ and the associated chemical
potential $\mue$ as introduced in (\ref{4.13}), which is equivalent to the
mutually dual pair of differential defining relations
\be d\Ue=\mue d\nue\ , \hskip 1 cm d\Te=-\nue d\mue \, .\eqn{5.15}\fe
Defining the surface duals of the current vector $\nue^\rho$ and momentum 
covector $\mue_\rho$ given by (\ref{4.15}) as
\be \stah\nue_\rho={\calE}_{\rho\sigma}\nue^\sigma=\nue \ve_\rho\ ,\hskip 1 cm
\stah\mue^\rho={\calE}^{\rho\sigma}\mue_\sigma=\mue \ve^\rho\, ,\eqn{5.16}\fe
one can use (\ref{4.17}) to obtain the force balance equation for a barotropic
string model in the self dual form
\be \mue_\rho\onab_\sigma\nue^\sigma+\stah\nue_\rho
\onab_\sigma\stah\!\mu^\sigma+\ag^{\!\sigma}_{\,\rho}
\big(\Ue\dot \ue^\sigma-\Te \ve^{\prime\sigma}\big)
=\of_\rho\, , \eqn{5.17}\fe
whose orthogonally projected part gives back the self dual extrinsic force
balance equation (\ref{5.12}), while its tangentially projected part gives a
mutually dual pair of surface current source equations of the form
\be (\Ue-\Te)\onab_\sigma\nue^\sigma=-\nue^\rho\of_\rho\, ,\hskip 1 cm
(\Ue-\Te)\onab_\sigma\stah\!\mue^\sigma=\stah \mue^\rho\of_\rho \, .
\eqn{5.18}\fe
It can be seen from this that, like the surface number current density
$\nue^\sigma$, the dual current $\stah\mue^\sigma$ will also be conserved
in a free barotropic string, i.e. when the external force $\of_\rho$ is 
absent.

It is useful \citeA{Carter89} to introduce a dimensionless state function called 
the ``characteristic potential" $\varthe$ say that is constructed in such 
a way as to be self dual according to the differential relation
\be d\varthe=\sqrt{d\mue\, d\nue\over \mue\nue} \, .\eqn{5.19}\fe
This function is convenient for the purpose of writing the internal
equations of motion obtained by tangential projection of (\ref{5.5}) in a
characteristic form that will be the analogue of the characteristic version
(\ref{5.7}) of the extrinsic equations of motion obtained by the orthogonal
projection of (\ref{5.5}). The longitudinal bicharacteristic unit vectors
$\elle_{_\pm}{^{\!\mu}}$ say (the analogues for the internal sound type waves
of the extrinsic bicharacteristic unit vectors $\ue_{_\pm}{^{\!\mu}}$
introduced above) are defineable by
\be \elle_{_\pm}{^{\!\mu}}=\big(1-\cL{^2}\big)^{-1/2}\big(\ue^\mu
\pm \cL\ve^\mu\big) \ ,\hskip 1 cm  \elle_{_\pm\mu} 
\elle_{_\pm}{^{\!\mu}}=-1\, , \eqn{5.20}\fe
where  $\cL$ is the longitudinal characteristic speed as given by
(\ref{4.20}). In terms of these, the tangentially projected part of the force
balance equation (\ref{5.5}) is convertible \citeA{Carter89} to the form of a pair
of scalar  equations having the characteristic form
\be (\Ue-\Te)\elle_{_\pm}{^{\!\rho}}\big(\onab_\rho\varthe \mp 
\ve_\nu\onab_\rho \ue^\nu\big)=\pm{\calE}_{\mu\nu}\,\elle_{_\mp}{^{\!\mu}}\,
\of{^\nu}\, , \eqn{5.21}\fe
from which it is directly apparent that, as asserted in the Section
\ref{Section4}, the state function $\cL$, as given by (\ref{4.20}),
is indeed the characteristic speed of longitudinal ``woggle" propagation
relative to the preferred frame specified by the timelike eigenvector $u^\mu$.
When rewritten in terms of the unit bicharacteristic vectors
$\ue_{_\pm}{^{\!\mu}}$ that are the extrinsic analogues of
$\elle_{_\pm}{^{\!\mu}}$, the corresponding extrinsic equations (\ref{5.7})
take the form
\be (\Ue-\Te)\ag^{\!\mu}_{\,\nu}\ue_{_\pm}{^{\!\rho}}\onab_\rho\,
\ue_{_\mp}{^{\!\nu}}=\ag^{\!\mu}_{\,\nu}\of{^\nu}\, ,\eqn{5.22}\fe
(It is to be noted that the corresponding equation (32) at the end of the
original discussion \citeA{Carter89} of the characteristic forms of the equations
of string motion contains a transcription error whereby a proportionality
factor that should have been $\sqrt{(\Ue-\Te)\Te}$ was replaced by its 
non-relativistic -- i.e. low tension -- limit, namely $\sqrt{\Ue\Te}$.)

\subsection{*The Hookean prototype example of an elastic 
string model} \label{5-4}

What essentially distinguishes different physical kinds of ``perfectly 
elastic", i.e. barotropic, string model from one  another is the form of 
the equation of state, which depends on the microscopic internal details of 
the vacuum defect (or other structure) that the string model is supposed to
represent. In a cosmological context, the most important special case,
which will be discussed in detail later on, is given by the constant
product formula (\ref{4.21}), but for terrestrial applications the most
useful kind of equation of state for general purposes is the one named
after Robert Hooke, who is recogniseably the principal founder of string
theory as a branch of physics in the modern sense of the word.
(As for superstring theory, it remains to be seen whether it will become
established as a branch of physics or just as a topic of mathematics.)
Hooke's famous law to the effect that the tension, $\Te$ is proportional 
to the extension, i.e. the change in $1/\nue$ in the notation used 
here, provides a very good approximation in the low tension limit that 
is exemplified by applications such as that of an ordinary violin string.
This law is expressible within the present scheme as
\be \Ue(\Ye+\Te)=\Uzero \Ye+{_1\over^2} \Te^2\, ,\eqn{5.23}\fe
where $\Uzero$ is a constant interpretable as the rest frame energy in the
fully relaxed (zero tension) limit, and $\Ye$ is another constant 
interpretable as what is known (after another pioneer worker on the 
subject) as Young's modulus.

\subsection{General elastic string models}
\label{5-5}

A more general category of non-barotropic string models is of course
required, not only in such terrestrial applications as (above ground)
electric power transmission cables (the subject of my first research project
as an undergraduate working working for commercial industry during a
vacation),  but also for many cosmologically relevant cases that might be
envisaged (such as that of a ``warm superconducting string" \citeA{Carter94b}) in
which several independent currents are present. Nevertheless most of the
cosmological string dynamical studies that have been carried out so far have
been restricted (not just for simplicity, but also as a very good
approximation in wide range of circumstances) to models of the the
``perfectly elastic" barotropic type, and more specifically to the category
(including the Hookean and Goto Nambu examples as opposite extreme special
cases) of string models describable by a Lagrangian of the form
\be \olag= \Lambde+\oj{^\mu}\bA_\mu+{_1\over^2}\oW{^{\mu\nu}}\bB_{\mu\nu}\, 
, \eqn{5.24}\fe
in which the only independent field variable on the world sheet
can be taken to be a stream function $\psie$, which enters only via
the corresponding identically conserved number density current
\be \ce^\mu={\calE}^{\mu\nu}\pe_\nu \, , \hskip 1 cm  
\pe_\nu=\onab_\nu\psie\, ,\eqn{5.27}\fe
whose squared magnitude,
\be \ch=\ce_\mu \ce^\mu= -\pe^\mu \pe_\nu= - \hg^{\ii\ji}\psie_\ii\psie_\ji\, ,\eqn{5.28}\fe
is the only argument in a (generically non-linear) {\it master function},
\be \Lambde=\Lambde\{ \ch \}\, .\eqn{5.29}\fe
(It is to be noted that there was a transcription error whereby the factor $^1/_2$
in the final term of (\ref{5.24}) was left out in the preceeding version of
these notes \citeA{Carter95}.) In order to satisfy the vorticity flux conservation
requirement (as derived in Section \ref{Section2} from the condition of
invariance with respect to gauge changes in the background Kalb-Ramond field
$B_{\mu\nu}$) one must have
\be \oW{^{\mu\nu}}=\okappa{\calE}^{\mu\nu} \, ,\eqn{5.25}\fe
for some fixed (quantised) circulation value $\okappa$ say -- a sort of
helicity constant associated with the string. In the simplest ``local''
string models \citeA{Carter89}, \citeA{Larsen93} this quantity
$\okappa$ can simply be taken to vanish, while in the in the simplest
``global'' models \citeA{VilenkinVachaspati87},
\citeA{Sakellariadou91}, \citeA{BenYaacov92} it will be directly
identifiable with Planck's constant, i.e. one will have
$\okappa=2\pi\hbah$.  (In more general cases it might be an integral
multiple of this.) The electromagnetic current density is specified in
terms of the gradient of the stream function in the form
\be \oj{^{\mu}}=\qe \ce^{\mu}\, ,\eqn{5.26}\fe
where $\qe$ is another fixed parameter which is interpretable as the
electromagnetic charge coupling constant per idealised particle of the
current, which means that its value will depend on the convention used to fix
the normalisation of $\psie$ or equivalently of $\ce^\mu$, a question that will
be taken up in the discussion following equation (\ref{5.41}) below.

Previous studies of cosmic strings have mostly been restricted to cases
involving allowance for a Kalb-Ramond coupling  (which is relevant for vortex
defects of ``global type"
 \citeA{VilenkinVachaspati87}, \citeA{DavisShellard89}, \citeA{Sakellariadou91}, \citeA{Witten85a}, \citeA{DavisShellard88a} 
in the absence of an internal current, or alternatively to cases (such as
string models of the kind \citeA{Carter89}, \citeA{CarterPeter95} needed to describe local
superconducting vortices arising from the Witten \citeA{Witten85}) mechanism
involving allowance for an internal current in the absence of Kalb-Ramond
coupling. However there is no particular difficulty in envisaging the presence
of both kinds of coupling simultaneously as is done here. A new result
obtained thereby (in \ref{Section6}) is that for stationary and other
symmetric configurations, the effect of the  coupling to the Kalb Ramond
background is expressible as that of a fictitious extra electromagnetic field
contribution.

It is the specific form of the master function that determines the relevant
equation of state -- or if its range of validity is sufficiently extensive,
the pair of equations of state, corresponding to the qualitatively different
regimes distinguished by the positivity or negativity of $\ch$ -- by which
the dynamics of the model is governed. The derivative of the master function
provides a quantity 
\be {\calK}=2 {d\Lambde\over d\ch}\, , \eqn{5.30}\fe
in terms of which the surface stress momentum energy density tensor 
defined by (\ref{2.7}) will be given by
\be \oT{^{\mu\nu}}={\calK} \pe^\mu \pe^\nu+\Lambde\og^{\mu\nu}\, . 
 \eqn{5.31}\fe
This only depends on the internal field gradient and on the imbedding and the
background geometry: it involves neither the background electromagnetic and
Kalb Ramond fields nor their respective coupling constants $\qe$ and $\okappa$.
The role of the latter is just to determine the background force density,
which will be given, according to the general formula (\ref{2.26}), by
\be \of_\rho=\qe \bF_{\rho\mu} \ce^\mu+ {\okappa\over 2}\bN_{\rho\mu\nu}{\calE}
^{\mu\nu} \, .\eqn{5.32}\fe
Unlike the force density, the stress momentum energy density tensor, as given
by (\ref{5.31}), is formally invariant under a duality transformation 
\citeA{Carter90}, \citeA{Carter89} whereby the variational momentum covector
$\pe_\mu$, the current $\ce^\mu$, and the master function, $\Lambde$ itself, are
interchanged with their duals, as denoted by a tilde, which are definable by
\be \tilde\ce^\mu={\calK} \pe^\mu \ ,\hskip  1 cm \tilde\pe_\mu={\calK} \ce_\mu
\, , \hskip 1 cm \tilde\ch= \tilde\ce_\mu\tilde\ce^\mu =-\tilde\pe^\nu
\tilde \pe_\nu \, ,\eqn{5.33}\fe
and by the relation
\be \ch{\calK}=\Lambde-\tilde\Lambde=-\tilde\ch \tilde{\calK}\, , 
\eqn{5.34}\fe
with                     
\be \tilde{\calK}=2{d\tilde\Lambde\over d\tilde\ch}={\calK}^{-1}\, ,\hskip
1 cm \tilde \ch=- {\calK}^2\ch \, .\eqn{5.35}\fe
In terms of these quantities the stress momentum energy density tensor can 
be rewritten in either of the equivalent mutually dual canonical forms
\be \tilde \ce^\mu \pe_\nu+\Lambde\og^\mu_{\ \nu} = \oT{^\mu}{_{\!\nu}}
=\ce^\mu\tilde\pe_\nu+\tilde\Lambde\og^\mu_{\ \nu} \, . \eqn{5.36}\fe
Substitution of these respective expressions in (\ref{5.5}) then gives the 
force balance equation in the equivalent mutually dual forms
\be \pe_\rho\onab_\mu\tilde \ce^\mu + 2\tilde \ce^\mu\onab_{[\mu} \pe_{\rho]}
+\Lambde \Ke_\rho= \of_\rho = \tilde \pe_\rho\onab_\mu \ce^\mu + 2 \ce^\mu
\onab_{[\mu}\tilde \pe_{\rho]}+\tilde \Lambda \Ke_\rho \, .\eqn{5.37}\fe
The independent internal equations of motion of the system are obtainable by
contraction with the independent mutually orthogonal tangent vectors $\pe^\rho$
and $\tilde \pe^\rho$, the latter giving what is just a kinematic identity,
since, by construction according to (\ref{5.32}) one automatically has
\be \tilde\pe^\rho \of_\rho= 0\, \eqn{5.38a}\fe
and hence 
\be \onab_\mu \ce^\mu= 0\ \ \Leftrightarrow\ \ {\calE}^{\rho\mu}\onab_\mu \pe_\rho\ =0
.\eqn{5.38}\fe
The only part of the internal equations of motion that is properly dynamical
(from the point of view of the variation principle  with respect to
$\Lambda$) is thus just the part got by contraction with $\pe^\rho$:
since (\ref{5.32}) gives
\be \pe^\rho\of_\rho=\qe \ch {\calE} ^{\mu\rho} \bF_{\mu\rho}\, . \eqn{5.39}\fe
one obtains the equivalent alternative forms
\be \onab_\mu\tilde \ce^\mu= \qe{\calE} ^{\mu\rho} \bF_{\rho\mu}\ \ \Leftrightarrow\ \
{\calE}^{\mu\rho}\big( 2\onab_{[\mu}\tilde\pe_{\rho]}+\qe \bF_{\mu\rho}\big)
=0\, .\eqn{5.40}\fe
It is to be noted that unlike the electromagnetic force, the Joukowski force
(interpretable as a manifestation of the Magnus effect) arising from the
Kalb-Ramond coupling in (\ref{5.32}) will always act orthogonally to the world 
sheet and thus does not affect the internal dynamics of the string. 
 
\subsection{Standard normalisation of current and charge}
\label{5-6}

The preceeding equation (\ref{5.40}) can be construed as the
(Poincar\'e type) integrability condition for the local existence on
the string worldsheet of a scalar $\varphi$  in terms of which the dual
momentum will simply be given as the tangentially projected gauge
covariant derivative:
\be \tilde \pe_\rho=\she\oD_\rho\phie\ ,\hskip 0.6 cm \oD_\rho=
\og^\mu_{\ \rho}\bD_\mu\ , \hskip 0.6 cm \bD_\mu=\nabl_\mu-\ee \bA_\mu
 \ ,\eqn{5.41}\fe 
where $\she$ is a normalisation constant and $\ee$ is a new, implicitly
more fundamental, charge coupling constant in terms of which the
original charge per unit idealised particle of the current is given by
\be \qe=\she \ee\, .\eqn{5.41a}\fe 
The scalar $\phie$ can be used as the primary independent variable,
instead of $\psie$, in an alternative (dynamically conjugate)
variational formulation \citeA{Carter90}, \citeA{Carter89} using a
Lagrangian constructed from $\tilde\Lambde$ instead of $\Lambde$.

By adjusting the scale of calibration used for the measurement of either the
current $\ce^\mu$ or the field $\phie$ it would of course always be possible,
for purposes of mathematical convenience to simply set the dimensionless scale
factor $\she$ to unity. However other choices may be more useful for
purposes of physical interpretation. In particular, for the kind of string
model \citeA{CarterPeter95} that provides a macroscopic descrition of Witten type
syperconducting vortices, the scaling of $\phie$ can most naturally be
chosen in such a way as to allow this field to be interpreted as the phase of
the complex scalar field characterising the underlying bosonic condensate, so
that the latter will be expressible in the form $\bPhi=\vert\bPhi\vert {\rm
e}^{i\phie}$. On the other hand it has been found mathematically convenient
in the same context to fix the normalisation of the current $\ce^\mu$ in a
standard manner by requiring that the coefficient ${\calK}$ introduced in
(\ref{5.30}) should tend to unity in the null current limit, i.e. as $\ch$
tends to zero. The condition for this standard current normalisation to be
consistent with the natural normalisation of $\phie$ is expressible as the
requirement that the ratio $\she$ (interpretable as the number of
fundamental particle units per idealised particle unit of the current $\ce^\mu$)
should just be a square root of  the model dependent constant $\kappazero$
that was used in previous work \citeA{CarterPeter95}, i.e. $\she^2=\kappazero$.
The corresponding fundamental particle current vector $z^\mu$ say 
will thus be given by
\be \ze^\mu=\she \ce^\mu \ ,\eqn{5.41b}\fe
so that if the particles are charged, with fundamental charge coupling
constant $e$, the corresponding electromagnetic surface current vector
(\ref{5.26}) will be given simply by
\be \oj{^{\mu}}=\ee \ze^{\mu}\, .\eqn{5.41c}\fe
 
If its range of definition is sufficiently extended, the master function
$\Lambde$ will determine not just one but a pair of distinct equations of
state, one applying to the ``magnetic" regime where the current $\ce^\mu$ is
spacelike so that $\ch$ is positive, and the other applying to the
``electric" regime where the current $\ce^\mu$ is timelike so that the $\ch$ is
negative.  In either of these distinct regimes -- though not on the critical
intermediate ``null state" locus where the current is lightlike so that $\ch$
vanishes -- the mutually dual {\it canonical} forms (\ref{5.36}) of the stress
momentum energy density tensor are replaceable by the equivalent more
manifestly symmetric {\it standard} form
\be \oT^{\mu\nu}={\Lambde\over\ch}\, \ce^\mu \ce^\nu+{\tilde\Lambde
\over\tilde\ch}\,\tilde \ce^\mu\tilde \ce^\nu\, .\eqn{5.42}\fe
By comparison with the corresponding expression (\ref{5.11}) it can thus be seen
that the state functions introduced in the variational function that has just
been specified will be phyiscally interpretable in terms of the corresponding
energy density $\Ue$ and  tension $\Te$, and also of the number density $\nue$ and
effective mass $\mue$  associated with the corresponding physically normalised
particle flux (meaning that for which the charge per particle is  $\ee$)  will
be given parametrically in the ``electric" regime where $\ce^\mu$ is timelike,
by
\be \Lambde=- \Ue\, ,\hskip 0.6 cm \tilde\Lambde = - \Te\, , \hskip 0.6 cm
\she^2\ch=-\nue^2 \ ,\hskip 0.6 cm \tilde\ch=\she^2\mue^2\, ,\hskip 0.6 cm
{\calK}=\she^2{\mue\over\nue} \, ,\eqn{5.45}\fe
with
\be \she\ce^\rho=\nue \ue^\rho\, ,\hskip 0.6 cm \tilde \ce^\rho=\she\mue \ve^\rho
\, ,\hskip 0.6 cm \she\pe_\rho =\nue \ve_\rho\, ,\hskip 0.6 cm 
\tilde\pe_\rho=\she \mue \ue_\rho \, .\eqn{5.46}\fe
(For the corresponding mathematically normalised flux, whose idealised
particles are to be considered as each containing $\she$ of the fundamental
particles -- so that, in accordance with (\ref{5.41a}) the corresponding
charge per particle is $\she \ee$ -- the corresponding effective mass would be
$\she\mue$ and the corresponding number density would be $\nue/\she$.) The
analogous formulae for the ``magnetic" regime where $\ce^\mu$ is spacelike (the
first possibility to be considered in the early work on the Witten
mechanism \citeA{Witten85}), will be given by 
\be \Lambde=- \Te\ ,\hskip 0.6 cm \tilde\Lambde = - \Ue\, , \hskip 0.6 cm
\she^2\ch=\mue^2 \, ,\hskip 0.6 cm \tilde\ch=-\she^2 \nue^2\, ,
\hskip 0.6 cm {\calK}=\she^2{\nue\over\mue}\, ,\eqn{5.43}\fe
with
\be \she \ce^\rho=\mue \ve^\rho\, ,\hskip 0.6 cm \tilde \ce^\rho=
\she\nue \ue^\rho\, ,\hskip 0.6 cm \she \pe_\rho =\mue \ue_\rho\, ,
\hskip 0.6 cm \tilde \pe_\rho=\she\nue \ve_\rho \, .\eqn{5.44}\fe

Whether or not it is charged, and whatever  the local (spacelike, timelike, or
null) character of the current may be, the conservation law (\ref{5.31})
implies that its integral round any {\it closed} string loop will give a
corresponding conserved particle quantum  number
\be \Ze=\oint\! dx^\mu {\calE}_{\mu\nu}\ze^\nu  \, .\eqn{5.60}\fe
For such a loop there will also be a second independent topologically
conserved phase winding quantum number, $\Ne$ say,  defined by
\be 2\pi \Ne=\oint\! d\phie \, .\eqn{5.61}\fe
It can be seen from (\ref{5.41}) that this number will be given by 
\be 2\pi\she \Ne = \qe\Phie+\oint dx^\mu\, \tilde \pe_\mu\, ,
\eqn{5.61a}\fe
where $\Phie$ is the magnetic flux integral,
\be \Phie=\oint\! dx^\mu \bA_\mu\, .\eqn{5.62}\fe
It is to be noted that in the electromagnetically charged coupled case, i.e.
when the relevant fundamental particle coupling constant $\ee$ has a non zero
value (whose approximate value will be given in fundamental units by
$1/\sqrt{137}$ or a multiple thereof) the loop will have a conserved total
electric charge, $\Qe$ say, given  by
\be \Qe=\ee \Ze 
=\oint\! dx^\mu {\calE}_{\mu\nu}\oj{^{\nu}}  \, .\eqn{5.63}\fe

\subsection{Analytic string model for Witten vortex}
\label{5-7}

In realistic conducting string models one can not expect the appropriate
master function $\Lambde\{\ch\}$ to be given exactly by an explicit 
analytic formula, but only by the output of a detailed numerical computation
of the internal structure of the underlying vacuum 
defect \citeA{BabulPiranSpergel88}, \citeA{Peter92}, \citeA{Peter92a}, \citeA{Peter93}. Nevertheless it has 
recently been found \citeA{CarterPeter95} that a remarkably good approximated description
of the results of such computations for Witten's original ``superconducting''
bosonic vortex model can be provided by analytic expressions of a fairly
simple kind, involving only two model dependent mass parameters. The one that
is presumed to be the larger is the usual Kibble mass $m$ say, whose square
determines the energy and tension in the null current limit, and whose value
is expected to be of the order of the mass of the Higgs field responsible for
the spontaneous symmetry breaking that produces the vortex defect. The other
mass parameter, $\mast$ say, is expected to be of the order of the
considerably smaller mass scale that is presumed to characterise the
independent current carrier field in the relevant version of Witten's bosonic
field model \citeA{Witten85}. In all the cases that have been investigated in
detail \citeA{Peter92}, \citeA{Peter92a}, \citeA{Peter93} it has been found that to obtain a
reasonably accurate representation throughout the (``magnetic'') regime for
which $\tilde \ch$ is negative, and for moderate positive values as well, it
suffices \citeA{CarterPeter95} to use the purely polynomial formula
\be\tilde\Lambde=-\mag^2+{\tilde\ch\over 2}\Big(1-{\tilde\ch\over 
2\mast^{\,2}}\Big)^{-1} \, ,\eqn{poly} \fe
which is applicable within the range 
\be  -{1\over 3}<{\tilde\ch\over 2 \mast^{\,2}} <
1-{\mast^{\, 2}\over \mag^2+\mast^{\,2}}\, , \fe
which is limited below by the condition that $\cL{^2}$ should be
positive and limited above by the condition that $\cE{^2}$ should be
positive. For values near the upper end of this range (in the ``electric''
regime) the formula (\ref{poly}) does however, becomes inadequate, but to
obtain a very satisfactory description throughout the positive range of
$\tilde\ch$, it suffices to replace it by the still very simple though
non-polynomial formula
\be \tilde\Lambde=-\mag^2-{\mast^{\,2}\over 2}\,{\rm ln}\big\{1-{\tilde\ch
\over \mast^{\,2}}\big\} \, ,\eqn{witseq}\fe
for which the corresponding range of applicability is given by
\be  -1<{\tilde\ch\over \mast^{\,2}} < 1-{\rm e}^{-2\mag^2/ \mast^{\,2}}\, .
\fe

Since the underlying field theoretical model proposed by Witten \citeA{Witten85}
provides what can at best be only a highly simplified description of the
mechanisms that would be involved in any real vacuum vortex, there is no point
in investing too much effort in describing it with unduly high precision. To
provide just the moderate degree of accuracy that is physically reasonable,
the single formula (\ref{witseq}) is all that is needed to cover the entire
admissible range: it is more accurate than it need be for this purpose in the
positive range of $\tilde\ch$; it agrees perfectly with the preceeding
formula (\ref{poly}) in the null limit where $\tilde\ch$ tends to zero; and
finally even in the negative (``magnetic'') range, where (\ref{poly}) is
quantitatively more accurate, the newer formula (\ref{witseq}) still provides
a description that is in qualitatively adequate agreement with that of its
predecessor. In the (``electric'') range $\tilde\ch>0$ where the formula
(\ref{witseq}) is most accurate, the equation of state relating the energy
$\Ue$ to the tension $\Te$ will take the form
\be \Ue=\Te+\mast^{\,2}\big({\rm e}^{2(\mag^2-\Te)/\mast^{\,2}}-1\big) \, ,
\eqn{elecstate}\fe
while  in the (``magnetic'') range the corresponding relation is given by
\be \Te=\Ue-\mast^{\,2}\big(1-{\rm e}^{-2(\Ue-\mag^2)/\mast^{\,2}}\big)\, .
\eqn{magstate}\fe

It is evident that in the absence of background forces, knowledge of the two
parameters $\mag$ and $\mag_\ast$ is fully sufficient to characterise the dynamics
of such a string model -- indeed it is only the ratio $\mag/\mast$ that actually
matters for this purpose. However one will also have to evaluate the relevant
scaling factor $\she=\sqrt{\kappazero}$ (which, like $\mag$ and $\mast$ will
depend on the parameters characterising the underlying bosonic field theory)
if one wishes to be able to allow for the effect of any background
electromagnetic field that may be present, since for this purpose it is
necessary to know the value of the mathematically normalised coupling constant
$\qe$. The latter will be directly obtainable from the value of $\she$ (which
in typical cases may be expected to be of the order of unity) via
(\ref{5.41a}), on the assumption that the relevant value of the physical
charge coupling constant $\ee$ characterising the boson condensate is just the
usual electronic charge (as given in Planck units by $\ee^2\approx 1/137$) or a
simple integral multiple thereof.

\subsection{*The self-dual transonic string model}
\label{5-8}

Although the formulae \citeA{CarterPeter95} in the preceeding section provide the simplest
decription that one could reasonably desire for the purpose of describing
realistic superconducting strings arising from the Witten mechanism, it is
instructive to study an even simpler model that originally turned up in a
rather different physical context of a more artificial nature, namely that of
a Kaluza Klein type projection of a simple Goto-Nambu model in an extended
background with an extra space dimension in the manner first suggested by
Nielsen \citeA{Nielsen80}. It can easily be shown \citeA{Carter90} that this gives rise to a
particularly elegant illustration of the above formalism in which the relevant
master function has the rather unusual though by no means unique property of
being self dual in the sense that (for an appropriate choice of the scaling of
the stream function $\psie$) the dependence of the master function $\Lambda$ on
its argument $\ch$ is exactly the same as that of the corresponding function
$\tilde\Lambde$ on $\tilde\ch$, their form in this case being is given in
terms of a constant mass parameter $\mag$ simply by
\be \Lambde= \mag\,\sqrt{m^2-\ch}\ ,\hskip 1  cm \tilde\Lambde=
\mag\,\sqrt{\mag^2-\tilde\ch} \, .\eqn{5.47}\fe
This self duality property means that the equation of state relationship
between $\Ue$ and $\Te$ will the {\it same} form in the magnetic regime
$\ch>0>\tilde\ch$ as in the electric regime $\tilde\ch>0>\ch$. What is quite
unique about the equation of state obtained for particular self dual string
model is that it will have the non-dispersive -- i.e. permanently transonic --
``constant product" form (\ref{4.21}), which can be seen \citeA{Carter90a} to be
expressible parametrically in terms of a dimensionless self dual state
function $\varthe$ of the kind introduced in (\ref{5.19}) by
\be \Ue=\mag^2\sqrt{1+{\nue^2\over \mag^2}}=\mag^2{\rm coth\,}\varthe\ ,\hskip 1 cm
\Te=\mag^2\sqrt{1-{\mue^2\over \mag^2}}=\mag^2{\rm tanh\,}\varthe \, ,\eqn{5.48}\fe
which gives
\be \cE=\cL={\rm tanh\,}\varthe \ ,\eqn{5.49}\fe
so that by (\ref{4.21}) and (\ref{4.20}), the extrinsic (``wiggle") and
longitudinal (``woggle" or ``jiggle") bicharacteristic vectors will coincide,
having the form
\be \elle_{_\pm}{^{\!\mu}}=\ue_{_\pm}{^{\!\mu}} ={\rm cosh\,}\varthe\, 
\ue^\mu \pm {\rm sinh\,}\varthe\, \ve^\mu \ .\eqn{5.50}\fe

The revelation \citeA{Carter90} of the special transonicity property of the model
characterised by (\ref{5.47}) was sufficient to invalidate the over hasty
claim \citeA{NielsenOlesen87} that Nielsen's elegant artifice \citeA{Nielsen80} effectively
represents the outcome in the ``pure" string limit (in which effects of finite
vortex thickness are neglected) of the Witten mechanism \citeA{Witten85}. Long before
the derivation  \citeA{CarterPeter95} of the explicit formula (\ref{witseq}) it was already
known that in order to provide a satisfactory representation a Witten type
vacuum vortex it would be necessary to use a model \citeA{BabulPiranSpergel88} of generically
dispersive type \citeA{Carter89}. More specifically, it was known from the work of
Peter \citeA{Peter92}, \citeA{Peter92a}, \citeA{Peter93} that such a model would need to
be characterised typically by supersonicity, $\cE>\cL$ for
spacelike and moderate timelike values of the current provided the ratio of
the carrier mass to the Higgs mass is not too large. Now that the model
(\ref{witseq}) is available, it can be verified directly that it will indeed
exhibit supersonicity for all spacelike values and at least for small timelike
values of the current not just when the ratio $\mast^{\,2}/\mag^2$ is small
compared unity but so long as it does not exceed the implausibly large value
2. On the other hand it can be seen that such model will always become
subsonic for sufficiently large timelike current values (near the relaxation
limit in the ``electric'' regime).

It is to be remarked that despite its inability to reproduce these features,
the use of the mathematically convenient transonic model (\ref{5.47}), as
advocated by Nielsen and Olesen \citeA{NielsenOlesen87}, for the purpose of describing the
dynamics of a realistic superconducting string model was nevertheless a
considerable improvement, not just on the use \citeA{SpergelPressScherrer89} for this purpose of
an unmodified Goto-Nambu model, but also on the more commonly used description
provided by the naively linearised model
 \citeA{SpergelPiranGoodman87}, \citeA{VilenkinVachaspati87a}, \citeA{CopelandHindmarshTurok87}, 
\citeA{Copelandetal88}, \citeA{HawsHindmarshTurok88}, \citeA{Amsterdamski89} 
with equation of state given by constancy of the trace (\ref{5.3}) which, by
(\ref{4.10}) and (\ref{4.20}), is evidently characterised by permanent
subsonicity, $\cE<\cL=1$.

The intrinsic equations of motion (\ref{5.18}) for the transonic model can be
usefully be recombined  \citeA{Carteretal94}, \citeA{Carter94b} as an equivalent pair of
divergence  relations that are expressible in terms of the bicharacteristic
vectors (\ref{5.50}), in the form
\be \onab_\rho\Big( (\Ue-\Te)\elle_{_\pm}{^{\!\rho}}\Big)
=-\elle_{_\mp}{^{\!\rho}}\of_\rho \, ,\eqn{5.51}\fe
which shows that the  ``left" and ``right" moving ``bicharacteristic
currents", $(\Ue-\Te)\elle_{_+}{^{\!\rho}}$ and $(\Ue-\Te)\elle_{_-}{^{\!\rho}}$ will
each be conserved separately in the free case, i.e. when $\of_\rho$
vanishes.  As an alternative presentation of the tangential force balance
equations for this permanently transonic model, a little algebra suffices to
show that the intrinsic equation of motion (\ref{5.21}) can be rewritten for
this case, in a form more closely analogous to that of the corresponding
extrinsic equation of motion (\ref{5.22}), as the pair of equations
\be (\Ue-\Te)\og^{\mu}_{\ \nu}\,\elle_{_\pm}{^{\!\rho}}\onab_\rho \elle_{_\mp}
{^{\!\nu}}=\big(\og^{\mu}_{\ \nu}+ \elle_{_\mp}{^{\!\mu}} 
\elle_{_\mp\nu}\big) \of{^\nu}\, .\eqn{5.52}\fe
This tangentially projected part of the dynamic equations can now be 
recombined \citeA{Carter90a} with its orthogonally projected analogue (5.22),
so as to give the {\it complete} set of force balance equations for the
non dispersive permanently transcharacteristic string model (4.21) as 
the extremely useful pair of bicharacteristic propagation equations
\be (\Ue-\Te)\elle_{_\pm}{^{\!\rho}}\onab_\rho \elle_{_\mp}{^{\!\mu}}=
\big(g^{\mu\nu} + \elle_{_\mp}{^{\!\mu}} \elle_{_\mp}{^{\!\nu}}\big)
\of{_\nu}\, .\eqn{5.53}\fe

\subsection{*Integrability and application of the transonic 
string model}\label{5-9}

It is evident that the bicharacteristic formulation (\ref{5.53}) of the
equations of motion for the transonic string model will take a particularly
simple form when translated into terms of the corresponding characteristic
coordinates $\sigme^{_\pm}$ on the worldsheet, where the latter are  defined
by taking the ``right moving" coordinate $\sigme^{_+}$ to be constant allong
``left moving" characteristic curves, and taking the ``left moving" coordinate
$\sigme^{_-}$ to be constant along ``right moving" characteristic curves, with
the convention that the correspondingly parametrised bicharacteristic tangent
vectors $\elle_{_\pm}^{\,\mu}\equiv{\partial x^\mu/\partial \sigme^{_\pm}}$
should be future directed. 

In the case of a {\it free} motion (i.e. when the force term on the right of
(\ref{5.53}) vanishes) in a {\it flat}  background, this simplification can be
used \citeA{Carter90a} to obtain the complete solution of the dynamical equations in
a very simple explicit form. The way this works is that with respect to
Minkowski coordinates in such a background  (\ref{5.53}) will reduces to the
simple form 
\be {\partial \elle_{_\pm}^{\,\nu}/\partial \sigme_{_\mp}}=0 \
,\eqn{5.54}\fe 
whose general solution is given in terms of a pair of generating curves
$x_{_\pm}^{\,\mu}\{\sigme\}$ as a sum of single variable functions by the
ansatz
\be x^\mu=x_+^{\,\mu}\{\sigme^{_+}\} +x_-^{\,\mu}\{\sigme^{_-}\}\ 
,\eqn{5.55}\fe
which gives $\elle_{_\pm}^{\,\mu}=\dot x_{_\pm}^{\,\mu}$
using the dot here to denote the ordinary derivatives of the single variable
functions with respect to the corresponding characteristic variables. The
solution (\ref{5.55}) generalises a result that is well known for more familiar
but degenerate Goto Nambu case, in which the tangents to the generating
curves are required to be null, $\elle_{_\pm}^{\,\mu}\elle_{_\pm\mu}=0$ (so
that with the usual normalisation their space projections lie on what is
known as the Kibble Turok sphere \citeA{KibbleTurok82}). The only restriction in the
non - degenerate case is that they should be non-spacelike and future
directed (the corresponding projections thus lying anywhere in the interior,
not just on the surface, of a Kibble Turok sphere) the unit normalisation
condition (\ref{5.20}) being imposable as an option, not an obligation, by
choosing the parameter $\sigme$ to measure {\it proper} time allong each
separate generating curve.

This special property of being explicitly integrable by an ansatz of the same
form (\ref{5.55}) as has long been familiar for the Goto Nambu case can
immediately be used to provide a new direct demonstration of the validity of
the claim I originally made several years ago \citeA{Carter90a} to the effect that
the non-dispersive constant product model (\ref{4.21}) is not just a
mathematical curiosity, but that it is usefully applicable for the purpose of
providing a realistic description of the average motion of a ``wiggly" Goto
Nambu string, effectively extending to the tension $\Te$ the concept of
``renormalisation" that had previously been introduced for the energy density
$\Ue$ by Allen and Shellard \citeA{AllenShellard90}, \citeA{ShellardAllen90}. 

The new justification \citeA{Carter95a} summarised here is needed because my original
argument \citeA{Carter90a} was merely of a qualitative heuristic nature, while
Vilenkin's mathematical confirmation \citeA{Vilenkin90} was based on indirect
energetic considerations, and was legitimately called into question \citeA{HongKimSikivie80}
on the grounds that it did not cover quite the most general class of
``wiggles'' that can be envisaged. Such an objection is effectively bypassed
by the alternative approach presented here, but there would still be a paradox
if allowance for more general ``wiggles'' really did produce what was alleged:
thus although it was based on manifestly muddled reasonning and
misinterpretation of preceding work, the purported demonstration \citeA{HongKimSikivie80} of
higher order ``deviations" from the constant product form  (\ref{4.21}) for
the effective equation of state gave rise to a controversy that was not
definitively resolved until a genuinely valid extension of Vilenkin's 
method \citeA{Vilenkin90} was finally produced by Martin \citeA{Martin95}, whose conclusion
was that the constant product equation of state (\ref{4.21}) will after all
provide an account that remains accurate for {\it arbitrary} ``wiggle"
perturbations of a Goto Nambu string (subject only to the restriction that
their amplitudes should not be so large as to bring about a significant rate
of self intersection).

As an alternative to the energy analysis originated by Vilenkin and completed
by Martin, the more direct justification \citeA{Carter95a} for the use of the elastic
string model characterised by (\ref{5.47}) as a model for the large scale
averaged behaviour of a Goto-Nambu string is as follows. The argument is
simply based on the observation that such a model implicitly underlies the
diamond lattice discretisation that, since its original introduction by Smith
and Vilenkin \citeA{SmithVilenkin87}, has been commonly employed by numerical
simulators \citeA{Albrecht90} as a very convenient approximation scheme -- of in
principle unlimited accuracy -- for the representation of a Goto-Nambu string
worldsheet in a flat background. As a way of replacing the exact continouous
description by a discrete representation such as is necessary for numerical
computation, the idea of the Smith Vilenkin method is simply to work with a
pair of discrete sets of sampling points  $x_{_\pm\rm r}^{\,\mu}=
x_{_\pm}^{\mu}\{\sigme_{\rm r}\}$ determined by a corresponding discrete set
of parameter values $\sigme_{\rm r}$ on the generating curves of the exact
representation (\ref{5.55}). This provides a ``diamond lattice" of sample
points given (for integral values of r and s) by 
\be x^\nu_{\rm rs}=x_{_+\rm r}^{\,\mu}+x_{_-\rm s}^{\,\mu}\, ,\eqn{5.56}\fe 
that will automatically lie
exactly on the ``wiggly"  Goto Nambu worldsheet (\ref{5.55}), which is thus
represented to any desired accuracy by choosing a sufficiently dense set of
sampling parameter values $\sigme_{\rm r}$ on the separate ``wiggly" null
generating curves $x_{_\pm}^{\,\mu}\{\sigme\}$. It is evident that the chosen
set of sample points $x_{_\pm\rm r} ^{\,\mu}=x_{_\pm}^{\mu}\{\sigme_{\rm r}\}$
on the separate ``wiggly" null generators can also be considered to be sample
points on a pair of {\it smoothed out}, and thus no longer null but {\it
timelike}, interpolating curves that, according to the result \citeA{Carter90a}
demonstrated above, can be interpreted according to (\ref{5.55}) as generating
a corresponding solution of the equations of motion for an elastic string
model of the kind governed by (\ref{4.21}).  The not so ``wiggly" elastic
string worldsheet constructed by this smoothing operation will obviously be an
even better approximation to the exact ``wiggly" Goto Nambu worldheet than the
original Smith Vilenkin lattice representation, which itself could already be
made as accurate as desired by choosing a sufficiently high sampling
resolution. No matter how far it is extrapolated to the future, the smoothed
elastic string worldsheet generated according to (\ref{5.54}) can never
deviate significantly from the underlying ``wiggly" Goto-Nambu worldsheet it
is designed to represent because the exact worldsheet and the smoothed
interpolation will always coincide precisely at each point of their shared
Smith Vilenkin lattice (\ref{5.56}). This highly satisfactory feature of
providing a potentially unlimited accuracy could not be improved but would
only be spoiled by any ``deviation" from the originally proposed \citeA{Carter90a}
form (\ref{4.21}) for the effective equation of state.

After thus conclusively establishing that the permanently transonic elastic
string model characterised by the simple constant product equation of state
(\ref{4.21}) (without any higher order corrections) provides an optimum
description of the effect of microscopic wiggles in an underlying Goto Nambu
model so long as self intersections remain unimportant (as was assumed in all
the discussions  \citeA{Carter90a}, \citeA{Vilenkin90}, \citeA{HongKimSikivie80}, \citeA{Carter95a}, \citeA{Martin95} 
cited above), it remains to be emphasised that the neglect of such
intersections will not be justified when the effective
temperature \citeA{Carter90a}, \citeA{Carter94b} of the wiggles is too high (as will
presumably be the case \citeA{Carteretal94} during a transient period immediately
following the string - forming phase transition). The result of such
intersections will be the formation of microscopic loops, of which some will
subsequently be reconnected, but of which a certain fraction will escape.
Estimation of the dissipative cooling force density that would be needed to
allow for such losses remains a problem for future work.  

\section{Symmetric configurations including rings and their vorton equilibrium
states}
\label{Section6}

\subsection{Energy-momentum flux conservation laws}
\label{6-1}

Whenever the background space time metric is invariant under the action of a
(stationarity, axisymmetry, or other) continuous invariance group
generated by a solution $\el^\mu$ Killing 
of the equation (\ref{2.31}), i.e. in the
notation of (\ref{2.20})
\be \vec{\ \el\Libra} \bg_{\mu\nu}= 0\, ,\eqn{6.1}\fe
then any string that is isolated (i.e. not part of the boundary of an 
attached membrane) will have a corresponding {\it momentum} current  
(interpretable, depending on the kind of symmetry involved, 
as representing a flux of energy, angular momentum, or whatever) given  by
\be \oP{^\mu}=\oT{^\mu}_{\!\nu} \el^\nu\, .\eqn{6.2}\fe
In accordance with (\ref{2.33}), this will satisfy a source equation of the 
form
\be \onab_\mu\oP{^\mu}= \of_\mu \el^\mu\, , \eqn{6.3}\fe
which means that the corresponding flux would be strictly conserved when
the string were not just isolated but {\it free}, i.e. if the
background force $\of_\mu$ were zero. When the string is subject to a
background force of the Lorentz-Joukowsky form (\ref{5.33}) that arises from
background electromagnetic and Kalb-Ramond fields, then provided these
background fields are also invariant under the symmetry group action
generated by $\el^\mu$, i.e. in the notation of (\ref{2.20})
\be \vec{\ \el\Libra} \bA_{\mu}= 0\ ,\hskip 1 cm 
\vec{\ \el\Libra} \bB_{\mu\nu}= 0 \, ,\eqn{6.4}\fe
it can be seen to follow that, although the physically well defined surface
current $\oP{^\mu}$ will no longer be conserved by itself, it still 
forms part of a gauge dependent generalisation, $\caloP{^\mu}$ say, 
that is strictly conserved, 
\be \onab_\mu\calP{^\mu}=0\, ,\eqn{6.5}\fe
and that is given, in terms of the gauge dependent generalisation
\be \caloT{^\mu}_{\!\nu}=\oT{^\mu}_{\!\nu}+\oj{^\mu}\bA_\mu+
\oW{^{\mu\rho}}\bB_{\nu\rho} \eqn{6.6}\fe
of the surface stress momentum energy density tensor, by
\be \caloP{^\mu}=\caloT{^\mu}_{\!\nu}\el^\nu=\oP{^\mu}+
\big(\qe \ce^\mu \bA_\nu +\okappa{\calE}^{\mu\rho}\bB_{\nu\rho}\big)\el^\mu
\, . \eqn{6.7}\fe

\subsection{Bernoulli constants for a symmetric string 
configuration} \label{6-2}

Let us now restrict attention to cases in which string configuration itself
shares the background symmetry under consideration. This category includes
the  astrophysically interesting case of circular configurations in an
axisymmetric background -- which is of particular interest in the context of
vorton formation in a flat background -- and of course it also includes many
familiar terrestrial examples of static string configurations in a stationary
background -- one of the most obvious being that of the overhead telephone and
power cables that festoon our environment. Any string configuration
that is symmetric in this sense will be characterised by the condition
\be \ag^{\!\mu}_{\,\nu}\el^\nu=0\, ,\eqn{6.8}\fe
meaning that the relevant symmetry generator $\el^\mu$ is tangential to the
worldsheet, and the corresponding Lie invariance condition on its surface
stress momentum energy density tensor will have the form
\be \el^\mu\nabl_\mu\oT{^{\nu\rho}}=2 \oT{^{\mu(\nu}}\nabl_\mu \el^{\rho)}
\, .\eqn{6.9}\fe
Under such conditions, as well as the ordinary momentum flux $\oP{^\mu}$, 
what may be termed the {\it adjoint momentum} flux,
\be ^\dagger\!\oP{^\mu}=\el^\nu {^\dagger}\oT{_\nu}{^\mu}\, ,
\hskip 1 cm ^\dagger\oT{_\nu}{^\mu}={\calE}_{\nu\rho}
{\calE}^{\mu\sigma}\oT{^\rho}_{\!\sigma} \, ,\eqn{6.10}\fe
will also obey an equation of the form (\ref{6.3}), i.e.
\be \onab_\mu^\dagger\oP{^\mu}= \of_\mu \el^\mu\, , \eqn{6.11}\fe
and  hence would also be conserved if the string were free.
In the presence of an electromagnetic or Kalb Ramond background, it
can be seen that, like the ordinary momentum flux, this adjoint momentum flux
has a gauge dependent extension,
\be ^\dagger\!\caloP{^\mu}=\el^\nu {^\dagger}\caloT{_\nu}{^\mu}\, ,
\hskip 1 cm ^\dagger\oT{_\nu}{^\mu}={\calE}_{\nu\rho}
{\calE}^{\mu\sigma}\caloT{^\rho}_{\!\sigma} \, ,\eqn{6.12}\fe
that will share with the generalised momentum flux
$\caloP{^\mu}$ (that would be conserved even if the string did not
share the symmetry of the background) the property of obeying
a strict surface current conservation law, namely
\be \onab_\mu\, ^\dagger\!\caloP{^\mu}=0\, .\eqn{6.13}\fe

The group invariance conditions 
\be \el^\nu\nabl_\nu \pe_\mu+\pe_\nu\nabl_\mu \el^\nu=0\ ,\hskip 1 cm
\el^\nu\nabl_\nu \tilde \pe_\mu+\tilde \pe_\nu\nabl_\mu \el^\nu=0 \, ,\eqn{6.14}\fe
that are the analogues of (\ref{6.9}) for the separate mutually dual pair of
internal momentum covectors $\tilde\pe_\mu$ and $ \pe_\mu$
associated with the internal current within the string, can be rewritten,
with the aid of the corresponding electromagnetic background invariance 
condition (\ref{6.4}), in the form
\be \el^\rho{\calE}^{\mu\nu}\big(\nabl_\nu\tilde \pe_\mu+
{\qe\over 2}\bF_{\nu\mu}\big)={\calE}^{\rho\nu}\nabl_\nu\Beta\, , 
\hskip 1 cm\el^\rho{\calE}^{\mu\nu}\nabl_\nu \pe_\mu=
{\calE}^{\rho\nu}\nabl_\nu\bbeta \, ,\eqn{6.15}\fe
in terms of a pair of scalars $\Beta$ and $\bbeta$, that can be considered as
generalisations of the well known Bernoulli constant in stationary classical
perfect fluid dynamics, and that are defined by
\be \Beta=\tilde\bbeta +\qe \bA_\mu \el^\mu\ ,\hskip 1 cm 
\tilde\bbeta=\tilde\pe_\mu 
\el^\mu \ , \hskip 1 cm \bbeta =\pe_\mu \el^\mu \ .\eqn{6.16}\fe
If $\el^\mu$ is timelike, the corresponding symmetry will be interpretable
as {\it stationarity}, while the more restrictive case \citeA{Carter90b} in
which the string is actually {\it static} (in the sense that there is no
transverse current  component relative to the background rest frame determined
by $\el^\mu$) will be given by the condition that the second Bernoulli
constant, $\bbeta$ should vanish.

Comparing the conditions (\ref{6.15}) with
(\ref{5.38}) and (\ref{5.40}) it can be seen that the {\it internal}
dynamical equations are equivalent in this group invariant case
to the corresponding pair of Bernoulli type conservation laws to the
effect that $\Beta$ and $\bbeta$ (but, unless the electromagnetic field is
absent, not $\tilde\bbeta$) should both be {\it constant} over the worldsheet.
This observation allows the problem of solving the dynamical for a symmetric
configuration of the kind of (``perfectly elastic" , i.e. barotropic) string
model under consideration to one of solving just the extrinsic equations
governing the location of the worldsheet. 

\subsection{Generating tangent vector field for symmetric 
solutions}\label{6-3}

A recent investigation \citeA{CarterFrolovHeinrich91} based on the systematic use of variational
methods in the restricted case for which the Kalb-Ramond coupling was absent
has drawn attention to the interest of extrapolating the Bernoulli constants
outside the supporting worldsheet as a pair of scalar fields defined over the
entire background spacetime by the uniformity conditions
\be \nabl_\mu\Beta=0\, , \hskip 1 cm \nabl_\mu\bbeta=0\, ,\eqn{6.17}\fe
and formulating the problem in terms a certain preferred generating
vector field that is constructed in such a manner as
to be everywhere tangent to the worldsheet according to the specification
\be \Xx^\mu={\calE}^{\mu\nu}{^\dagger}\oP_\nu=\el^\nu{\calE}_{\nu\rho}
\oT{^{\rho\mu}}\, .\eqn{6.18}\fe

Using (\ref{5.42}) and (\ref{6.16}), it can be seen that the generating vector
defined in this way will be expressible in terms of the variables introduced
in Section \ref{Section5} as
\be \Xx^\mu={\bbeta\Lambde\over\ch} \ce^\mu+{\tilde\bbeta\tilde\Lambde
\over\tilde\ch}\tilde\ce^\mu\, ,\eqn{6.19}\fe
while the Killing vector itself will be expressible in analogous form by
\be \el^\mu={\tilde\bbeta\over\Lambde-\tilde\Lambde}\,\ce^\mu+{\bbeta\over\tilde\Lambde
-\Lambde}\,\tilde\ce^\mu \, .\eqn{6.20}\fe
By further use of the formulae in  Section \ref{Section5} it can be seen that
the squared amplitude of the latter (which in the case of stationary symmetry
can be interpreted as an effective gravitational potential field)
will be given by an expression of the form
\be \el^2=
\el^\mu \el_\mu=-{\bbeta^2\over\ch}-{\tilde\bbeta^2\over\tilde\ch}
\ .\eqn{6.21}\fe

To obtain a solution of the dynamical equations governing a symmetric string
configuration of he kind under consideration it will evidently be sufficient
to obtain an integral trajectory of the generating tangent vector $\Xx^\mu$,
since when this is available the complete worldshhet can then be swept out in
a trivial manner by dragging this trajectory along under the symmetry action
generated by $\el^\mu$. A very useful approach to the problem of finding sych a
tangent trajectory is based on a procedure that exploits of the preceeding
equation (\ref{6.21}) in the following manner. 

In view of (\ref{6.16}), and of the state functional relationship (\ref{5.35})
between $\ch$ and its dual, $\tilde\ch$,  the equation (\ref{6.21}) can be
solved for any particular choice of the constant``tuning parameters" $\Beta$
and $\bbeta$ to determine the internal variables $\ch$ and $\tilde \ch$ as
functions of  the  (geometrically determined) potential $\el^2$ and the
(electric type) potential $\el^\nu \bA_\nu$, and hence by implication as scalar
fields over the entire background space, not just on the worldsheet where they
were originally defined.

Again using the formulae in Section \ref{Section5}, it can be seen  from
(\ref{6.19}) and (\ref{6.20}) that the contraction of the worldsheet
generating vector $\Xx^\mu$ with the Killing vector field will be given simply
by
\be \el^\mu \Xx_\mu=\bbeta\tilde\bbeta\, ,\eqn{6.22}\fe
and that the squared amplitude of $\Xx^\mu$ will given by an expression of the
form
\be \Xx^2= \Xx^\mu \Xx_\mu={\bbeta^2\Lambde^2\over \ch}+{\tilde\bbeta^2
\tilde\Lambde^2\over\tilde\ch} \, . \eqn{6.23}\fe
By the latter equation, $\Xx^\mu \Xx_\mu$ is also
 implicitly defined as a function of $\el^2$ and $\el^\nu \bA_\nu$, 
so that it too can be considered defined scalar field, not just on the worldsheet
but also over the background as a whole. The gradient of the field
constructed in this way will be given by the expression
\be \nabl_\mu \Xx^2=-\Lambde\tilde\Lambde\, 
\nabl_\mu \el^2+2\qe\tilde\bbeta(\Lambde
-\tilde\Lambde){\tilde\Lambde\over\tilde\ch} \nabl_\mu(\el^\nu \bA_\nu)
\, .\eqn{6.24}\fe

Previous experience \citeA{CarterFrolovHeinrich91} with the case in which only the electromagnetic
but not the Magnus force contribution is present suggests the interest of
formulating the problem in terms of the propagation of the special generating
vector $\Xx^\mu$, which (using the formulae (\ref{5.5}) and (\ref{5.42}) of the
previous section) can be seen from (\ref{6.19}) and (\ref{6.20}) to be given
by
\be \Xx^\nu\nabl_\nu \Xx_\mu=\Lambda\tilde\Lambda\, \el^\nu\nabl_\nu \el_\mu
+\Xx^\nu{\calE}_{\nu\rho}\el^\rho \of_\mu \, ,\eqn{6.25}\fe
where $\of^\mu$ is the background force as given by (\ref{5.32}). It can now 
be seen that the two preceeding equations can be combined to give the 
equation of motion for the worldsheet generating vector $\Xx^\mu$ in
the very elegant and convenient final form
\be \Xx^\nu\nabl_\nu \Xx_\mu-{_1\over^2}\nabl_\mu \Xx^2={\cal F}_{\mu\nu}
\Xx^\nu\ ,\eqn{6.26}\fe
in terms of a pseudo Maxwellian field given by
\be {\psF}_{\mu\nu}=\qe\bbeta \bF_{\mu\nu}+\okappa \bN_{\mu\nu\rho}\el^\rho
=2\nabl_{[\nu}{\psA}_{\rho]}\ ,\eqn{6.27}\fe
where ${\psA_\mu}$ is a gauge dependent pseudo-Maxwellian potential
covector given by
\be {\psA}_\mu=\qe\bbeta \bA_\mu+\okappa \bB_{\mu\nu}\el^\nu\ .\eqn{6.28}\fe

If $\el^\mu$ is timelike so that the corresponding symmetry is interpretable
as {\it stationarity}, then the equation of motion (\ref{6.26}) will be 
interpretable as the condition for the string to be in {\it equilibrium} 
with the given values of the constant ``tuning" parameters $\Beta$ and 
$\bbeta$, (of which, as remarked above, the latter, $\bbeta$, will vanish
in the case of an equilibrium that is not just stationary but 
{\it static}, \citeA{Carter90b}). The new result here is that the Joukowski type
``lift" force (which was not allowed for in the previous analysis \citeA{CarterFrolovHeinrich91})
due to the Magnus effect on the string, as it ``flies" (like an aerofoil) 
through the background medium represented by the current 3-form 
$\bN_{\mu\nu\rho}$, has just the same form as an extra Lorentz type 
electromagnetic (indeed in the stationary case purely magnetic) force 
contribution. 

\subsection{Hamiltonian for world sheet generator}
\label{6-4}

When the genuine electromagnetic background coupling, and the similarly acting
Kalb Ramond coupling  are both absent, then (as pointed out
previously \citeA{CarterFrolovHeinrich91}) the equation of motion (\ref{6.26}) (that is the
equilibrium condition for ``steady flight" in the stationary case) is just a
simple  geodesic equation with  respect to, not the actual background
spacetime metric $\bg_{\mu\nu}$, but the conformally modified metric $\Xx^2
\bg_{\mu\nu}$, with the conformal factor $\Xx^2$ determined as a field over the
background by (\ref{6.23}) in conjunction with (\ref{6.17}) and (\ref{6.21}).
Even when the Lorentz and Joukowski force contributions are present, the
equation (\ref{6.26}) governing the propagation of the world sheet generator
$\Xx^\mu$ retains a particularly convenient Hamiltonian form, given by
\be \Xx^\mu={dx^\mu\over d\tu}={\partial \Ha\over\partial{\mPi}_\mu}
\ ,\hskip 1 cm {d{\mPi}_\mu\over d\tu}=-{\partial \Ha\over\partial x^\mu}
\, ,\eqn{6.29}\fe
for the quadratic Hamiltonian function
\be \Ha={_1\over^2}\bg^{\mu\nu}({\mPi}_\mu-\psA_\mu)
({\mPi}_\nu-\psA_\nu) -{_1\over^2}\Xx^2 \, ,\eqn{6.30}\fe
subject to a restraint fixing the (generically non-affine) parametrisation 
$\tu$ of the trajectory by the condition that the numerical value of the
Hamiltonian (which will automatically be a constant of the motion)
should vanish, 
\be \Ha=0\ , \eqn{6.31}\fe
together with a further momentum restraint, determining the (automatically 
conserved) relative transport rate ${\mPi}_\mu \el^\mu$ in accordance 
with the relation (\ref{6.22}) by the condition
\be {\mPi}_\mu \el^\mu=\Beta\bbeta\, .\eqn{6.32}\fe
The Hamiltonian momentum covector itself can be evaluated as
\be {\mPi}_\mu=\Xx_\mu+\psA_\mu= \el^\nu{\calE}_{\nu\rho}\caloT
{^\rho}_{\mu}\, .\eqn{6.33}\fe
In terms of the original conserved generalised momentum fluxes
given by  (\ref{6.7}) and (\ref{6.12}, and the gradients of the scalar (stream 
function and phase) potentials introduced in (\ref{5.27}) and (\ref{5.41}), 
this Hamiltonian covector will be given by 
\be {\mPi}_\mu -\psA_\nu\ag^{\!\nu}_{\,\mu}
={\calE}_{\mu\nu}{^\dagger}\caloP^\nu
= \caloP{^\nu}{\calE}_{\nu\mu}+\Beta\onab_\mu\psie
+\bbeta\she\onab_\mu\phie\, . \eqn{6.34}\fe
(It is to be noted that there were omissions due to transcription errors
in the analogues of the latter formulae in the preceeding version \citeA{Carter95}
of these notes.)

The advantage of a Hamiltonian formulation is that it allows the problem
to be dealt with by obtaining the momentum covector in the form
${\mPi}_\mu=\nabl_\mu \Sa$ from a solution of the corresponding Hamilton 
Jacobi equation, which in this case will take the form
\be \bg^{\mu\nu}\big(\nabl_\mu \Sa-\psA_\mu\big)\big(\nabl_\nu \Sa-\psA_\nu
\big)=\Xx^2\ ,\eqn{6.35}\fe
with $\Xx^2$ given by (\ref{6.23}) via (\ref{6.21}), while the restraint
(\ref{6.32}) gives the condition
\be \el^\nu\nabl_\nu \Sa=\bbeta\Beta\ .\eqn{6.36}\fe

It is to be remarked that the square root $\Xx$ of $\Xx^2$ may be purely imaginary
since the generating vector may be timelike. Instead of working with $\Xx$ it
has been found convenient \citeA{CarterPeterGangui96} to work with a related 
field, $\Upsilo$,  that will be real if the Killing vector is spacelike so
that $\el$ is real, and that is given explicitly by either of the mutually
dual (by (\ref{5.34}) and (\ref{6.21}) identically equivalent) alternative
formulae
\be \Upsilo={\bbeta^2 {\calK}\over \el}-\tilde\Lambde=
{\tilde\bbeta^2\tilde{\calK}\over \el}-\Lambde\el\, .\eqn{6.37}\fe
Considered as a function of the field $\el$ defined by (\ref{6.21}) the
derivative of $\Upsilo$ will be given by 
\be {d\Upsilo\over d\el}=-{\bbeta^2{\calK}\over \el^2}-\Lambde
=-{\tilde\bbeta^2\tilde{\calK}\over\el^2}-\tilde\Lambde\, ,\eqn{6.38}\fe
which can be seen to be equivalent to
\be \el^2 {d\Upsilo\over d\el}= \oT{^{\mu\nu}}\el_\mu \el_\nu \eqn{6.38a}
\, \fe
It can be seen that the field $\Xx^2$ in (\ref{6.35}) will be expressible in
terms of this new self-dual field $\Upsilo=\tilde\Upsilo$ in the form 
\be \Xx^2= {\bbeta^2\tilde\bbeta^2\over\el^2}-\Upsilo^2\, .\eqn{6.39}\fe

\subsection{Free evolution of circular loops}
\label{6-5}

A particularly simple and important application of the formalism that has just
been presented is to case \citeA{Larsen93a}, \citeA{CarterPeterGangui96} of a small circular
string loop undergoing motion that is free in the strong sense, meaning that
there are no external forces of axionic, electromagnetic or even gravitational
origin, i.e for which $\bB_{\mu\nu}$ and $\bA_\mu$ can be taken to vanish, and for
which $\bg_{\mu\nu}$ can be taken to be flat, which will be a realistic
approximation under a wide rang of circumstances, including cosmological
applications in which the loop scale is small compared with the relevant
Hubble length scale. There will be no loss of generality in taking such a
circular loop to lie in the equatorial plane of a spherical coordinate system
with origin at its center of symmetry, in terms of which the metric will have
the standard form 
\be \bg_{\mu\nu}dx^\mu dx^\nu=d\rad^2+\rad^2(d\theta^2 +\sin^2\theta
d\phi^2) -dt^2\ .\eqn{6.40}\fe
This means that 
the loop lies in the hyperplane $\theta=\pi/2$,
with a worldsheet whose internal coordinates could be taken to be $t$ and
$\phi$. The extrinsic location of such a worldsheet will be specifiable
just by giving its radius $\rad$ as a function $t$. 
The Killing vector $\el^\mu$ generating the symmetry of the world
sheet can be conveniently taken to be given by
\be   \el^\mu{\partial\over\partial x^\mu}=2\pi{\partial\over\partial\phi}
\, ,\eqn{6.41}\fe
so that its magnitude $\el$ as given by (\ref{6.21}) will be directly
identifiable with the circumference
\be \el=2\pi \rad \eqn{6.42} \fe
of the loop. The flux ${\calJ}^\mu$ of angular momentum as defined with 
respect to the usual angular momentum generator $\partial/\partial\phi$
will thus be given by the proportionality relation
\be \oP^\mu=2\pi{\calJ}^\mu \, ,\eqn{6.43} \fe
where  $ \oP^\mu$ is the momentum current (\ref{6.2}) associated with $\el^\mu$
(which in this case is the same as that given by (\ref{6.7}), since we are
neglecting the effects of electromagnetic and axionic fields). The
stationarity of the background (though not of the worldsheet) with respect to
the Killing vector $k^\mu$ that generates time translations according to the
specification
\be   \kil^\mu{\partial\over\partial x^\mu}={\partial\over\partial t}
\, ,\eqn{6.44}\fe
ensures that as well as the conserved angular momentum flux (\ref{6.43}) there
will also be a non-trivially conserved mass-energy flux vector that is specified by
the corresponding analogue of (\ref{6.7}) as
\be {\calM}^\mu = -\oT{^\mu}_{\!\nu}\kil^\nu \, .\eqn{6.45}\fe
For each of these independent conserved surface fluxes there will be a 
corresponding conserved constant characterising the loop, namely
its total angular momentum
\be \Ja=\oint dx^\mu\,{\calE}_{\mu\nu}{\calJ}^\nu\, ,\eqn{6.46}\fe
and its total mass
\be \Ma=\oint dx^\mu\,{\calE}_{\mu\nu}{\calM}^\nu\, .\eqn{6.47}\fe

As well as these two quantities, whose conservation depends on the
postulated summetry, there will be two other integrals that will be
conserved in any case, namely the winding number $\Ne$ and the particle
quantum number $\Ze$ that are defined by (\ref{5.60}) and (\ref{5.61}),
which can be seen to be determined in the present case by the Bernoulli
constants (\ref{6.16}), in terms of which they are given by
\be \Ze=\she\bbeta\, \hskip 1 cm 
2\pi\she \Ne=\Beta\, .\eqn{6.48}\fe

The angular momentum and mass integrals can also be expressible by purely
local  formulae, in terms of the the momentum covector ${\mPi}_\mu$ for the
worldsheet generator. Since the effective gauge field $\psA_\mu$ vanishes
in the present application this momentum covector will be given directly by
the formula (\ref{6.18}) as
\be {\mPi}_\mu=\el^\nu{\calE}_{\nu\rho}\oT{^\rho}_\mu
 \, ,\eqn{6.49}\fe
from which it can be seen that the corresponding total angular momentum and 
mass constants will be given by
\be \el^\mu{\mPi}_\mu= 2\pi \Ja\, ,\hskip 1 cm 
k^\mu{\mPi}_\mu=-\Ma\, .\eqn{6.50}\fe
It is to be noticed that only three of the four conserved integrals $\Ja$, $\Ma$,
$\Ze$, $\Ne$ are actually independent: by substituting (\ref{5.31}) in
(\ref{6.49}) it can be seen that the first one is just the product of the last
two, i.e.
\be \Ja=\Ze\Ne\, .\eqn{6.51}\fe
(It it to be remarked that yet another way of obtaining a conserved integral
would to take the adjoint analogue $^\dagger\!\Ja$ of the angular momentum that
is constructed from $^\dagger \oP^\mu$ as given by (\ref{6.10}) instead of
from $\oP^\mu$ as given by (\ref{6.2}) according to the specification
$2\pi\, ^\dagger\!\Ja = \el^\mu{\calE}_{\mu\nu}\,^\dagger\oP^\nu$, but the
result in this case is not independent: it can be seen that the angular
momentum is self adjoint in the sense that $^\dagger \Ja=\Ja$.)

The mass-energy constant (\ref{6.50}) is interpretable in the present case as
the rate of variation of the ordinary coordinate time $t$ with respect to the
Hamiltonian parameter $\tu$ introduced in (\ref{6.29}):
\be {dt\over d\tu}=\Ma\, .\eqn{6.52}\fe
Using a dot to denote differentiation with respect to the coordinate
time $t$, the radial momentum component ${\mPi}_{_1}$ will correspondingly
be given by
\be {\mPi}_{_1}={d\rad\over d\tu}=\Ma\dot \rad\, .\eqn{6.53}\fe
In terms of this quantity the Hamiltonian (\ref{6.30}) will be expressible as
\be \Ha={_1\over ^2}\big({\mPi}_{_1}^{\, 2}+{\Ja^2\over \rad^2}
-\Ma^2-\Xx^2\big)
\, .\eqn{6.54}\fe
It is at this point that the identity (\ref{6.39}) is useful, since it can 
be seen from (\ref{6.48}) and (\ref{6.51}) that it provides the convenient
simplification
\be {\Ja^2\over \rad^2}-\Xx^2=\Upsilo^2\, .\eqn{6.55}\fe
The normalisation constraint (\ref{6.31}) to the effect that the
Hamiltonian should vanish thus provides the required equation of motion
for the loop radius $r$ in the conveniently first integrated form
\be \Ma^2 \dot \rad^2=\Ma^2-\Upsilo^2\, ,\eqn{6.56}\fe
where $\Upsilo$ is the function specified -- independently of $\Ma$ -- by the
Bernoulli constants (i.e. by $\Ze$ and $\Ne$) according to the formula
(\ref{6.37}), whose explicit form depends of course on the equation of state. 

For the case of a string model representing a Witten type
superconducting vortex an appropriate approximation for the equation of
state is provided by the formula (\ref{witseq}) whose application to
this problem has recently been the subject of detailed
investigation \citeA{CarterPeterGangui96} (following a more specialised study of the
zero angular momentum case for which $\Ze$ or $\Ne$ vanishes so that
collapse is unavoidable  \citeA{LarsenAxinedes96}): the function $\Upsilo$ has no
maximum, but for moderate values of the ratio of $\Ze$ to $\Ne$ it has a
finite minimum, and in that case, depending on the particular values of
the three independent constants $\Ze$, $\Ne$, and $\Ma$ the loop may either
oscillate smoothly between a finite maximum and minimum value of the
radius or else it may reach a critical minimum radius at which the
model breaks down due to overconcentration of the current which results
in a local instability:  in the case of a timelike current such an
instability arises because the ``wiggle'' speed $\cE$ given by
(\ref{4.10}) tends to zero, whereas in the case of a spacelike current
it arises because the longitudinal characteristic speed $\cL$
given by (\ref{4.20}) tends to zero.

\subsection{Vorton equilibrium states}
\label{6-6}

It can be seen from the final equation (\ref{6.56}) of the Subsection
\ref{6-5} that the circular loop will have a stationary equilibrium
state characterised by the minimum admissible value of $\Ma$ for given
values of $\Ze$ and $\Ne$ at the radius $r$ for which $\Upsilo$ takes
its minimum value. This happens for $\Ma=\Upsilo$ where
$d\Upsilo/d\rad$ vanishes, which according to (\ref{6.21}) and
(\ref{6.38}) occurs where
\be \bbeta^2\tilde\Lambde\tilde\ch+\tilde\bbeta^2\Lambde\ch=0\, .
\eqn{6.57}\fe
According to (\ref{6.38a}) this is equivalent to
\be \oT{^{\mu\nu}}\el_\mu\el_\nu=0  \eqn{6.57a}\fe
which is interpretable as the condition that the stress in the direction of
the axisymmetry generator should vanish.

Such  centrifugally supported ring states are simple examples of the
phenomenon for which Davis and Shellard \citeA{DavisShellard88}, \citeA{DavisShellard89a} have introduced
 the term ``vorton''. The possibility of forming ``vortons'', meaning
centrifugally supported but not necessarily circular equilibrium states
of cosmic string loops is of considerable potential importance for
cosmology, but it was largely overlooked during the early years of
cosmic string theory.

The formalism embodied in the generator equation (\ref{6.26}) and the
corresponding Hamiltonian (\ref{6.30}) can be applied to the general study of
stationary equilibrium states by replacing the axisymmetry vector $\el$ to
which it was applied in the preceeding subsection by  the Killing vector
$\kil^\mu$ that generates  the stationary symmetry of the background, which, when
substituted for $\el^\mu$ in (\ref{6.18}), will determine a corresponding
world sheet tangent vector 
\be  \sima^\mu=\kil^\nu{\calE}_{\nu\rho}\oT^{\rho\mu}\, ,\eqn{6.58}\fe 
instead of the vector $\Xx^\mu$.

Such an approach provides results of considerable mathematical (as
opposed to physical) interest when applied to the case of stationary
strings in a black hole background. Generalising results obtained
previously \citeA{Frolovetal89}, \citeA{CarterFrolov89} for the Goto Nambu limit case, it has
been shown ~ \citeA{CarterFrolovHeinrich91} that for a string model of the non-dispersive
permanently trasonic type with the constant product equation of state
(\ref{4.21}) that is governed by the Lagrangian (\ref{5.47}) (as
obtained \citeA{Carter90} both from the Nielsen dimensional reduction
mechanism and also \citeA{Carter90a}, as explained in Section 5,  from the
more physically realistic ``wiggly" string approximation) the
stationary Hamilton Jacobi equation is {\it exactly soluble by
separation of variables} in a Kerr black hole spacetime, not just of
the ordinary asymptotically flat kind but even of the generalised
asymptotically De Sitter kind \citeA{Carter73}, \citeA{GibbonsHawking76}. Except in the
Schwarschild-De Sitter limit, where it could of course have been
predicted as a consequence of spherical symmetry, this separability
property still seems rather miraculous, reflecting a ``hidden symmetry"
of the Kerr background that is still by no means well understood. The
newly discovered separability property \citeA{CarterFrolovHeinrich91} is not just an
automatic consequence of the simpler, though when first discovered
already surprising, property of separability for the ordinary geodesic
equation \citeA{Carter68} but depends on a more restrictive  requirement of
the kind needed for the more delicate separability property of the
scalar wave equation \citeA{Carter68a}. (It is however more robust than the
separability properties that have turned out to hold for higher spin
bosonic \citeA{Teukolsky73} and fermionic  \citeA{Unruh73}, \citeA{Chandrasekhar76}, \citeA{Guven80}
wave equations, and other related systems \citeA{Marck83}, \citeA{Carter87}.)

From a physical point of view the mathematically interesting configurations
discussed in the preceeding paragraph are rather artificial. What is much more
important from a cosmological point of view is the the equilibrium of {\it
small  closed string loops} in the mathematically relatively trivial case for
which the background gravitational, electromagnetic, and axionic forces are
{\it negligible}. In the absence of  electromagnetic, or axionic forces, the
relevant generator equation, which is obtained (by substituting $\sigma^\mu$
for $\Xx^\mu$ in (\ref{6.26}) with the source term on the right set to zero)
will have the form
\be \sima^\nu\nabl_\nu \sima_\mu-{_1\over^2}\nabl_\mu 
(\sima_\nu\sima^\nu)= 0\, .\eqn{6.59}\fe
It is evident that the worldsheet generators obtained as solutions of this
equation in a Minkowski background will all just be {\it straight lines} for
the case of a stationarity Killing vector $k$, since in this case (unlike that 
of axisymmetry) $\kil^\nu \kil_\nu$ and hence also $\sima^\nu \sima_\nu$ will
be constant. Since in this case the Killing vector trajectories are also
straight, it might at first seem to follow that the worldsheet of a stationary
string in an empty Minkowski background would necessarily be flat. This
conclusion would exclude the possibility, when on background field is present,
of any closed loop equilibrium states -- such as the circular configurations
characterised by (\ref{6.57}). In fact cosmologists did indeed (albeit for
other reasons) entirely overlook the possibility that such states might exist
until the comparitively recent publication of an epoch making paper by Davis
and Shellard \citeA{DavisShellard88} provided the first counterexamples (the only
previously considered equilibrium
states \citeA{OstrikerThompsonWitten86}, \citeA{CopelandHindmarshTurok87}, \citeA{Copelandetal88}, \citeA{HawsHindmarshTurok88} having been based on a
magnetic support mechanism that was was finally judged to be too feeble to be
effective except \citeA{Peter92b} as a minor correction).

The loophole in the deduction that if the trajectories generated by
$\sima^\mu$ and by $\kil^\mu$ are both straight then the worldsheet must be flat
is that it is implicitly based on the assumption that the two kinds of
trajectories cross each other transversly. However there will be no
restriction on the curvature in the transverse direction in the critical case
for which the two kinds of trajectory coincide, i.e. for which $\sima^\mu$
and $\kil^\mu$ are parallel. The condition for criticality in this sense is
expressible as
\be \sima^\mu{\calE}_{\mu\nu}\kil^\nu=0 \, ,\eqn{6.60}\fe
which can be seen from the definition (\ref{6.58}) to be interpretable as
meaning that the extrinsic characteristic equation (\ref{3.10})
is satisfied by the tangent covector $\chie_\mu={\calE}_{\mu\nu} \kil^\nu$,
which is proportional to $\el_\mu$. This is equivalent to the condition that
the Killing vector $\kil^\mu$ itself should be bicharacteristic, in the sense
of being directed allong the propagation direction of extrinsic perturbations
of the world sheet. The criticality condition (\ref{6.60}) -- which can be
seen to be equivalent to (\ref{6.57}) -- is thus interpretable as a condition
of {\it characteristic flow}. It means that the ``running velocity'', $\vv$ say,
of relative motion of the intrinsicly preferred rest frame of the string (as
determined by whichever of $\ce^\mu$ and $\tilde\ce^\mu$ is timelike) relative to
the background frame specified by the (in this case necessarily timelike)
Killing vector $\kil^\mu$ is the same as the extrinsic propagation velocity
$\cE$ given by (\ref{4.10}), i.e.
\be \vv^2={\Te\over \Ue}\, .\eqn{6.61}\fe

In the presence of generic gravitational, electromagnetic and Kalb Ramond
forces, the criticality condition (\ref{6.60}) can be satisfied at particular
positions, such as where there is a transition from a subcharacteristic
running velocity, $\vv<\cE$ to a supercharacteristic running velocity
$\vv>\cE$ (as will occur for instance on a string in a steady state of
radial flow into a Schwarzschild black hole) or where there is a cusp, with
$\vv=\cE$ but with subcharacteristic flow $v<\cE$ on both
sides. However in view of the constancy of the Bernoulli ``tuning" parameters
$\Beta$ and $\bbeta$, the absence of any background field will always allow,
and generically (the special integrable case (\ref{5.48}) being an exception)
will ensure, {\it uniformity} of the state of the string, so that the
transcharacteristic flow condition (\ref{6.60}) can be satisfied throughout
its length. The space configuration of such a uniformly transcharacteristic
steady string state can have arbitrarily variable curvature, and so is
compatible with a closed loop topology. 

The  value of the mass $\Ma$ at which the equilibrium condition (\ref{6.61})
is attained will of course be proportional to the circumference $\el$ of the
vorton: it can easily be seen that its mass to length ratio will be given by 
\be {\Ma\over\el}= \Ue+\Te= -\Lambde-\tilde\Lambde\, . \eqn{6.62} \fe 
For given values of the conserved quantum numbers $\Ne$ and $\Ze$, the
absolute scale of the vorton will be given by the formula for its length which
takes the comparatively simple form\cite{Carter90z} \be
\el^2={2\pi\vert\Ne\Ze\vert\over\sqrt{\Ue\Te}}
={2\pi\vert\Ne\Ze\vert\over\sqrt{\Lambde\tilde\Lambde}} \eqn{6.63} \, .\fe 
In the ``chiral'' case when the current is approximately null, and more
generally in the magnetic case for which it is spacelike, one expects (see
Subsection \ref{5-7}) the energy density and tension to be given by
$\Ue\approx \Te\approx \mag^2$, where $\mag$ is the relevant Higgs mass scale.
On this basis the two preceeding formulae are sufficient by themselves to
provide reasonably accurate estimates of the vorton length and circumference.
However if the current is timelike, $\Ue$ may become relatively large and
$\Te$ very small compared with $\mag^2$. To estimate their values in a case of
this more general kind, it is necessary to solve the (non linear) equilibrium
equation (\ref{6.57}) which (since our assumption that electromagnetic effects
are unimportant implies $\bbeta=\Beta$) determines the local state as a
function of the ratio of the particle number $\Ze$ to the winding number $\Ne$
by a relation of the form 
\be 2\pi\she^2\calK\sqrt{ {\tilde\Lambde\over\Lambde} } =\Big\vert{\Ze\over\Ne}\Big\vert
\eqn
{6.64}\, .\fe

\subsection{*Allowance for the electromagnetic ``spring'' effect.}
\label{6-7}

Instead of working with the generator tangent vector $\sigma^\mu$ that is
given by the (not so obvious) specification (\ref{6.58}), another way
of deriving the conclusions of Subsection {6-6}, and in particular
the {\it centrifugal equilibrium} condition (\ref{6.61}), is to use an
approach whereby the quantities in the fundamental dynamical
equation (\ref{2.35}) are directly evaluated in terms of the worldsheet
tangent vector $\ie^\mu$ that is given modulo a choice of sign by the (more
obvious) specification that it be orthogonal to the Killing vector $k^\mu$
with unit normalisation,
\be \ie^\nu \kil_\nu=0\ , \hskip 1 cm  \ie^\nu \ie_\nu= 1\, .\eqn{6.66} \fe
Since the Killing vector generator of ordinary stationary smmetry in flat
space is covariantly constant  and can be taken to have unit normalisation
\be \nabl_\mu \kil_\nu=0\ , \hskip 1 cm \kil^\nu \kil_\nu=-1\, ,\eqn{6.67}\fe
the first fundamental of the stationary string worldsheet will be expressible
by
\be \og^{\mu\nu}= -\kil^\mu \kil^\nu +\ie^\mu \ie^\nu\, ,\eqn{6.68}\fe
so that by (\ref{1.10}) the second fundamental tensor will be obtainable
simply as
\be \Ke_{\mu\nu}^{\, \ \ \rho}= \ie_\mu \ie_\nu  \Ke^\rho\, ,\eqn{6.69}\fe
where
\be \Ke^\rho = \ie^\sigma\nabl_{\!\sigma} \ie^\rho \,  .\eqn{6.70}\fe
Under these conditions, the preferred orthonormal diad introduced in 
Section \ref{5-2} will be expressible in the form
\be \ue^\mu=(1-\vv^2)^{-1/2}\big(\kil^\mu + \vv \ie^\mu\big)\, \hskip 1 cm
\ve^\mu=(1-\vv^2)^{-1/2}\big(\ie^\mu + \vv \kil^\mu\big)\, ,\eqn{6.71}\fe
where $\vv$ is the longitudinal ``running velocity'' characterising the rate
motion of relative to the stationary background of the internally preferred
rest frame of the string that is determined generically by whichever of $\ce^\mu$
and $\tilde\ce^\mu$ is timelike. Using the expression (\ref{2.15}) for
the surface stress-energy tensor, the left hand side of the fundamental
dynamical equation (\ref{2.35}) can thus be evaluated as
\be \oT{^{\mu\nu}} \Ke_{\mu\nu}^{\, \ \ \rho}=
\Big( {\Ue \vv^2-\Te  \over 1-\vv^2}\Big) \Ke^\rho
\, . \eqn{6.72}\fe

In the absence of external -- electromagnetic or Magnus type -- forces,
the condition for equilibrium is that the vector given by (\ref{6.68})
should vanish. This requirement can be satisfied in one or other of two
alternative manners. One way is for the  configuration is to be {\it
straight}, so that $\Ke^\rho=0$, in which case any value (including the
{\it static} case $\vv=0$) is possible for $\vv$.  However a straight
configuration is possible only if the string is infinite or else
terminates at fixed endpoints. For a closed string loop in flat space
the curvature vector $\Ke^\rho$ must necessarily be non-zero at least
along part of the configuration, and in this case it is evident that
the only way for the equilbrium condition to be satisfied is for the
running velocity to satisfy the transcharacteristic condition
(\ref{6.61}), in which case there will be no restriction at all on the
curvature vector $\Ke^\rho$.  Since by (\ref{4.10}), the centrifugal
equilibrium condition (\ref{6.61}) is interpretable as meaning that the
running velocity must be equal to the velocity of transverse
(``wiggle'') type perturbations, it can be seen to follow that
relatively backward moving perturbations will have their flow speed
exactly canceled out, i.e.  they will just be static deformations,
which explains why, in this transcharacteristic case the curvature
$\Ke_\rho$ is unrestricted.

The formula (\ref{6.72}) will of course still be applicable
in more general cases, for which the dynamical equation (\ref{2.35})
has a non vanishing right hand side expressing the effect of 
an electromagnetic or Magnus type force (the latter being responsible
for the equilibrium of an ordinary ``smoke ring'', which is the
most familiar laboratory example of a vorton type equilibrium state).
In a cosmological context particular interest attaches to the case
of electromagnetic self interaction arising in the case of a current
with a non negligible electromagnetic charge coupling. Unlike the effect of
a Magnus force or a purely external magnetic field which are 
straightforwardly tractable within the framework developed here, 
allowance for electromagnetic self interaction raises an awkward and
delicate problem because the self induced field is singular on
the worldsheet, so that its treatment (as in the more familiar case
of a particle self interaction) requires the use of some sort of
renormalisation procedure, typically involving a cut off length representing
the finite thickness of the worldsheet in a microscopic description.

No satisfactory treatment of electromagnetic (or gravitational) self
inetaction has yet been developed for the general, dynamically evolving case,
but for the very special case of a stationary circular string loop an adequate
analysis of the  corrections arising from ``spring'' effect due to
electromagnetic self interaction has been provided by Peter \citeA{Peter93a}. 
This treatment involves the introduction of a cut off length, $\rast$, say that
represents the effective thickness of the microscopic current destribution in
the string and that is expected to be of the order of magnitude of the Compton
wavelength associated with the Witten carrier mass scale $\mast$ that was
introduced in Section \ref{5-7}, which (in Planck units) means $\rast\approx
1/\mast$. In the limit when this radius goes to zero the electromagnetic
field is logarithmically divergent, its effective mean value within the string
worldsheet being given in terms of the surface current vector  $\oj{^\mu}$
by an expression of the form
\be  \bF_{\mu\nu}=- 2 \ln\Big\{ {\rad\over \rast}\Big\}
\big(\Ke_{[\mu} \kil_{\nu]}\kil_\sigma
+\Ke_{[\mu} \ie_{\nu]} \ie_\sigma\big)\oj{^\sigma} \, ,\eqn{6.73}\fe
where $\rad$ is the the radius of the circular configuration under 
consideration, which will be given in terms of the corresponding curvature
vector by 
\be \Ke^\rho \Ke_\rho ={1\over \rad^2}\, .\eqn{6.74}\fe 
In terms of the number density current vector $\ce^\mu$ and the
associated momentum vector $\pe_\mu$ as given by (\ref{5.27}) and
(\ref{5.26}) the corresponding electromagnetic force vector will be
given by
\be \bF_{\mu\nu}\oj{^\nu}=-\qe^2 \, {\rm ln}\Big\{ {r\over \rast}\Big\}
\big( (\ce^\nu \ie_\nu)^2 + (\pe_\nu \ie^\nu)^2) \Ke_\mu \, ,\eqn{6.75}\fe
where $q^2$ is the effective charge coupling constant which will be
given in terms of the charge per fundamental particle, $\ee^2\simeq 1/137$
by $\qe^2=\she^2 \ee^2$ where $\she$ is the normalisation factor
introduced in Subsection \ref{5-6}, which is expected to be of order unity.

It is convenient to rewrite this in terms of the modulus $\vert \ch\vert$
of the squared current magnitude $\ch
=\ce^\mu \ce_\mu $ of the current vector, which, according to
the results described in Subsection \ref{5-7}, is expected to be subject
to a saturation limit given in terms of the Witten mass scale
$\mast$ by $\ch\lta \mast^{\,2}$ . Setting the resulting expression
\be   \bF_{\mu\nu}\oj{^\nu}=- \qe^2\vert \ch\vert\Big(
{1+\vv^2\over 1-\vv^2}\Big)\ln \Big\{ {\rad\over \rast}\Big\} \Ke_\mu\, , 
\eqn{6.76}\fe
equal to the term (\ref{6.72}) in accordance with the fundamental
dynamical equation (\ref{2.35}), it can be seen that the curvature
covector $\Ke_\rho$ cancels out again, as in the electromagnetically
uncoupled case. The non-trivial equilibrium condition that remains is
Peter's generalisation \citeA{Peter93a} of the preceeding formula
(\ref{6.61}) for the running velocity $\vv$, which is expressible in the
form
\be \vv^2={\Te- \crec^2\over \Ue+\crec^2} \, ,\eqn{6.77}\fe
where the correction term allowing for the electromagnetic self interaction
is given by
\be \crec^2=\qe^2\vert \ch \vert\, {\rm ln}\Big\{ {r\over\rast}\Big\} \, .
\eqn{6.78}\fe

In any realistic cosmic string model of the kind described in
Subsection \ref{5-7} it is to be anticipated that we shall always have
\be \ch\lta \mast^2\lta \mag^2\lta \Ue \, .,\eqn{6.79}\fe
and one also expects to have
\be \qe^2\approx \ee^2\simeq 1/137 \, .\eqn{6.80}\fe 
More specifically, in the ``magnetic'' range, $\ch>0$, for any such
model, it is to be expected that the tension $\Te$ will remain
comparable with the energy density $\Ue$, which implies that the
condition \be \crec^2\ll \Te \approx \mag^2\approx \Ue \, \eqn{6.81}\fe
will be satisfied -- unless the logarithmic factor is enormous, which
is only conceivable for implausibly large values of the radius (too
large for the ring to be treated as an isolated system unaffected by
external matter).  The electromagnetic force contribution will
therefore be relatively negligible, not only in the ``chiral'' case,
i.e.  when current is approximately null, but also in the ``magnetic''
case considered in early work, i.e.  when the current is spacelike.
This means that the original notion \citeA{CopelandHindmarshTurok87},
\citeA{Copelandetal88}, \citeA{HawsHindmarshTurok88} of non rotating
magnetostatically supported ``cosmic spring'' configurations was
unrealistic\citeA{Peter93}.

On the other hand however,  as Peter has pointed out \citeA{Peter93a},
if the definition of a ``cosmic spring'' is extended to include
non-rotating configurations with {\it electrostatic} as opposed to
magnetostatic support, then the preceeding objection no longer
applies.  This is because, unlike the spacelike case, the tension can
become relatively small, $\Te\ll \Ue$, in the case of a timelike
current for a cosmic sting model of the kind described in Subsection
\ref{5-7}.  While the existence of exactly non rotating equilibrium
states which, according to (\ref{6.77}), would be characterised by the
very special electrostatic support condition
\be \Te=\crec^2 \, .\eqn{6.82}\fe
will in principle be possible,  nevertheless the formation of such
``spring'' type vortons would presumably be statistically rare under
natural conditions.  There will however be a more reasonable chance for
vortons to be formed in the extended range of nearby slowly rotating
states in which, if not actually dominant, electrostatic support does
provide a contribution that may be non negligible compared with the
purely centrifugal support mechanism considered Subsection \ref{6-6}.
Although one would expect that such very slowly rotating states would
initially represent only a small fraction of the total vorton
population, they might come to be of dominant importance in the long
run, due to the greater vulnerablity of supersonically rotating states
to secular and even dynamical instability \citeA{CarterMartin93},
 \citeA{Martin94}, \citeA{MartinPeter95}.

\subsection{*Vortons in cosmology}
\label{6-8}

The question of closed loop equilibrium states did not arise in the earliest
studies \citeA{Kibble80} of cosmic strings, which were restricted to the Goto-Nambu
model whose bicharacteristics are always null and so can never be aligned with
the timelike Killing vector generating a stationary symmetry. However in a generic 
string model \citeA{Carter89} for which the 2-dimensional longitudinal Lorentz of
the internal structure is broken by a current, whether of the neutral kind
exemplified in an ordinary violin string or the electromagnetic kind exemplified 
 \citeA{BabulPiranSpergel88}, \citeA{Peter92}, \citeA{Peter92a}, \citeA{Peter93} by 
Witten's superconducting vacuum vortex model, the bicharacteristic directions will
generically be timelike \citeA{Carter89a} (spacelike bicharacteristics being
forbidden by the requirement of causality) so there will be no obstacle to
their alignment with a timelike Killing vector in accordance with the
criticality condition (\ref{6.60}), i.e. to having a running velocity given by
$\vv= \cE$. The simple ``toy" complex scalar field model on which the
pioneering cosmic string studies \citeA{Kibble76} were based had longitudinally
Lorentz invariant vortex defects (of ``local" or ``global" type depending on
the presence or absence of coupling to a gauge field) that were describable at
a microscopic level by string models (with Kalb Ramond coupling in the
``global" case) that were indeed of the special Goto-Nambu type. However the
extra degrees of field freedom (starting with the additional scalar field
introduced by Witten in his original superconducting example \citeA{Witten85}) that
are needed in successively more realistic models \citeA{Peter92b}, \citeA{PerkinsDavis93} make
it increasingly difficult \citeA{Peter94}, \citeA{DavisPeter95} -- though it may still be 
possible \citeA{GoodbandHindmarsh96} -- to avoid the formation of internal structure
breaking the longitudinal Lorentz invariance and reducing the extrinsic
characteristic velocity to the subluminal range $\cE<1$ at which
stationary equilibrium with the critical running velocity $\vv=\cE<1$
becomes possible.

The cosmological significance of this is that whereas Goto Nambu string
loops cannot ultimately avoid gravitationally or otherwise radiating away
all their energy \citeA{Turok84}, \citeA{VachaspatiVilenkin85}, \citeA{Durrer89}, since they have no
equilibrium states into which they might settle down, on the other hand more
general kinds of strings, whose occurrence would now seem at least as
plausible \citeA{Peter92b}, \citeA{PerkinsDavis93}, \citeA{Peter94}, \citeA{DavisPeter95}, can leave a relic 
distribution of stationary loop configurations that may survive indefinitely
\citeA{DavisShellard88}, \citeA{DavisShellard89a}, \citeA{Davis88}, \citeA{Carter90c}, \citeA{Carter91}, \citeA{Carter95b}, \citeA{Brandenbergeretal96}. 
One would expect such configurations to be those that minimise the energy for
given values of the relevant globally conserved quantities, of which
there might be a considerable number in the more complicated multiply
conducting models\citeA{Carter94b} that might be considered, but of
which there are only a single pair in the ``barotropic" type string
considered here, namely the quantities $\Ze$ and $\Ne$ defined in
Subsection \ref{5-6}, which are proportional to the stream function
winding number $\oint d\psie$ and the phase winding number $\oint
d\phie$, whose respective constancy results from the conservation of
the mutually dual pair of currents $\ce^\mu$ and $\tilde\ce^\mu$.  If
the equilibrium is predominantly due to a single current with
corresponding particle and winding numbers $\Ze$ and $\Ne$, then
according to the formulae (\ref{6.62}) and (\ref{6.63}), the
corresponding vorton mass $\Ma$ and length $\el$ will have an order of
magnitude given in terms of the relevant Higg's mass scale $\mag$ by
rough estimates of the form
\be \Ma\approx\vert\Ze\Ne\vert^{1/2}\mag\, , \hskip 1 cm
\el\approx \vert\Ze\Ne\vert^{1/2}\mag^{-1} \, ,\eqn{6.83}\fe
whenever the current is spacelike (the ``magnetic'' case), or approximately
null (the ``chiral'' case), so that one has $\Te\approx\Ue\approx \mag^2$. The
exceptions -- for which considerably higher values may be obtained -- are very
slowly rotating low tension states,  with a strong timelike current, which
according to (\ref{6.64}) will occur only for $\Ze^2\gg\Ne^2$: this would
presumably be comparatively rare on the basis of random statistical
fluctuations, though it might ultimately be favoured by selection if, as is
not implausible, the initially more common (supersonically rotating) states
with $\Ze^2$ comparable to $\Ne^2$, turn out to be relatively unstable in the
long run.

The potential cosmological importance of such a distribution of stationary
relic loops, referred to as ``vortons", was first pointed out by Davis and
Shellard \citeA{DavisShellard88}, \citeA{DavisShellard89a}  \citeA{Davis88}, who emphasised that in the
case of the ``heavyweight" cosmic strings whose formation during G.U.T.
symmetry breaking had been postulated to account for galaxy formation, the
density of the ensuing ``vortons" would be more that sufficient to give rise
to a catastrophic cosmological mass excess (of the kind first envisaged as
arising from the formation of monopoles) even if they were formed with very
low efficiency. This prediction was implicitly based on the assumption that
the Witten carrier mass scale, represented by the parameter $\mast$ in
(\ref{witseq}), was of the same order of magnitude as the relevant Kibble mass
scale, $\mag$, determined by the Higgs field, as would be the case if both the
string formation and the current condensation were part of the same G.U.T.
symmetry breaking phase transition, for which the corresponding gravitational
coupling constant would be given by $\sqrt{\NG\mag^2}\approx 10^{-3}$. 

Subject to the foregoing assumption -- i.e. that the Witten mass scale
$\mast$ was not much smaller than the Kibble mass scale $\mag$ -- early
quantitative estimates suggested \citeA{Carter90c}, \citeA{Carter91}
that, to avoid a cosmological mass excess if the vortons were formed
with relatively high efficiency, the relevant phase transition energy
scale could not have been much greater than a limit given very roughly
by $\sqrt{\NG\mag  ^2}\approx 10^{-13}$, which is not so very much
larger than the scale of electroweak unification. This limit has
however been subject to upward revision by more recent and detailed
investigations, including allowance for more realistic -- i.e. lower --
estimates of the efficiency of vorton formation \citeA{Carter95b},
\citeA{Brandenbergeretal96}; the limit is further relaxed if one drops
the supposition that the Witten mass $\mast$ should be close to the
Kibble mass scale $m$, so that provided $\sqrt{\NG\mast^{\,2}}\gta
10^{-10}$ string formation at the G.U.T.  scale
$\sqrt{\NG\mag^{2}}\approx 10^{-3}$ is cosmologically admissible after
all even if the ensuing vortons are stable enough to survive to the
present epoch.  
%As can be seen from the diagram\ref{FigureBCDM96a}
%plotting the relative mass contribution of vortons against the
%cosmological epoch of the current condensation as determined by the
%carrier mass scale $\mast$, 
If the vortons are only stable enough to
last for a few minutes, the condition that the standard nucleosynthesis
process should not have been disrupted provides a somewhat weaker limit
given \citeA{Brandenbergeretal96} roughly by $\sqrt{\NG\mast^{\,2}}\gta
10^{-12}$. It follows from this work that if the current condensation
does not occur until the epoch of electroweak symmetry breaking then
the ensuing vortons would contribute at most a very small fraction of
the present cosmological closure density. This conclusion applies in
particular to scenarios in which the relevant strings are themselves
formed during electroweak symmetry breaking -- which would happen only
in a non-standard (e.g. supersymmetric) theory, since the standard
Glashow Weinberg Salam version, does not give rise to stable string
like vortex defects \citeA{JamesPerivolaropoulosVachaspati93}.  However
even if their density is low by cosmological standards it does not
necessarily follow that the vortons would be undetectable: for example
it has recently been suggested \citeA{BonazzolaPeter97} that they might
account for otherwise inexplicable cosmic ray observations.

The potential importance of such cosmological and astrophysical effects
provides the motivation for more thorough investigation of equilibrium states
that may be involved, a particularly important question being that of their
stability. Prior to the derivation of the general symmetric string generator
equation (\ref{6.26}), the only closed loop equilibrium states to have been
considered were the circular kind to which the use by Davis and 
Shellard  \citeA{DavisShellard88}, \citeA{Davis88} of the term ``vorton" was
originally restricted. The first general investigation \citeA{Carter90}, \citeA{Carter90z} 
of such circular ``cosmic ring" states showed that under conditions of purely
centrifugal support (neglecting possible electromagnetic corrections of the
kind evaluated more recently \citeA{Peter93a}) the condition (\ref{6.61}) for
equilibrium, namely the requirement of a transcharacteristic rotation speed
$\vv=\cE$, is such that the ring energy is minimised with respect to
perturbations preserving the circular symmetry. However a more recent
investigation of non-axisymmetric perturbations has shown that although there
are no unstable modes for states of subsonic rotation \citeA{CarterMartin93} (as
exemplified by a cowboy's lassoe loop) with $\vv=\cE<\cL$,
instability can nevertheless occur for rotation in the supersonic 
regime \citeA{Martin94}, \citeA{MartinPeter95} that (contrary to what was implicitly assumed 
in earlier work
 \citeA{SpergelPiranGoodman87}, \citeA{VilenkinVachaspati87a}, \citeA{CopelandHindmarshTurok87}, \citeA{Copelandetal88}, \citeA{HawsHindmarshTurok88}, \citeA{SpergelPressScherrer89}, \citeA{Amsterdamski89} 
using the subsonic type of model given by a linear equation of state for which
the sum $\Ue+\Te$ is constant) has been shown by
Peter \citeA{Peter92}, \citeA{Peter92a}, \citeA{Peter93} to be relevant in the kind of cosmic
vortex defects that have been considered so far. Although the first category
of string loop equilibrium states to have been studied systematically has been
that of circular ring configurations \citeA{Carter90z}, it has been made clear by
recent work \citeA{CarterFrolovHeinrich91}, \citeA{CarterMartin93} that, as explained above, arbitrary non
circular equilibrium states are also possible. The stability of such more
general equilibrium states has not yet been investigated. While it seems
plausible that some kinds of ``vorton" relic loops may be destroyed by the
recently discovered classical instability mechanism \citeA{CarterMartin93}, \citeA{Martin94},
and also the kind of quantum tunnelling instability mechanism considered by
Davis \citeA{Davis88}, it does not seem likely that such mechanisms could be so
consistently efficient as to prevent the long term survival of a lot of other
``vorton" equilibrium states.

\section{Acknowledgements}

I wish to thank B. Allen, C. Barrab\`es, U. Ben-Ya'acov, A-C. Davis, 
R. Davis, V. Frolov, G. Gibbons, R. Gregory, T. Kibble, K. Maeda, X. Martin, 
P. Peter, T. Piran, D. Polarski, M. Sakellariadou, P. Shellard, 
P. Townsend, N. Turok, T. Vachaspati, and  A. Vilenkin, for many
stimulating or clarifying discussions.

%\bibliographystyle{unsrt}
%\bibliographystyle{apacite}
%\bibliography{Ca}

%\end{document}

%\nonumsection
%{References}

\end{document}